%% file: article_main.tex
\documentclass[useAMS,usenatbib]{mn2e}

\usepackage{graphicx}
\usepackage{color}

\input{article_title}

\def\LaTeX{L\kern-.36em\raise.3ex\hbox{a}\kern-.15em
    T\kern-.1667em\lower.7ex\hbox{E}\kern-.125emX}

\begin{document}

\label{firstpage}

\maketitle

\input{article_abstract}

\input{article_introduction}

\input{article_models}

\input{article_simulations}

\input{article_metallicity_profiles}

\input{article_stellar_migration}

\input{article_conclusion}

\section*{Acknowledgements}
We wish to thank the anonymous referee for the many helpful comments and
suggestions which improved this paper. We also thank Volker Springel for
making publicly available the {\sc Gadget-2} simulation code. JS would
like to thank the Fund for Scientific Research - Flanders (FWO). MK is a
postdoctoral Marie Curie fellow (Grant PIEF-GA-2010-271780) and a fellow
of the Fund for Scientific Research- Flanders, Belgium (FWO11/PDO/147).


\input{article_bibliography}

\label{lastpage}

\end{document}

%% file: article_title.tex
\title[Survival of metallicity gradients in dwarfs] {Stellar orbits and the survival
  of metallicity gradients in simulated dwarf galaxies} \author[J. Schroyen et al.]
  {J.~Schroyen$^1$\thanks{JS thanks the Fund for Scientific Research -
  Flanders, Belgium (FWO) for financial support, E-mail:
  Joeri.Schroyen@UGent.be.}, S.~De Rijcke$^1$\thanks{SDR, ACO, and BV
  thank the Special Research Fund (BOF) of Ghent University for
  financial support}, M.~Koleva$^1$\thanks{MK is a Marie Curie fellow},
  A. Cloet-Osselaer$^1$\footnotemark[2],
  B. Vandenbroucke$^1$\footnotemark[2]\\ $^1$Sterrenkundig
  Observatorium, Ghent University, Krijgslaan 281, S9, 9000 Gent,
  Belgium} \date{Accepted. Received ; in original form}

\pagerange{\pageref{firstpage}--\pageref{lastpage}} \pubyear{2012}

%% file: article_abstract.tex
\begin{abstract}

We present a detailed analysis of the formation, evolution, and possible
longevity of metallicity gradients in simulated dwarf galaxies. 
Specifically, we investigate the role of potentially orbit-changing
processes such as radial stellar migration and dynamical heating in
shaping or destroying these gradients. We also consider the influence
of the star formation scheme, investigating both the low density star
formation threshold of $ 0.1\, \mathrm{amu\,cm^{-3}} $, which has been
in general use in the field, and the much higher threshold of $ 100~%
\mathrm{cm^{-3}} $, which, together with an extension of the cooling
curves below $10^{4}$\,K and and increase of the feedback efficiency,
has been argued to represent a much more realistic description of star
forming regions.

The Nbody-SPH models that we use to self-consistently form and evolve
dwarf galaxies in isolation show that, in the absence of significant
angular momentum, metallicity gradients are gradually built up during
the evolution of the dwarf galaxy, by ever more centrally concentrated
star formation adding to the overall gradient. Once formed, they
are robust and can easily survive in the absence of external
disturbances, with their strength hardly declining over several Gyr, and
they agree well with observed metallicity gradients of dwarf galaxies in
the Local Group. The underlying orbital displacement of stars is quite
limited in our models, being of the order of only fractions of the half
light radius over time-spans of 5 to 10\,Gyr in all star formation
schemes. This is contrary to the strong radial migration found in
massive disc galaxies, which is caused by scattering of stars off the
corotation resonance of large-scale spiral structures. In the dwarf
regime the stellar body only seems to undergo mild dynamical heating,
due to the lack of long-lived spiral structures and/or discs.

The density threshold, while having profound influences on the star
formation mode of the models, has only an minor influence on the
evolution of metallicity gradients.  Increasing the threshold
1000-fold causes comparatively stronger dynamical heating of the stellar
body due to the increased turbulent gas motions and the scattering of
stars off dense gas clouds, but the effect remains very limited in
absolute terms.

\end{abstract}

\begin{keywords}
galaxies: dwarf -- galaxies: evolution -- galaxies: formation --
methods: numerical.
\end{keywords}

%% file: article_introduction.tex
\section{Introduction}

Comparing observations to simulations is a powerful approach to studying
the physical processes involved in the formation and evolution of
galaxies. Dwarf galaxies, in particular, are ideal probes for
this. Their low masses and small sizes allow hydrodynamical simulations
to reach high spatial resolutions. For the same reasons, they are also
very sensitive to the effects of star formation (e.g. supernova
explosions), contrary to massive galaxies. And at least the dwarfs
within the Local Group can be studied in great depth, thus providing
sufficiently detailed data to compare the high resolution simulations
to. The results derived from studying dwarf galaxies are of
direct relevance to galaxy evolution in general.


Stellar population gradients, i.e. the radial variation of metallicity
and age projected along the line of sight, offer direct insights into
the past star-formation and metal enrichment histories of galaxies.
Observationally, there is an ever growing collection of dwarf galaxies
with gradients, found both in the Local Group (\citealt{alard01,
harbeck01, tolstoy04, battaglia06, battaglia11, battaglia12, bernard08,
kirby11, kirby12, monelli12}) and in nearby galaxy clusters and groups
(\citealt{koleva09, koleva11, chilingarian09, crnojevic10, lianou10,
brok11}). These encompass objects of different masses and star formation
modes (spheroidals, ellipticals, star-forming, quiescent, transitional
type, ...), in different environments (from isolated to densely
populated), with different gradient formation histories (slowly built-up
gradients, \citealt{battaglia06,battaglia12}; or already present in the
oldest populations, \citealt{tolstoy04, bernard08, koleva09}), which are
investigated with different techniques. Likewise, on the
theoretical/numerical side, there is a physical foundation for the
existence of metallicity gradients in both massive galaxies \citep[][and
references therein]{ bekki99, hopkins09, pipino10} and for dwarf
galaxies \citep{sander:dgmodels, stinson2009, joeri:angmomentum,
lokas12}, but see \citet{revaz12}.

This prompts the question of how and when these gradients are
formed and how, once formed, they can be maintained. These are the
topics we want to investigate in the current paper. More specifically,
we want to address whether radial displacements of stars through
orbit-changing processes (such as dynamical heating by scattering of
stars or radial stellar migration through interactions with spiral-like
structures) play a role in potentially erasing or weakening any
pre-existing population gradients in dwarf galaxies.

This paper is structured as follows. We give a description of our
numerical models and methods in Section \ref{section_models}, and
present the simulations in Section \ref{section_simulations}. General
influences of the model parameters are summarized in Section
\ref{section_ldtvshdt} before going on to the more specific results. In
Section \ref{section_metallicity_profiles} we analyse the evolution of
the metallicity profiles in our model dwarf galaxies, and compare them
to observed metallicity gradients in the Local Group. Section
\ref{section_stellar_migration} looks into the underlying orbits and
kinematics in the stellar body of our simulated dwarf galaxies, and
tries to connect these to the findings on metallicity gradients. We
summarize and conclude in Section~\ref{section_conclusion}.

%% file: article_models.tex
\section{Models and methods}
\label{section_models}

\subsection{Codes}

For this work, we rely mainly on two codes. Firstly, we use the
Nbody-SPH code {\sc Gadget-2} \citep{springel05} for the actual
simulating, providing us with snapshots of the model dwarf galaxies -
which contain positions, masses and velocities for all particles, and
properties specific to stellar and gaseous particles such as ages,
metallicities and densities. Secondly, we developed our own in-house
analysis tool {\sc Hyplot} for analysing and visualizing these data
files, which is able to calculate any desired physical quantity of the
dwarf galaxy model, and also transform them to mimic observational
quantities.

\subsubsection{Simulation code}
The simulation code we use is a modified version of the Nbody-SPH code
{\sc Gadget-2} \citep{springel05}. To the freely available basic
version, which only incorporates gravity and hydrodynamics, we added
several astrophysical extensions. These consist of radiative cooling,
star formation, feedback processes and chemical enrichment - which
altogether give rise to the self-regulating chemical evolution cycle in
galaxies. We did not include a cosmic UV background. In order not
to add more free parameters to the formalism, we opted not to include a
prescription for metal diffusion \citep{shen10,gibson13}. Judging from
figure 9 from \citet{shen10}, the gas enriched by metal diffusion
achieves metallicities that are more than an order of magnitude smaller
than the ``enriching'' material. We therefore expect that the inclusion
of diffusion would only have a minor effect on our results. More
detailed information about the basic implementations of these processes
can be found in \citet{sander:dgmodels,joeri:angmomentum}, and the
relevant modifications we have made to them are described further in
this section.

\subsubsection{Analysis and visualisation}
For the analysis, we used our own {\sc Hyplot} package, which is freely
available on
SourceForge\footnote{http://sourceforge.net/projects/hyplot/}. It is an
analysis/visualisation tool especially suited for Nbody-SPH simulations
(currently only for {\sc Gadget-2} datafiles, but easily extensible to
any data format), written mainly in Python and C++. PyQT and Matplotlib
are used for the GUI and the plotting, and it is also fully scriptable
in Python. All analyses, plots and visualisations in this paper have
been made using {\sc Hyplot}.

\subsection{Initial conditions}

As in \citet{sander:dgmodels,joeri:angmomentum,annelies}, the
simulations start at redshift 4.3 from an isolated spherically symmetric
initial setup, consisting of a dark matter halo with an initial density
profile, and a gas sphere that can be given an initial rotation
profile. See \citet{annelies} for a detailed account of the cosmological
motivation for our initial conditions.  Subsequently, the gas cools,
collapses into the gravitational well of the dark matter, and is able to
form the stellar body of the galaxy through a self-consistent
description of star formation, feedback and metal enrichment.

Stellar feedback alone is unable to unbound the gas from the
potential well of the simulated dwarf galaxies so they retain at least
part of their gas reservoirs throughout the simulations. In
\citet{joeri:angmomentum}, it was shown that high-angular-momentum dwarf
galaxies foster continuous star formation and are dwarf-irregular-like
while low-angular-momentum dwarf galaxies have long quiescent periods in
between star-formation events. During these lulls, these dwarfs would be
classified as dwarf spheroidals or dwarf ellipticals (although
containing gas for future star formation). We do not aim specifically at
simulating either late-type dwarfs or early-type dwarfs:~their
classification depends on when during their star-formation they are
observed, a view which is also advocated in \citet{koleva13}. Within
this unified picture, a meaningful comparison of our simulations with
observations of both early-type and late-type dwarf galaxies is
possible.

\subsubsection{Dark matter halo}
\label{section_dm_halo}

The dark matter (DM) halo in our dwarf galaxy models, which is implemented as
a ``live'' halo, serves as the background potential in which the
baryonic matter collapses. The density profile is set to a cusped NFW
profile \citep{nfw} dark matter halo:
\begin{equation}
\rho_{dm}(r) = \frac{\rho_{s}}{\frac{r}{r_{s}}(1+\frac{r}{r_{s}})^{2}},
\end{equation}
with radius $r$ and characteristic parameters $\rho_{s}$ and $r_{s}$
(see Fig.~\ref{IC_profiles}). In \citet{annelies} the details of the
implementation are discussed, with the difference that instead of using
the parameter correlations from \citet{wechsler02} and \citet{gentile04}
we now use those from \citet{strigari2007}, since they are derived
specifically for low-mass systems.

It is convincingly shown in \citet{annelies} how the central dark-matter
cusp is naturally and quickly flattened through gravitational
interactions between dark and baryonic matter: 
 \begin{enumerate}
\item when collapsing into the DM potential well the central density of
      the gas gradually increases, hereby adiabatically compressing the
      center of the DM halo
\item when the gas density reaches the threshold for star formation (see
      Section \ref{section_hdt}), stellar feedback causes a fast removal
      of gas from the central regions of the DM halo, to which the DM
      halo responds non-adiabatically by a net lowering of the central
      density.
 \end{enumerate}
A cusped density profile will thus naturally evolve into a more cored
density profile if a self-consistent star formation cycle is included in
the model, as shown in the inset in Fig.~\ref{IC_profiles}. 

This flattening effect is seen in all our models, with its strength
related to the value of the density threshold (ranging from $
\mathrm{0.1~to~100\,cm^{-3}}$, see \ref{section_hdt}). While this cusp
flattening effect has also been found by other authors, such as
\citet{governato10}, its strength and the sizes of the formed cores may
vary between authors. This can be explained by the fact that different
authors use different star-formation density thresholds and that some,
like \citet{governato10}, select dwarf galaxies from fully cosmological
simulations of a larger volume of the universe. The latter type of
simulations includes the dwarf galaxies' full formation and merger
history. This is much more disruptive than what we employ here. Our
simulations start with an idealized, isolated setup where all the
baryonic matter is already present and free to collapse into the DM
potential.

\subsubsection{Gas sphere}
\label{section_gas_sphere}

\begin{figure}
 
\includegraphics[width=0.45\textwidth]{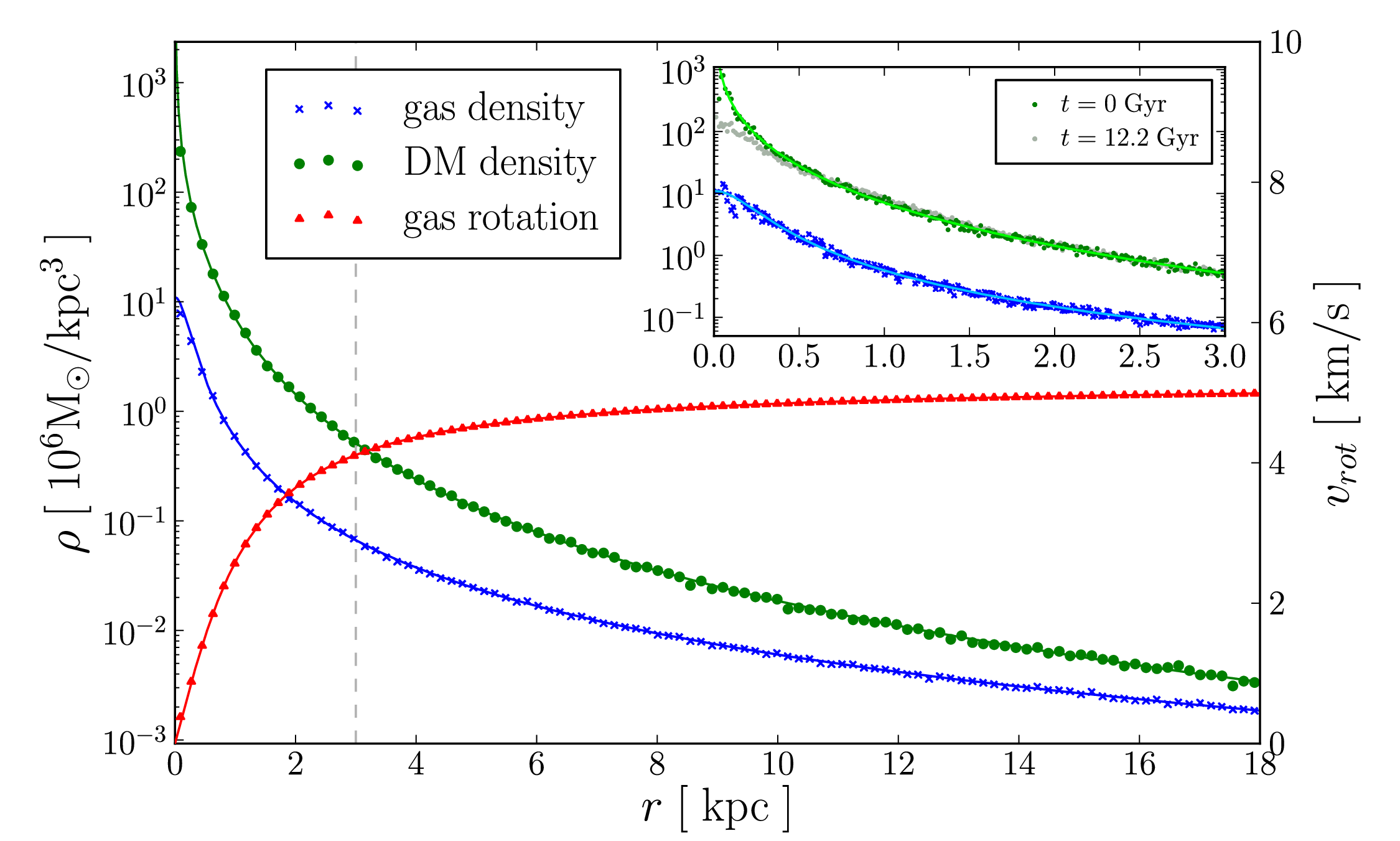}
 \caption{Initial conditions of our dwarf galaxy '05' model
 \citep{sander:dgmodels}. Several radial profiles are shown: the blue
 and green are the density profiles of, respectively, the gas and dark
 matter haloes (left $y$-axis) - red shows the rotational velocity
 profile of the gas (right $y$-axis). Full lines represent the
 theoretical curves (found in Paragraphs \ref{section_dm_halo},
 \ref{section_gas_sphere}, \ref{section_rotation_curves}) while markers
 indicate the actual radial profiles. Inset in the upper right corner
 shows a zoom-in on the central part of the density profiles, with a
 higher number of sampling bins (the same units on the axes as the large
 plot, central part is indicated with a dashed grey line). The dark and
 light shades of green respectively are the dark matter density profiles
 at $0$ Gyr and $12.2$ Gyr, showing the flattening of the NFW cusp as in
 \citet{annelies}.}

\label{IC_profiles}
\end{figure}

Similar to \citet{re09} we opted for a pseudo-isothermal density profile
for the gas of the form
\begin{equation}
 \rho_{g}(r) = \frac{\rho_{c}}{1+\left(\frac{r}{r_{c}}\right)^{2}},
\end{equation}
with radius $r$ and the two model parameters $\rho_{c}$ and $r_{c}$,
respectively the characteristic density and scale radius. The
characteristic density of the gas density profile is related to the
characteristic density of the dark matter density profile (see Paragraph
\ref{section_dm_halo}) in the following way:
\begin{equation}
 \rho_{g,c} = \frac{\Omega_{b}}{\Omega_{dm}} \rho_{dm,c},
\end{equation}
where $\rho_{g,c}$ and $\rho_{dm,c}$ are the relevant gas and dark
matter densities, respectively, and $\Omega_{b}/\Omega_{dm} = 0.2115$
the fraction of baryonic to dark matter (values for the latter taken
from the 3-year WMAP results, \citealt{spergel2007}). The outer radius
of the sphere and the total gas mass contained within it are taken to be
the same as in \citet{sander:dgmodels}, which fixes the scale radius
$r_{c}$.  Fig.~\ref{IC_profiles} shows the initial setup for one of our
models.

\subsubsection{Rotation curves}
\label{section_rotation_curves}

For the models that are to receive angular momentum, we have chosen an
arctangens-shaped initial rotation profile, rising from zero in the
center to the maximum value $v_{rot}$ on the edge of the gas sphere (see
Fig.~\ref{IC_profiles}):
\begin{equation}
 v_{tan,i}(r) = \frac{2}{\pi} \arctan\left(\frac{r}{r_{s}}\right)~ v_{rot},
\end{equation}
with radius $r$ and scale radius $r_{s}$. The rotation axis of the
galaxy coincides with the $z$-axis, making the latter also the short
axis of the galaxy's oblate stellar body. The $x$ and $y$ axes lie
within the galaxy's equatorial plane.

\subsection{Density threshold for star formation}
\label{section_hdt} 

As in many other numerical simulations of galaxy evolution, we allow
star formation to occur only if the local gas density exceeds a certain
physically motivated threshold. Initially, this threshold was set to the
low value of $ 0.1\, \mathrm{amu\,cm^{-3}} $ (LDT), a value which has been in
general use by several authors
(e.g. \citealt{katz96,stinson2006,sander:dgmodels,re09,joeri:angmomentum};
and references therein). 
More recently, with numerical resolution following the steady increase
of computing power, it has become possible to follow the formation of
cold, high-density clouds in which star formation is supposed to occur.
\citet{governato10}, amongst others, therefore advocate using a much
higher density threshold of $ 100~ \mathrm{cm^{-3}} $ (HDT), which,
together with an extension of the cooling curves below $10^{4}$\,K, is
argued to represent a much more realistic description of star forming
regions. 

At this high threshold density, the SPH smoothing length, which
encompasses about 50-60 gas particles, exceeds the Jeans length down to
temperatures of $T_{\rm min} \sim 100$~K. At these high densities
and low temperatures, the gravitational softening is larger than both
the SPH smoothing and the Jeans lengths and artificial fragmentation is
suppressed (softening is set to 30 pc, while the SPH smoothing is
typically 8 pc at these densities). Hence, the cold, dense clumps
forming in the interstellar medium - with dimensions larger than the
gravitational softening scale - are real and they are the cradles of
star formation, which is the main aim of the employed star-formation
criteria.

In order to produce HDT models which lie on the observed photometric and
kinematic scaling relations, an increase of the star-formation density
threshold needs to be accompanied by a simultaneous increase in stellar
feedback efficiency, in order to counteract the stronger gravitational
forces in collapsed areas. This constitutes a fundamental degeneracy in
the parameter space of this branch of galaxy modelling, as discussed in
\citet{annelies}.

In this paper, we wil investigate how the density threshold affects the
formation and maintenance of stellar population gradients.

\subsection{Cooling curves}
\label{section_cooling_curves} 

As an integral part of the HDT scheme, an extension of the radiative
cooling curves for the gas below $10^{4}$\,K is necessary. The gas needs
to be able to cool further than in LDT models, in order to be able to
reach the required density threshold for star formation.

\begin{figure}

 \includegraphics[width=0.45\textwidth]{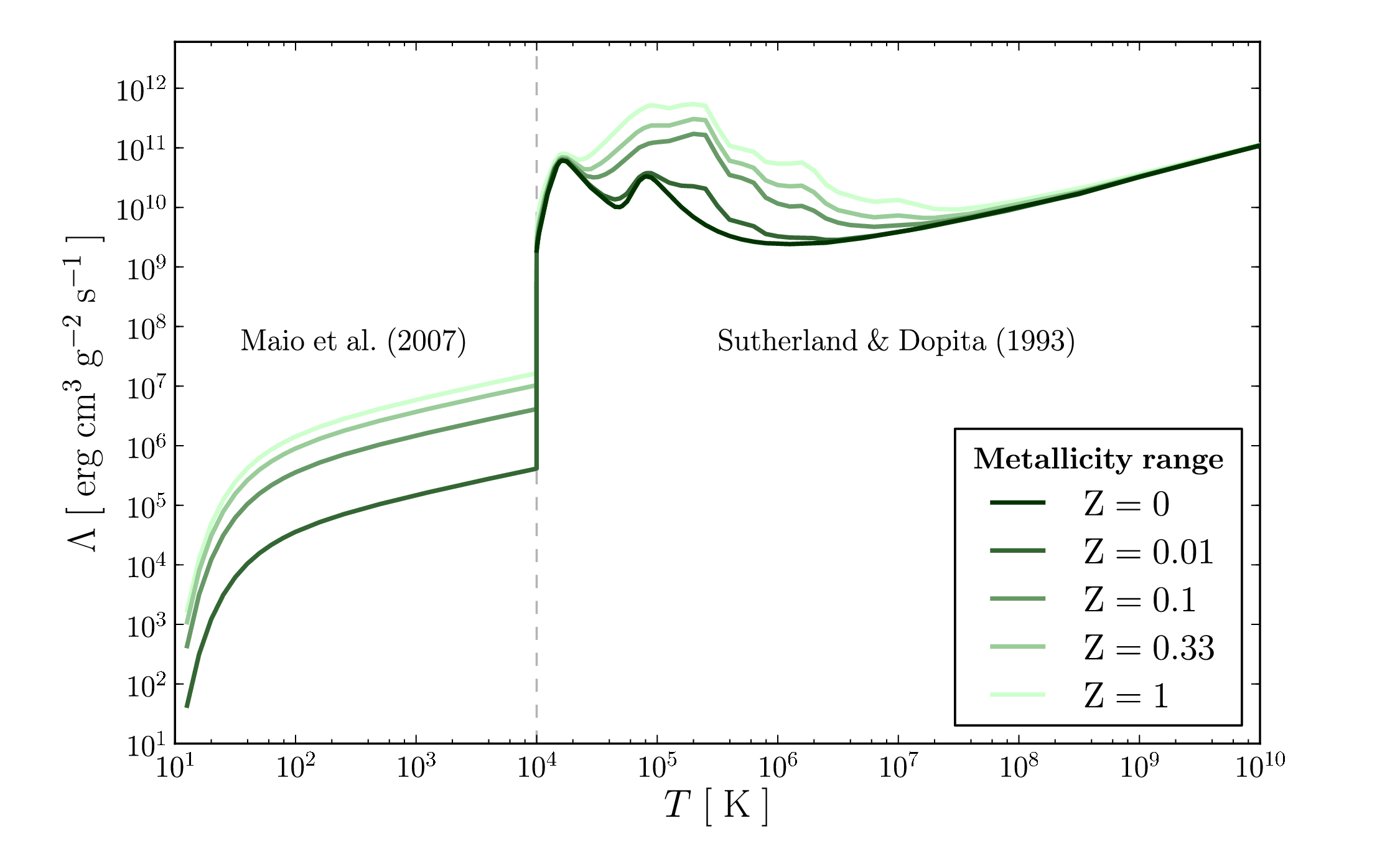}
 \caption{Plot of the first approach for extending the
 cooling curves below $10^{4}$\,K. Right of the dashed grey line are the
 cooling curves as calculated by \citet{suthdop}, left are the
 extensions as calculated by \citet{maio07}. In simulations these curves
 are interpolated in temperature and metallicity.}

 \label{dirty_cooling_curves}
\end{figure}

One way of reaching this goal is by supplementing the \citet{suthdop}
metallicity dependent cooling curves, which are valid for temperatures
above $10^{4}$\,K, with cooling curves calculated for temperatures below
$10^{4}$\,K, such as the ones presented by \citet{maio07}. This is an
often used approach in numerical simulations \citep{sa10, re09,
annelies}. However, besides the lack of consistency between different
cooling curve calculations and the chemical oversimplification of using
only metallicity as a compositional parameter, this approach also
features an unnaturally sharp drop by about 4 orders of magnitude of the
cooling rate at $10^{4}$\,K (see Fig.~\ref{dirty_cooling_curves}). This
leads to a pile-up of gas particles at precisely this temperature.

\begin{figure}
 
\includegraphics[width=0.45\textwidth]{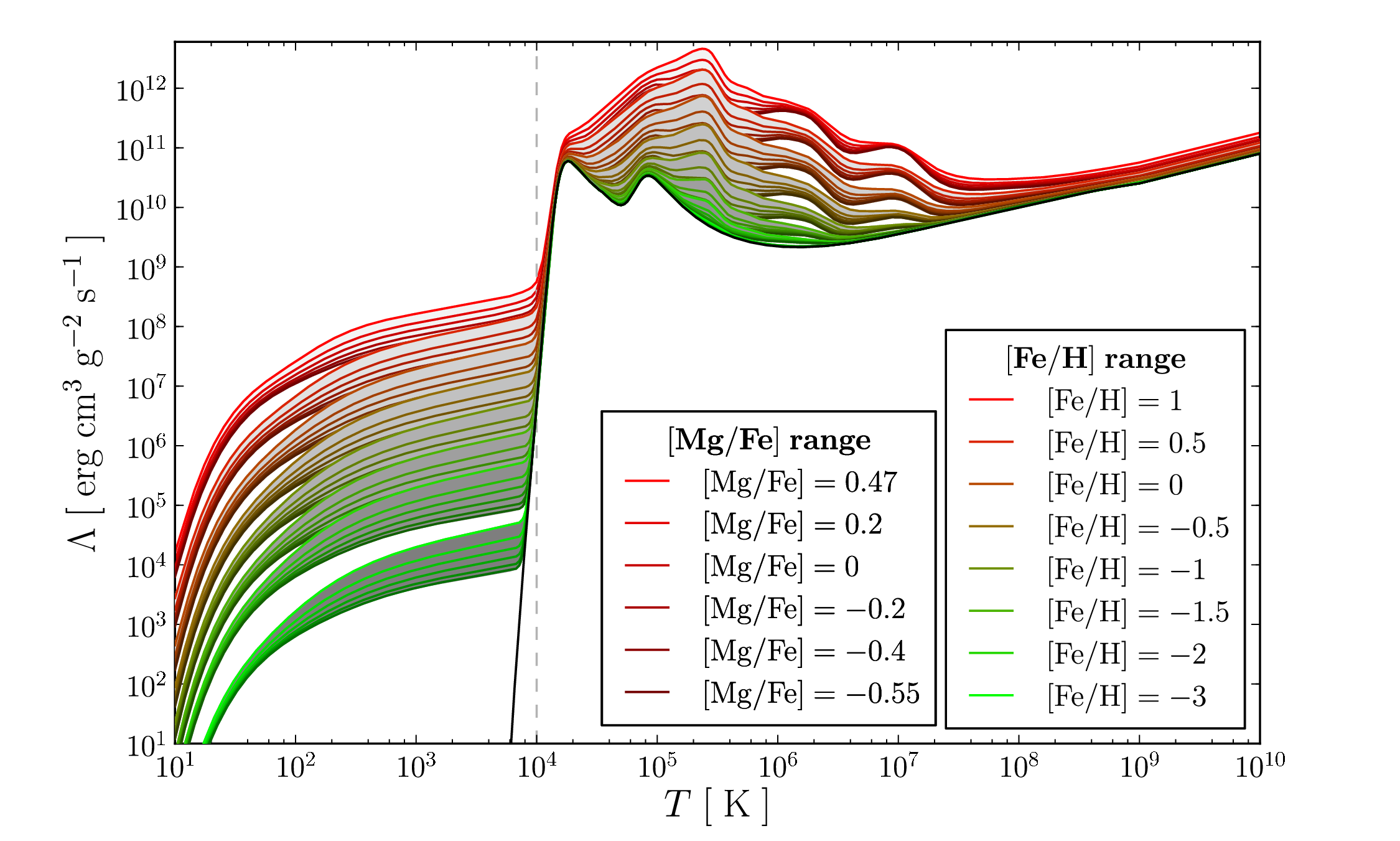}
 \caption{Similar plot as Fig.~\ref{dirty_cooling_curves}, but now for
 the ``first stage'' improved cooling curves. There is no discontinuity
 anymore around $10^{4}$\,K now, due to the consistency in the
 calculations across the entire temperature range. In simulations these
 curves are now interpolated in temperature, [Fe/H] and [Mg/Fe], where
 the latter two respectively encode the contribution of supernovae type
 Ia and type II to the composition of the gas. Different [Fe/H] are
 indicated by a colorscale from green to red, while different [Mg/Fe]
 are indicated through the intensity of the color (colorscale for the
 latter only shown for the highest [Fe/H]). The black line represents
 the zero metallicity curve.}

\label{sven_cooling_curves}
\end{figure}

We therefore embarked to produce fully self-consistent radiative cooling
curves, featuring a continuous temperature range from 10\,K to
$10^{9}$\,K in one consistent calculation scheme, incorporating all
relevant processes and elements. These fit into a larger effort of
expanding and improving the physical context of our simulations, which
is currently under way: radiative cooling and heating
\citep{sven:cooling_curves}; treating gas ionization
\citep{bert:ionization}. In this paper, however, we will only employ the
``first stage'' of the improved cooling curves. These not yet take into
account any UV background radiation, but they do feature the basis of
the new interpolation scheme. Both [Fe/H] and [Mg/Fe] are now used as
tracers for estimating, respectively, the contribution of supernovae
type Ia and type II to the total composition of the gas. The curves are
produced for a range of 8 [Fe/H] values and 6 [Mg/Fe] values, plus a
zero metallicity curve, resulting in the 49 continuous cooling curves
presented in Fig.~\ref{sven_cooling_curves}. This already represents a
significant improvement over the chemical composition scheme of
\citet{suthdop}, which at low metallicities only uses SNII abundance
ratios \citep{sven:cooling_curves}.

%% file: article_simulations.tex
\section{Simulations}
\label{section_simulations}

Here we describe the details of the simulation runs that have been performed
and used for this research. The methods and theory behind the models are
discussed in Section \ref{section_models}, and in
\citet{sander:dgmodels}, \citet{joeri:angmomentum} and
\citet{annelies}.

We investigate two types of models in this paper. On the one hand, we
run simulations according to the prescriptions from our previous
research \citep{joeri:angmomentum} which employ the, until recently
quite standard, low density threshold for star formation and
corresponding low feedback efficiency. On the other hand, we also run
simulations which feature the specifications discussed in Section
\ref{section_models}, such as a high density threshold and a high
feedback efficiency.  Henceforth, we will refer to them as the low
density threshold (LDT) and high density threshold (HDT)
models/simulations/runs, respectively.

Several properties of the simulations, however, are shared among the LDT
and HDT runs:
\begin{itemize}
 \item{all simulations start with an initial set of 200,000 gas
      particles and 200,000 dark matter particles.}
 \item{initial metallicity is set to $10^{-4}$ solar metallicities.}
 \item{initial gas temperature is $10^4$\,K.}
 \item{runtime is approximately $12$\,Gyr, starts at redshift 4.3.}
 \item{snapshots are made every $5$\,Myr, resolving the dynamical timescale.}
 \item{the models are isolated.}
 \item{the gravitational softening length is 30\,pc.}
\end{itemize}

\subsection{Low density threshold (LDT) runs}
\label{section_ldt_runs}

Here, we investigate simulations with initial dark-matter masses of
$660\times 10^6$~M$_\odot$ (these runs have labels that end with ``05'')
and $2476\times 10^6$~M$_\odot$ (these runs have labels that end with
``09''). The density threshold is set to $0.1~\mathrm{cm^{-3}}$, with a
feedback efficiency of 0.1 (i.e. 10~\% of the supernova energy is
absorbed by the interstellar medium), and cooling curves that do not go
below $10^{4}$\,K. Simulations with and without rotation have been
performed. Rotation was induced by adding a constant rotational velocity
of $v_{\rm rot}=5$~km~s$^{-1}$ to the gas particles. These simulations
are basically high time resolution reruns of some of our older models
and will serve as a ``reference sample'' to compare with the new
models. The details of these 4 simulations can be found in Table
\ref{table_ldt_sims}.

\input{table_LDT}

\subsection{High density threshold (HDT) runs}
\label{section_hdt_runs}

Simulations have also been performed that employ the HDT models as
described in Section \ref{section_models}. Again we use a lower mass and
a higher mass model, with the same initial masses as the LDT runs in
\ref{section_hdt_runs}, again one rotating and one non-rotating. The
density threshold for star formation is now set to $100~%
\mathrm{cm^{-3}}$ and employs the first stage novel cooling curves as
described in Paragraph \ref{section_cooling_curves}, which span a
temperature range from 10\,K to $10^{9}$\,K (and are extended to
$10^{10}$\,K with a Bremsstrahlung approximation). The feedback
efficiency has been increased to $0.7$, following the results of
\citet{annelies} (see Section \ref{section_hdt}). On top of the NFW dark
matter halo is placed a pseudo-isothermal gas sphere, which in the
rotating models is given an arctangens radial rotation profile, as in
\ref{section_rotation_curves}, with a $v_{\rm rot}$ of $5$km/s and
$r_{s}=1$~kpc. Further details are found in Table \ref{table_hdt_sims}.

\input{table_HDT}

\subsection{Truncated simulations}
For each of the abovementioned non-rotating simulations we also run a
``truncated'' version, where the star formation is shut off at a
specific time during the evolution. This allows us to assess most
clearly whether any population gradients present at the moment of
truncation can persist for an extended period.

The truncation can be done in two ways:
\begin{enumerate}
\item The star formation routines are shut off at a certain time, which
      for the HDT simulations also requires shutting off cooling below
      $10^{4}$\,K because otherwise the gas quickly becomes extremely
      dense, causing the code to crash. The gas remains present, but
      becomes inert, the ``gastrophysics'' are switched off.
\item All gas particles are removed from the simulation, without
      changing the physics. This is basically a poor man's version of
      ram-pressure-stripping, mimicking a dwarf galaxy that is stripped
      of its gas on short timescales by the intergalactic medium.
\end{enumerate}
We opted for the second option, since it stops star formation in a more
``natural'' way, without tinkering with the physics and switching
certain processes off. In practice we take a certain snapshot of the
existing simulations, remove the gas particles from it, and use it as
the initial condition for a simulation that restarts at the moment of
truncation. We have chosen the simulations to be truncated at 8\,Gyr,
allowing us to study the stability of any existing population gradients
over periods of time of the order of 4 billion years.

\section{LDT vs. HDT}
\label{section_ldtvshdt}
Several differences in the physical features of the LDT and HDT
simulations are worth highlighting here, before going on to the specific
research results in the next sections.

\begin{figure}
\includegraphics[width=0.45\textwidth]{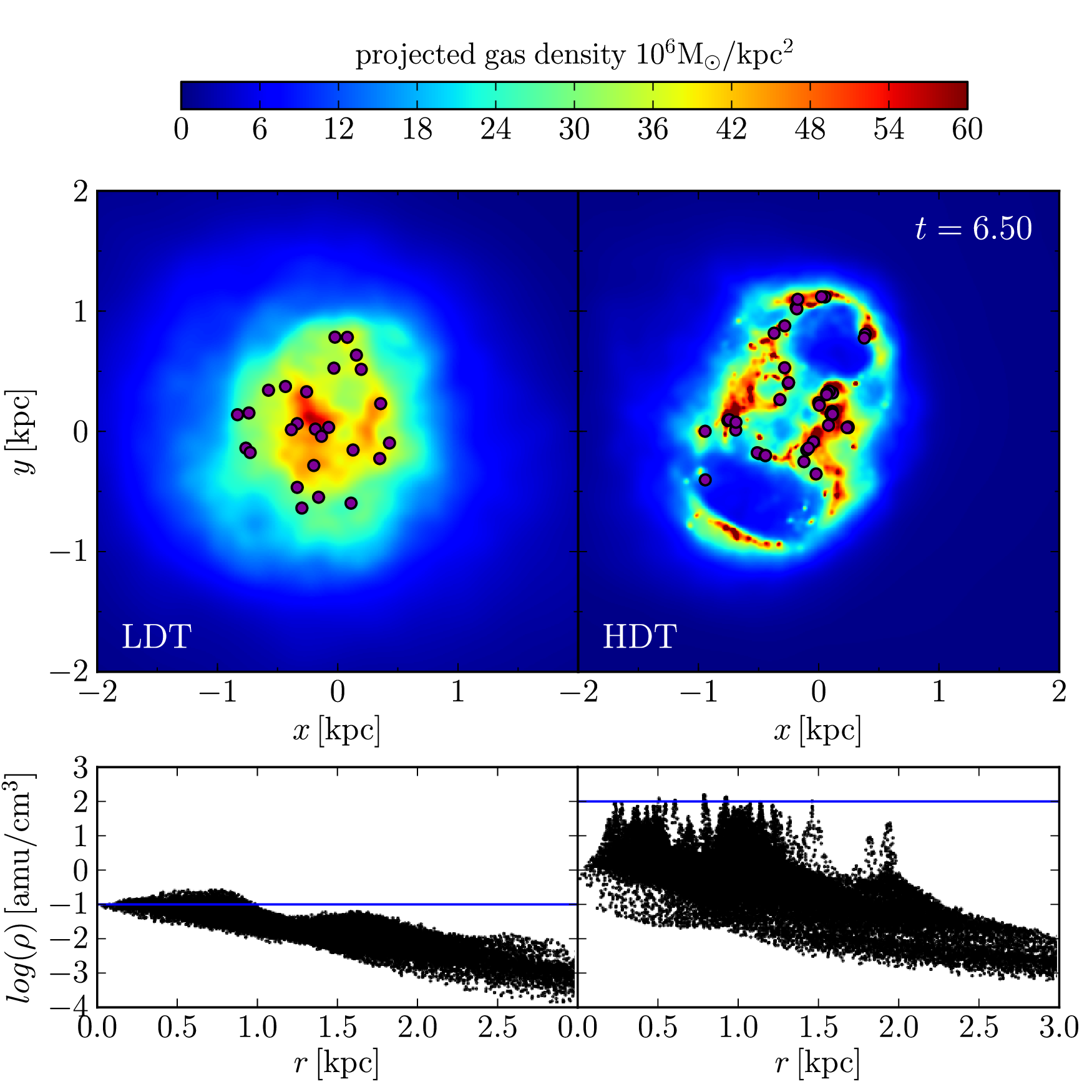}
\caption{The top row shows rendered images of a LDT (left) and HDT
(right) model. Superimposed on the color coded (projected) gas density
as purple dots are the newly formed stars, of ages 10\,Myr and
below. The bottom row shows density-radius scatter plots of the gas
particles, with a horizontal blue line indicating the density threshold
for star formation in the respective star formation schemes. Note: the
left rendered LDT image has actually been generated according to a color
scale with values scaled down with a factor 1/10, since otherwise none
of the gas would get out of the dark blue range. A full animation of
this figure can be found online\protect\footnotemark[2] \label{comparison_clumps}}
\end{figure}

The principal feature of the HDT models is obviously the formation of
dense and cold clumps in the gas in which star particles form
\citep{governato10}. In Fig.~\ref{comparison_clumps}, the difference
in gas structure between the LDT and the HDT simulation is
apparent. Whereas in the LDT case the gas is quite ``fuzzy'', and
collectively above the density threshold in a large area, in the HDT
case it is much more structured and fragmented in dense clumps, with
only localized individual density peaks reaching above the density
threshold.

\footnotetext[2]{Video: http://www.youtube.com/watch?v=lHDcFD6ok7c \\
YouTube channel of Astronomy department at Ghent University:
http://www.youtube.com/user/AstroUGent \\ Youtube playlist with all
additional material for this paper: \\
http://www.youtube.com/playlist?list=PL-DZsb1G8F\_lDNn3G-9ACGgrinen8aSQs}

\begin{figure}
\includegraphics[width=0.5\textwidth]{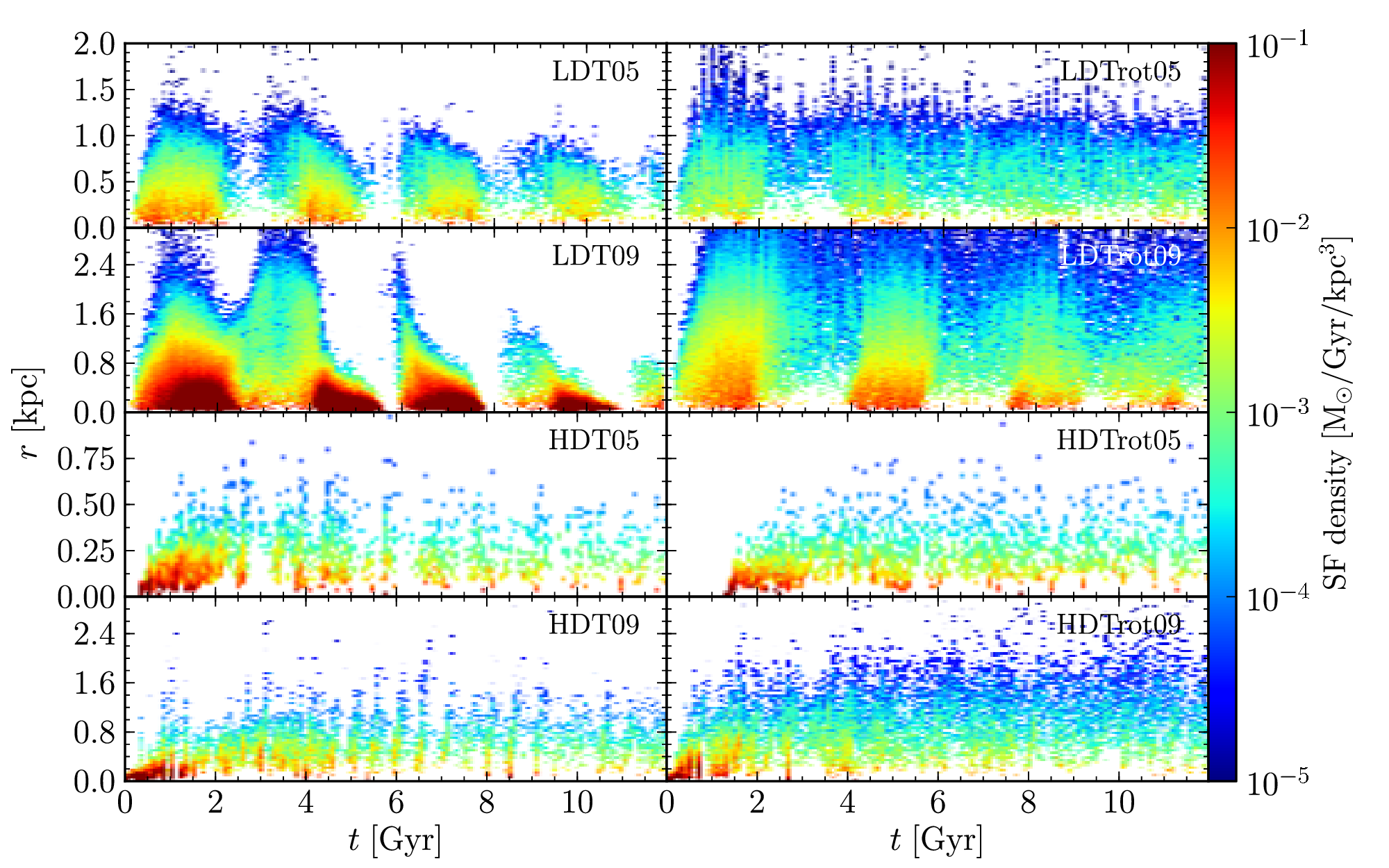}
\caption{The star formation density (in $\mathrm{M_\odot / Gyr / kpc^{3}}$, color
coded according to the color scale) plotted in function of time
($x$-axis) and spatial extent (radius, $y$-axis). The non-rotating
 models are in the left column, the rotating ones in the right
 column. The four upper plots show the LDT models, the four bottom
 models show the HDT models.
  \label{SFHistogram}}
\end{figure}

\begin{figure}
\includegraphics[width=0.45\textwidth]{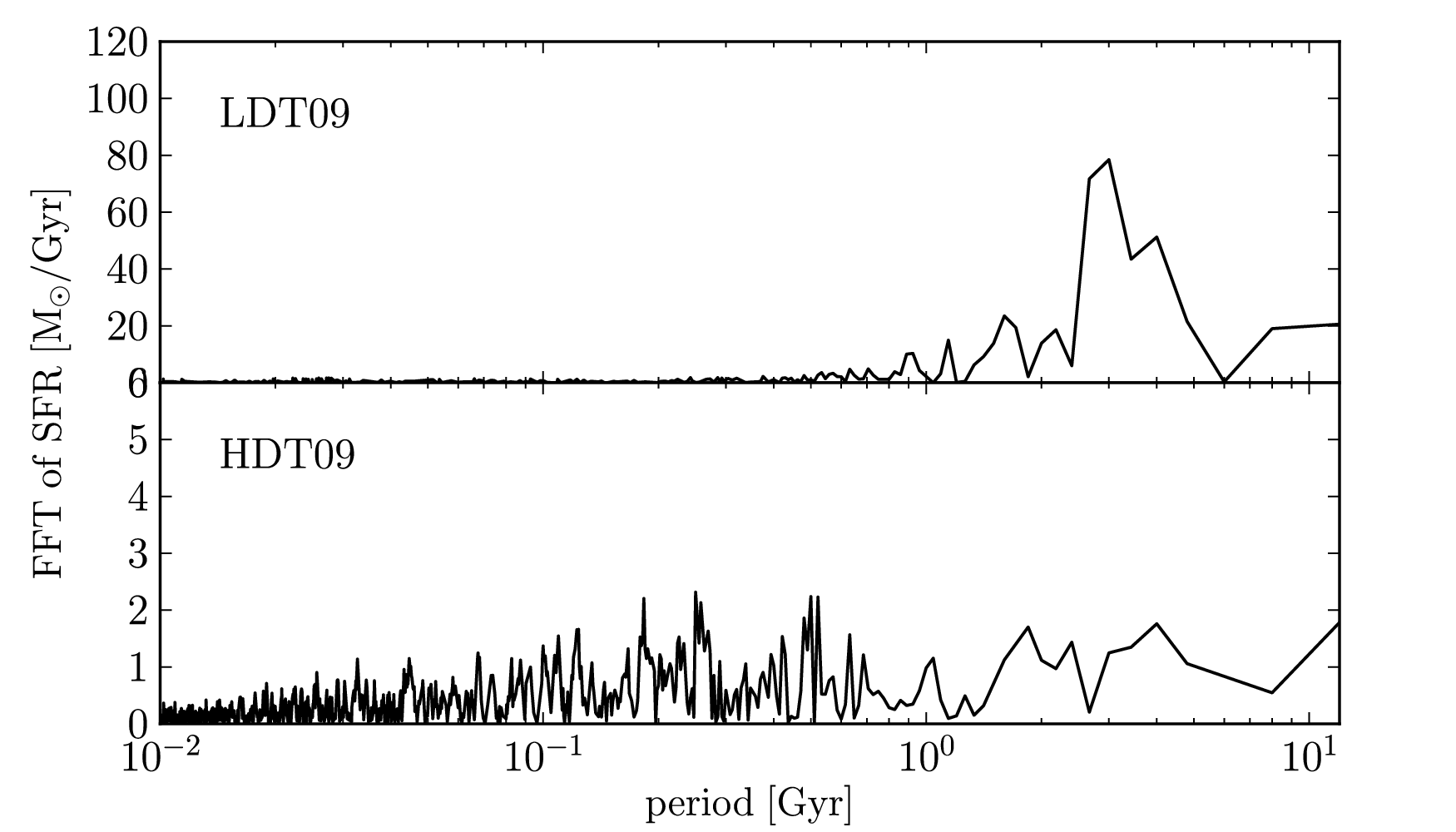}
\caption{Fourier transform of the star formation rate in function of
 time, shown as a function of the period of the mode on the $x$-axis.
\label{SFfft}}
\end{figure}

This distinction has immediate consequences for the star formation mode,
as can be seen in Fig.~\ref{SFHistogram}. The non-rotating LDT models
show clear star formation episodes of about 2\,Gyr long with
intermittent lulls of a Gyr or so, with an inwards shrinking of the SF
area between episodes and within each episode. The HDT models show much
shorter star formation bursts in faster sequention with no shrinking
within an episode, but still with a shrinking of the SF area over time
between episodes. We can see this change in star formation timescale
clearly in the Fourier transform of the star formation rate over time in
Fig.~\ref{SFfft}. The LDT model shows a peak at a period of 3\,Gyr,
which agrees well with the very noticeable 4 star formation episodes in
12\,Gyr seen in Fig.~\ref{SFHistogram}, while the HDT model shows a
clear shift to shorter periods, with 2 peak values at periods of 0.2 and
0.5\,Gyr. These match, respectively, with the sequence of star formation
events seen in Fig.~\ref{SFHistogram} in the first 2-3\,Gyr and the
last 6-7\,Gyr.

\begin{figure}
\begin{center}
\includegraphics[width=0.47\textwidth]{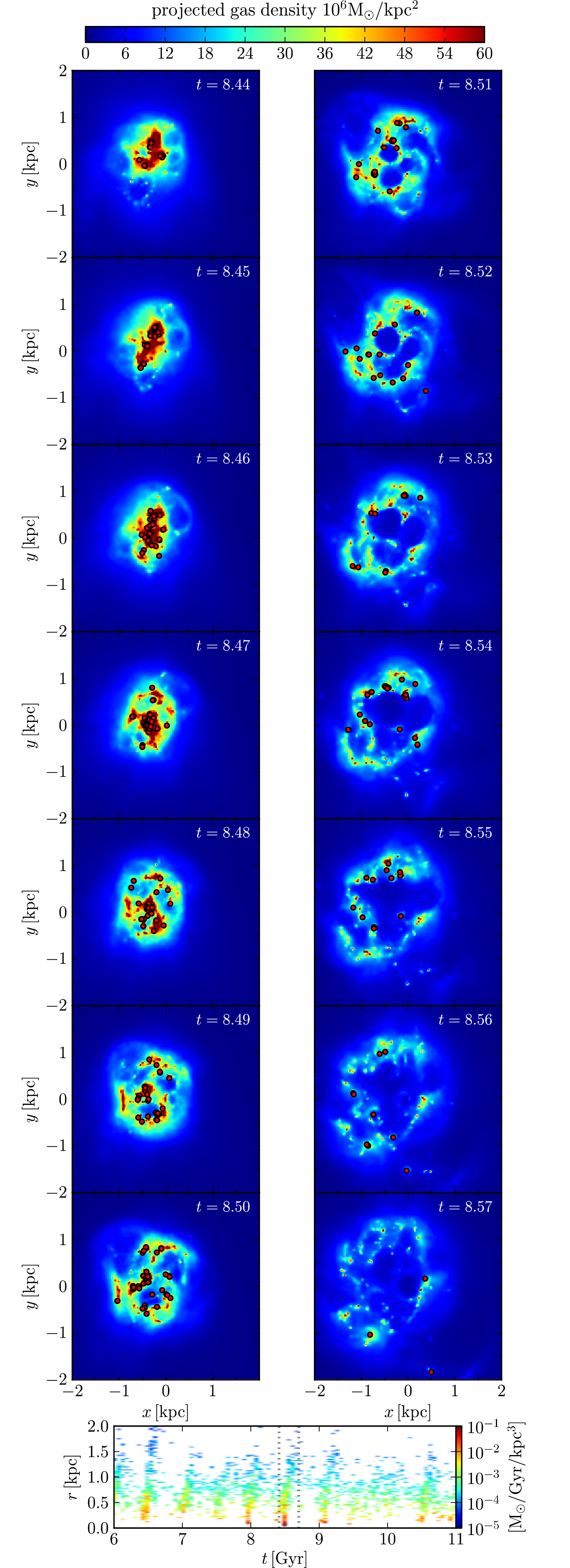}
\caption{A high time resolution sequence of snapshots capturing a single
star-formation event in a HDT model. Red dots indicate newly born star
particles. Below is a zoom of the SF density (Fig.~\ref{SFHistogram}),
with the dotted lines indicating the time range shown.
\label{starburst_sequence}}
\end{center}
\end{figure}

In Fig.~\ref{starburst_sequence}, we show the evolution of a HDT model
during a single short-duration star-formation event. The event starts
when several high-density peaks inside the already dense inner 1~kpc of
the galaxy start forming stars. Supernova feedback quickly disrupts the
individual star-forming clouds while triggering secondary star formation
throughout this dense inner region of the galaxy. Subsequently
accumulating feedback evacuates the dense central region about 30-40~Myr
after the start of the event, limiting star formation to condensations
on the edges of the expanding supernova-blown bubbles until it finally
peters out after about 150~Myr. The insets present a zoom on the
star-formation event in Fig.~\ref{SFHistogram}, which shows this rapidly
outward evolution by a slight tilt to the right of the corresponding
``plume'' of the burst. This chain of events is reminiscent of the
``flickering star formation'' observed in real dwarf galaxies with
bursty star-formation histories \citep{mcquinn10}. A full animation of
the simulated star burst event depicted in Fig.~\ref{starburst_sequence}
can be found online\footnote[3]{Video:
http://www.youtube.com/watch?v=0TB9LQaiKEs}.

As already concluded by \citet{joeri:angmomentum}, angular momentum
smears out star formation in time and space, making the major
star-formation events less conspicuous (Fig.~\ref{SFHistogram}).

Moreover, the HDT models produce significantly less stellar mass for the
same initial gas mass. However, since all other properties also scale
accordingly, the models remain in good agreement with the fundamental
observational characteristics:~they simply shift along the observed
photometric and kinematical scaling relations \citep{annelies}.


%% file: table_LDT.tex
\begin{table}
 \caption{Details of the LDT runs, re-run from \citet{joeri:angmomentum}
 (corresponding simulation number in this previous work shown in
 brackets, further details on the models can also be found there). All
 quantities are evaluated at the end of the simulation, except for the
 initial values indicated with index `i'. Rows: (1) initial gas mass
 [$10^{6} \mathrm{M_{\odot}}$], (2) dark matter mass [$10^{6}
 \mathrm{M_{\odot}}$], (3) stellar mass [$10^{6} \mathrm{M_{\odot}}$],
 (4) half-light radius [kpc], (5) luminosity-weighted metallicity
 (B-band), (6)(7) B-band and V-band magnitude, (8) initial rotation
 speed of gas [km/s].  }

 \label{table_ldt_sims}

 \begin{tabular}{lc@{\hspace{15pt}}cc@{\hspace{10pt}}c}
  \hline   
  &LDT05&LDT09&LDTrot05&LDTrot09 \\
  &(205)&(209)&(225)&(229)     \\                          
  \hline
  $M_{g,i}$ &140&524&140&524       \\  
  $M_{dm}$ &660&2476&660&2476        \\                        
  $M_{st}$ &18.93&468.57&14.35&325.33      \\ 
  $R_{e}$ &0.43&0.39&0.63&1.36        \\
  $[Fe/H]$ &-0.717&-0.053&-0.672&-0.281        \\
  $M_{B}$ &-11.87&-14.87&-12.2&-15.02         \\
  $M_{V}$ &-12.51&-15.62&-12.7&-15.62         \\
  $v_{i}$ &\multicolumn{2}{c}{0}&\multicolumn{2}{c}{5}         \\
  \hline                                
 \end{tabular}
\end{table}

%% file: table_HDT.tex
\begin{table}

 \caption{Details of the new HDT runs. All values in the first block
 refer to the initial conditions, the other values (besides those
 indexed with `i') are final values. Rows: (1) initial gas mass [$10^{6}
 \mathrm{M_{\odot}}$], (2)(3)(4) characteristics of pseudo-isothermal
 sphere: density [$10^{7} \mathrm{M_{\odot}/kpc^{3}}$] - scale radius
 [kpc] - cutoff radius [kpc], (5) dark matter mass [$10^{6}
 \mathrm{M_{\odot}}$], (6)(7)(8) characteristics of NFW halo: density
 [$10^{7} \mathrm{M_{\odot}/kpc^{3}}$] - scale radius [kpc] - cutoff
 radius [kpc], (9) stellar mass [$10^{6} \mathrm{M_{\odot}}$], (10)
 half-light radius [kpc], (11) luminosity-weighted metallicity (B-band),
 (12)(13) B-band and V-band magnitude, (14) initial rotation speed of
 gas [km/s].}

 \label{table_hdt_sims}

 \begin{tabular}{lc@{\hspace{15pt}}cc@{\hspace{8pt}}c}

  \hline   
  &HDT05&HDT09&HDTrot05&HDTrot09 \\
  \hline
  $M_{g,i}$ &140&524&140&524      \\                         
  $\rho_{pseudo-iso}$ &1.102&0.896&1.102&0.896 \\
  $r_{pseudo-iso}$ &0.234&0.403&0.234&0.403 \\
  $r_{g,max}$ &18.894&29.353&18.894&29.353 \\
  $M_{dm}$ &660&2476&660&2476       \\                        
  $\rho_{nfw}$ &5.211&4.236&5.211&4.236 \\
  $r_{nfw}$ &0.744&1.251&0.744&1.251 \\
  $r_{dm,max}$ &21.742&33.634&21.742&33.634 \\
  \hline
  $M_{st}$ &2.37&32.19&1.58&52.813      \\ 
  $R_{e}$ &0.23&0.58&0.22&1.06        \\
  $[Fe/H]$ &-0.98&-0.834&-0.922&-0.736        \\
  $M_{B}$ &-9.69&-12.79&-9.69&-13.44        \\
  $M_{V}$ &-10.3&-13.34&-10.24&-13.98        \\
  $v_{i}$ &\multicolumn{2}{c}{0}&\multicolumn{2}{c}{5}         \\
  \hline       
                         
 \end{tabular}

\end{table}

%% file: article_metallicity_profiles.tex
\section{Metallicity profiles}
\label{section_metallicity_profiles}

In this section we present and discuss the evolution of the radial
stellar metallicity profiles throughout the simulations, and compare
them qualitatively and quantitatively to observed metallicity gradients
of dwarf galaxies in the Local Group. As discussed in Section
\ref{section_simulations} we discern between the LDT/HDT scheme,
low/high total mass, and rotating/non-rotating models.  Furthermore, for
all of the non-rotating models we also present the results from
so-called ``truncated simulations'', where the star formation is shut
off at a specific time during the run. In all cases we consider both the
luminosity-weighted and mass-weighted metallicity - the former being the
quantity which mimics what is generally measured from observations,
while the latter reflects the actual physical distribution of metals.

The way of constructing the profiles from the star particles in our
simulations tries to mimic an observational configuration. We choose our
``line of sight'' along the $y$-axis and make bins along the $x$-axis,
mimicking a long-slit spectroscopic observation with the slit aligned
along the galaxy's major axis. In the $z$ direction we restrict the
particles to the range $-0.2~\mathrm{kpc} \leq z \leq
0.2~\mathrm{kpc}$. To reduce numerical scatter and give a clearer
picture, we ``stack'' the profiles in space and time. We use the same
procedure as before, now projecting along $x$ and binning $y$, and fold
all 4 profiles ($x$ and $y$ axes, both in positive and negative
direction) onto one profile of metallicity in function of projected
distance to the center. Furthermore, we stack subsequent profiles in
time, covering an interval of the order of the dynamical timescale
($\sim 50\mathrm{Myr}$). Adaptive binning is used at the end of the
procedure, to avoid erratic values at the edges of the profiles.

Luminosity weighting is done by multiplying the iron (Fe) and magnesium
(Mg) masses of a stellar particle with its B-band luminosity value,
which is obtained by interpolating in age and metallicity on the MILES
population synthesis data \citep{vazdekis96}. When summing over the
particles in a bin, the total metal masses of the bin are then divided
by the total B-band luminosity of the bin.


\subsection{LDT sims}

\begin{figure*}
\begin{minipage}[t]{2\columnwidth}
\hfill
\includegraphics[width=0.45\textwidth]{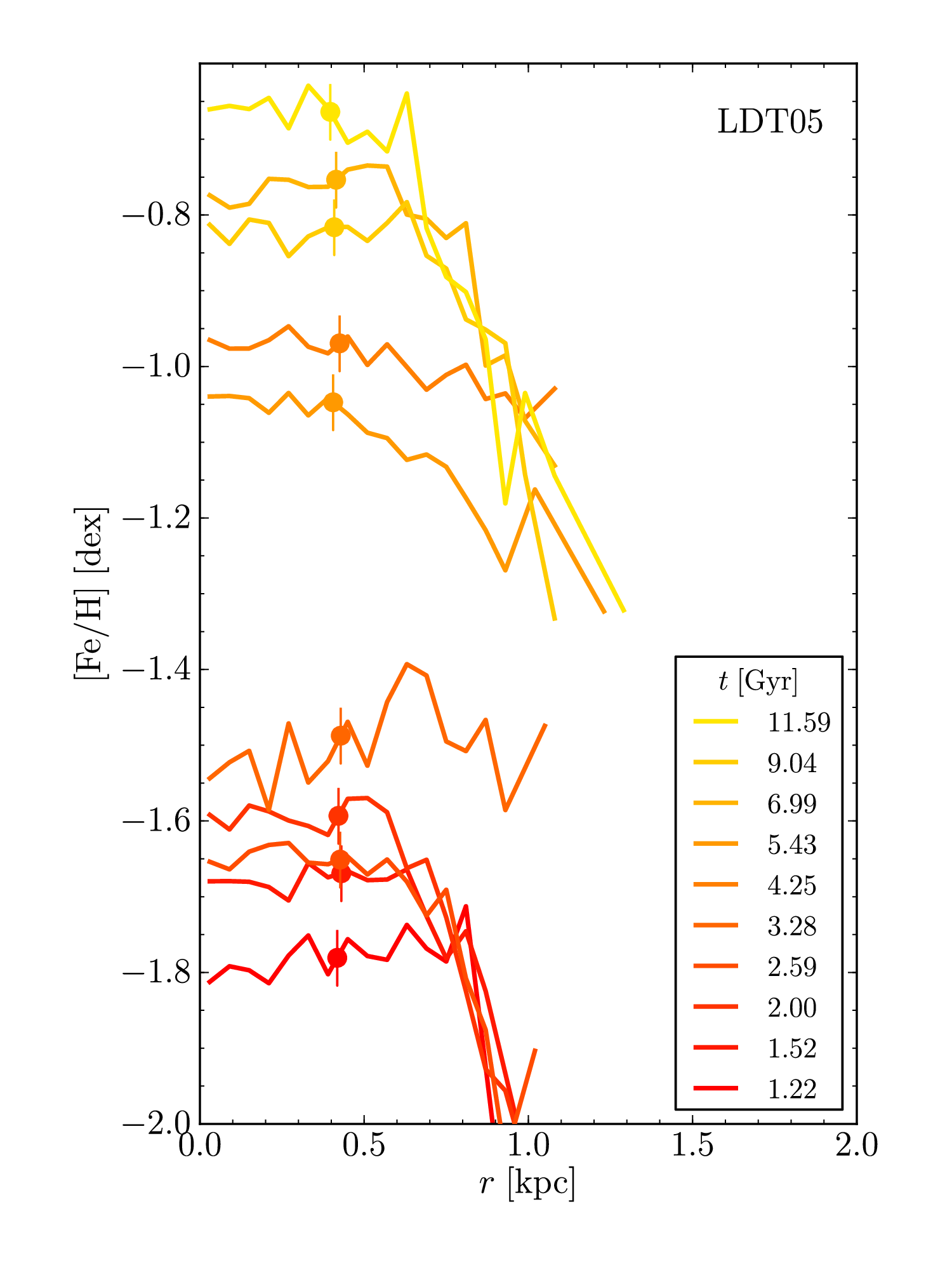}
\hfill
\includegraphics[width=0.45\textwidth]{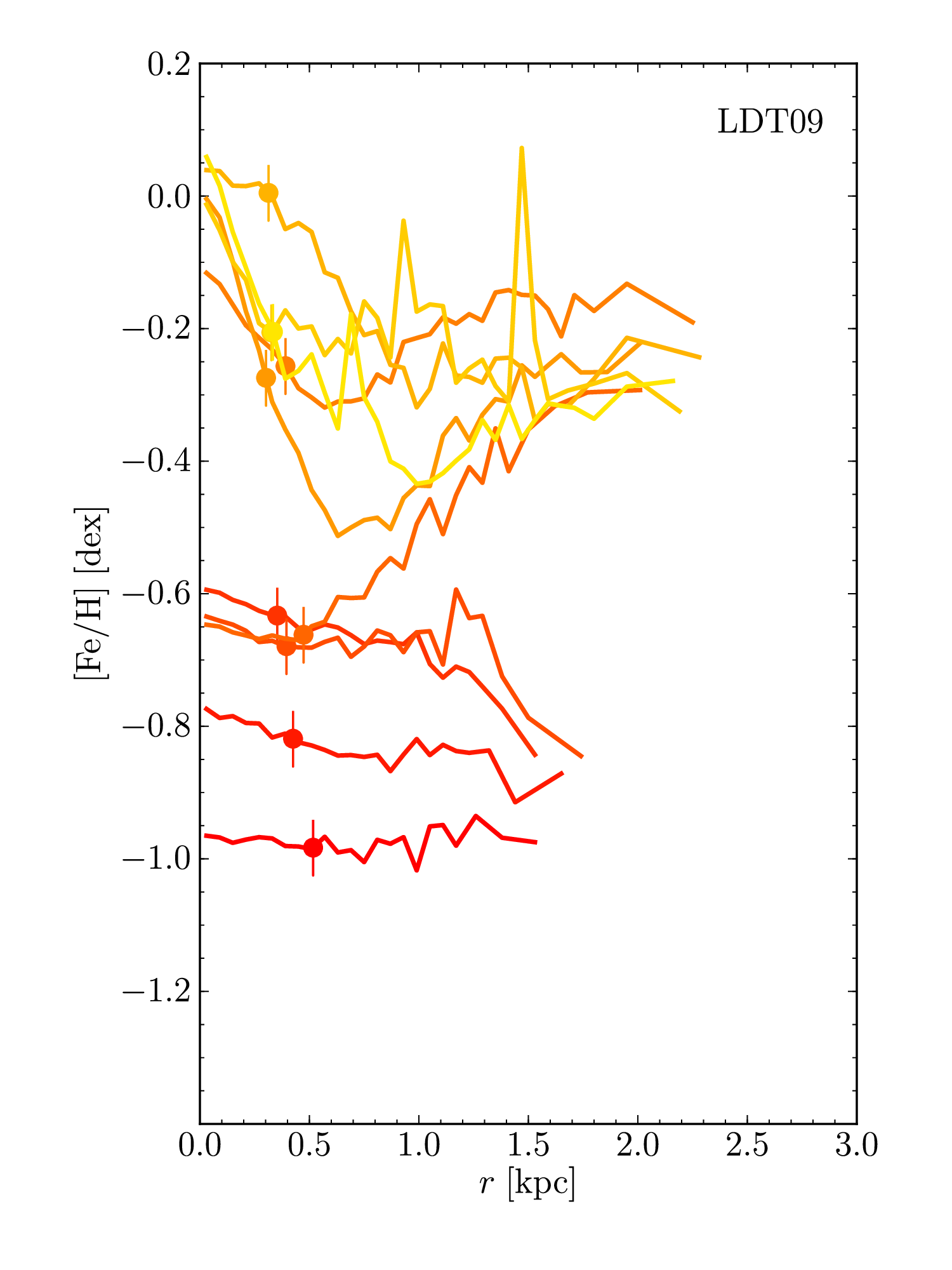}
\hfill
\\
 \hspace*{0mm} 
\hfill
\includegraphics[width=0.45\textwidth]{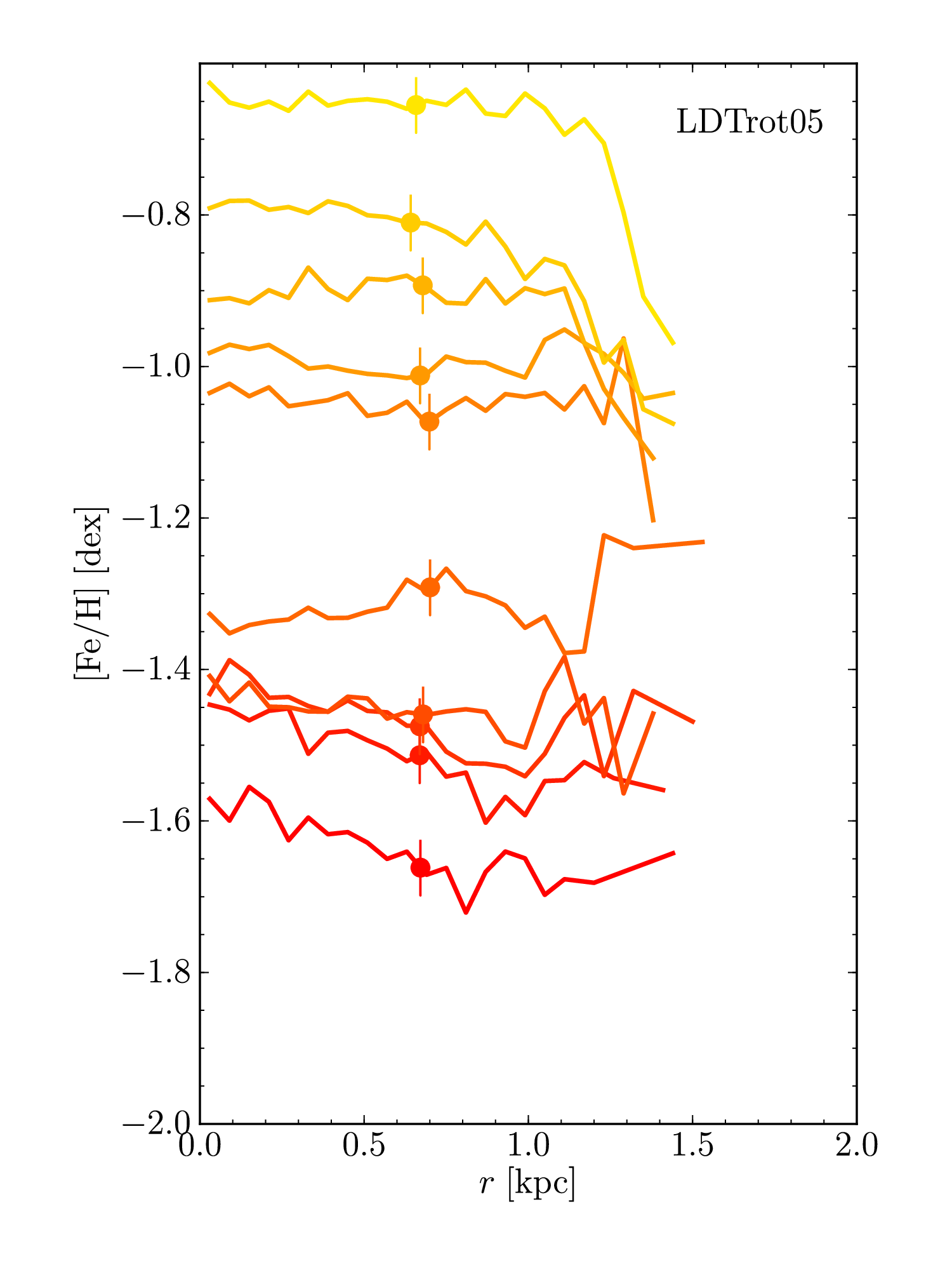}
\hfill
\includegraphics[width=0.45\textwidth]{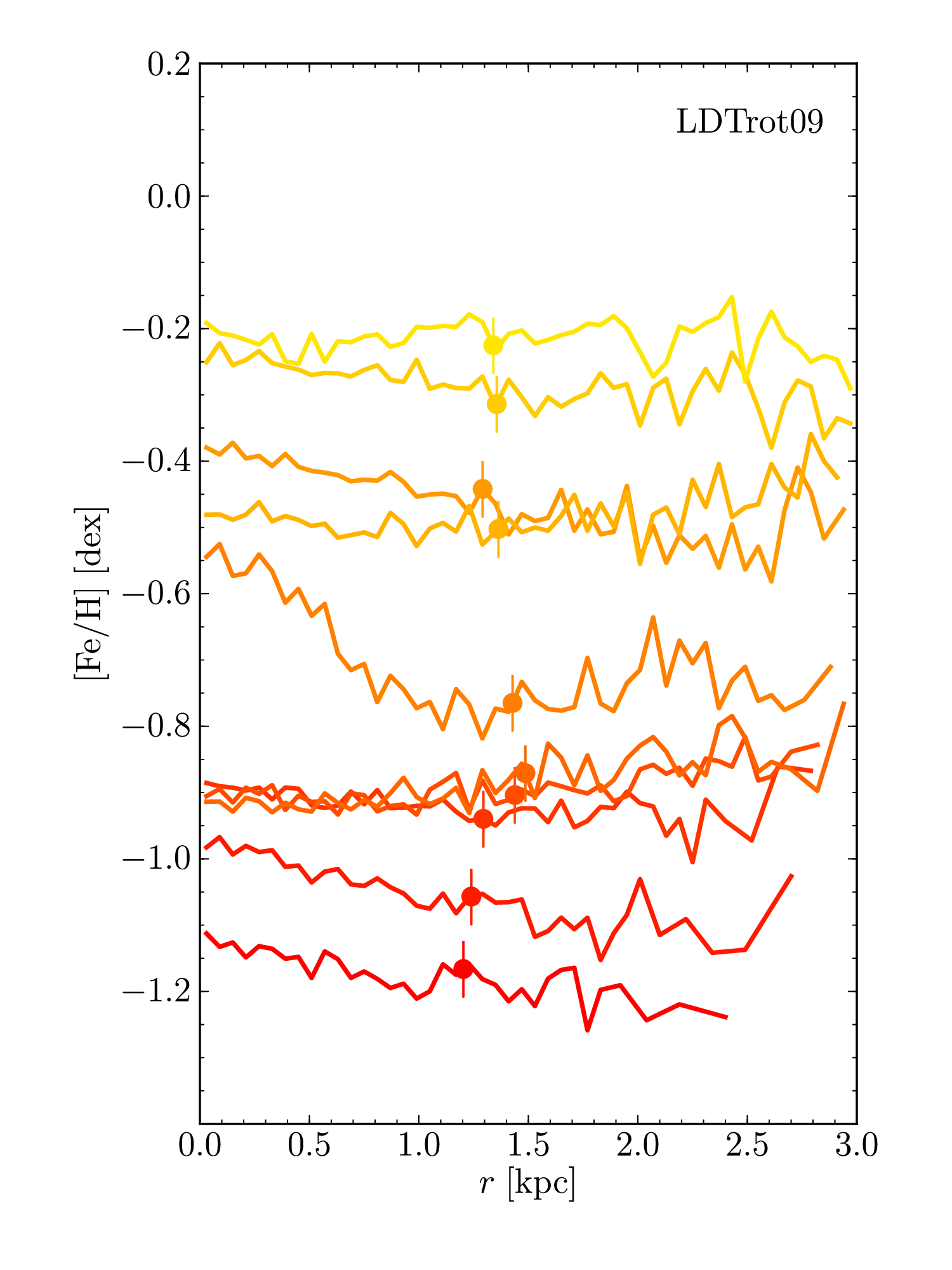}
\hfill 

\caption{Luminosity-weighted (B-band) metallicity profiles of the LDT
models. Evolution throughout the simulation is shown at logarithmic
intervals of time, with $0.11$ dex separation - the legend in the top
left plot shows the times. The left column shows the low-mass (05)
models, the right column shows the high-mass (09) models (see Table
\ref{table_ldt_sims}). Non-rotating models are on the top row, rotating
models on the bottom row. The metal contribution of each star particle
is weighted by its luminosity in the B-band, to mimick the actual
observed quantities. The vertical line symbol on each profile indicates
the half-light radius of the dwarf galaxy model at that time.
\label{LDT_metprof_lum}}

\end{minipage}
\end{figure*}

\begin{figure*}
\begin{minipage}[t]{2\columnwidth}
\hfill
\includegraphics[width=0.45\textwidth]{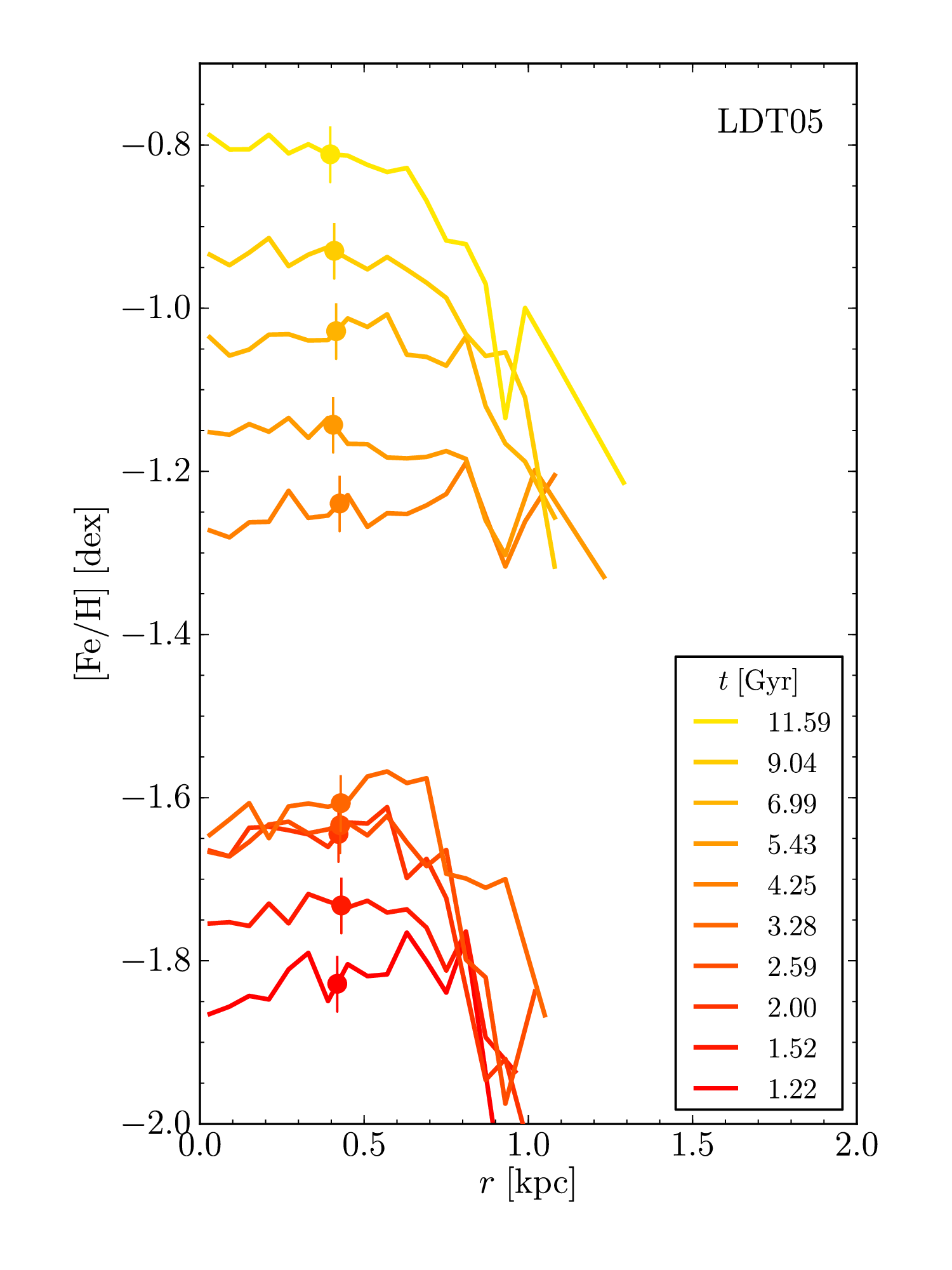}
\hfill
\includegraphics[width=0.45\textwidth]{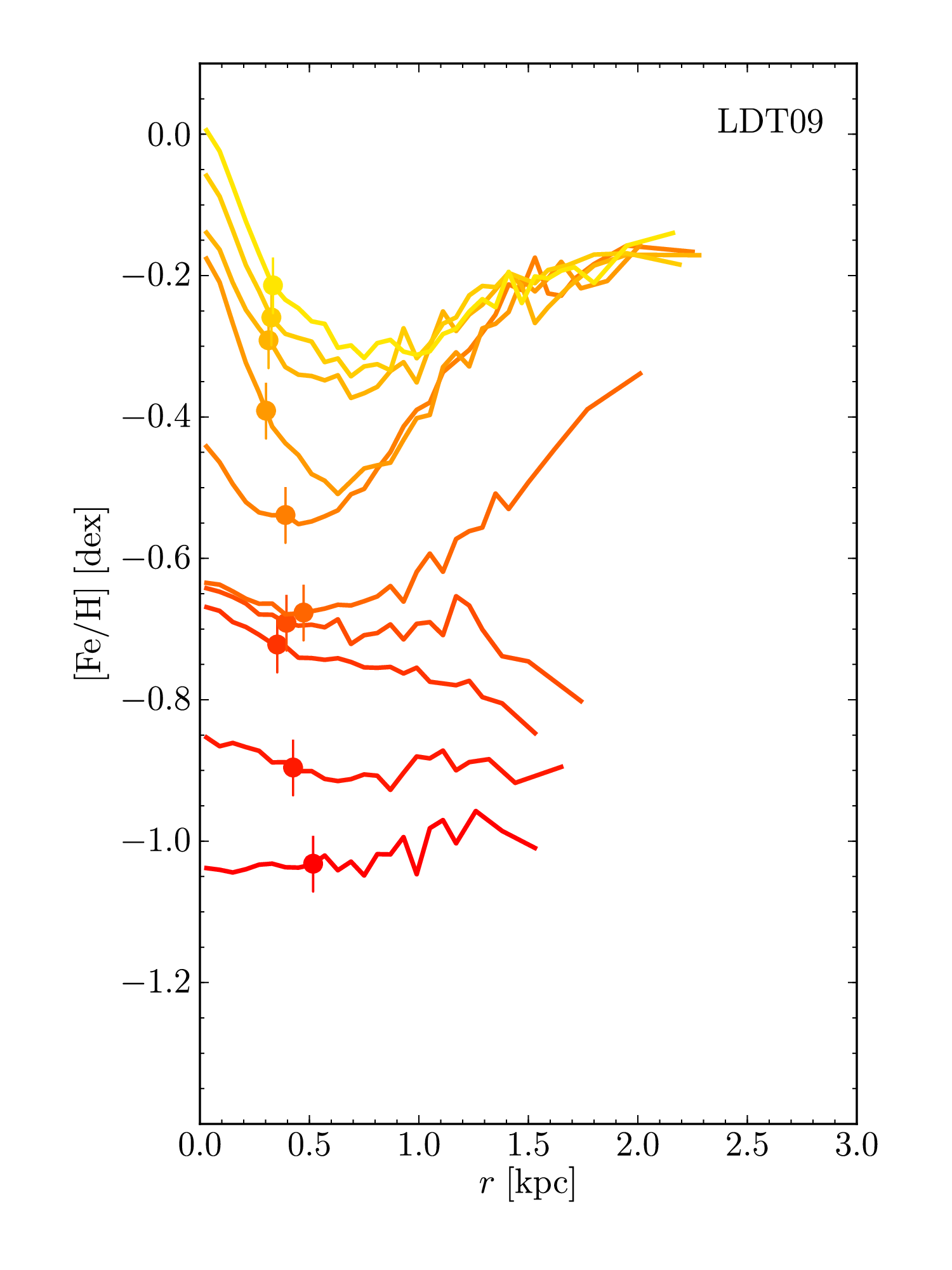}
\hfill
\\
 \hspace*{0mm} 
\hfill
\includegraphics[width=0.45\textwidth]{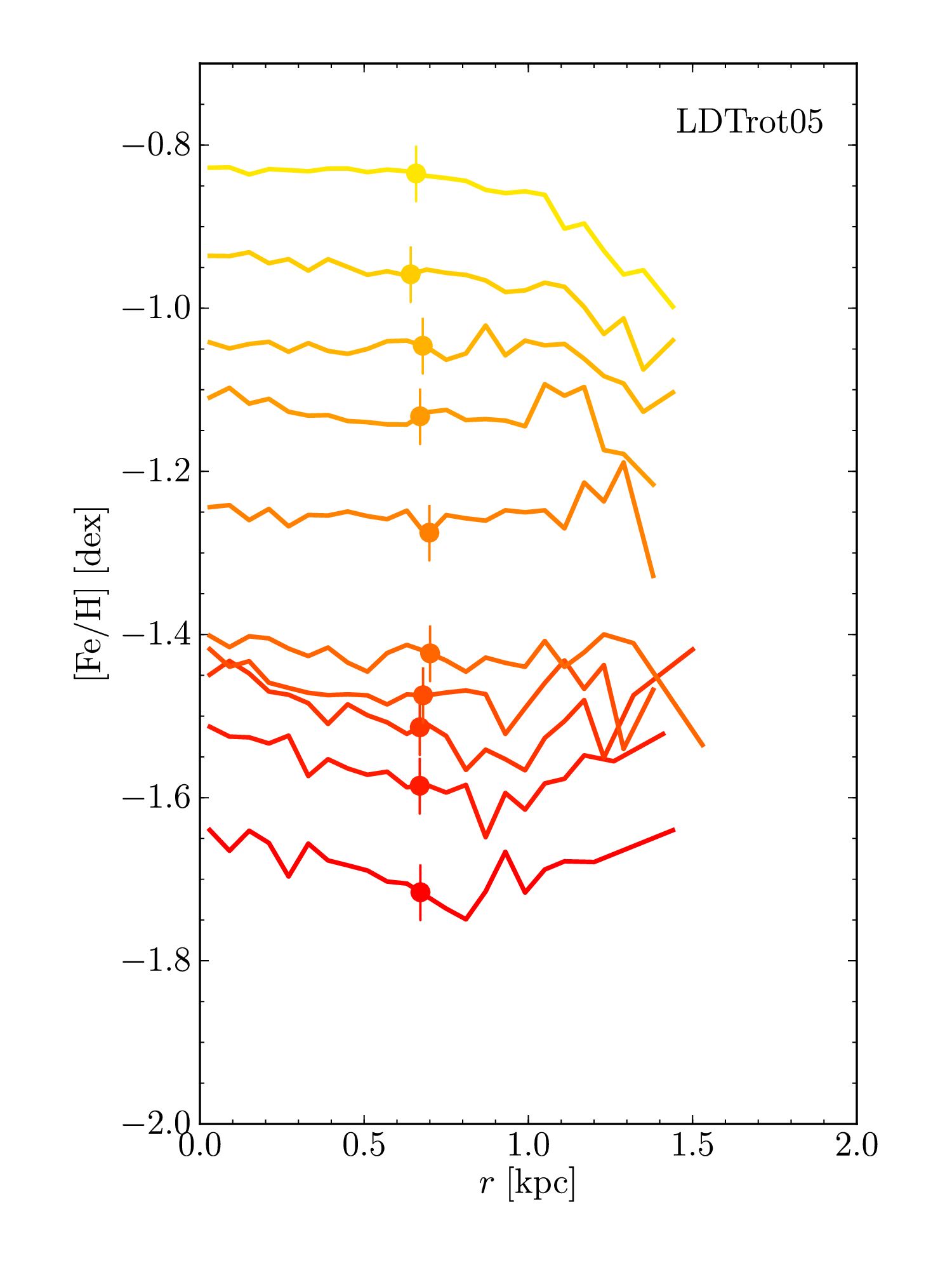}
\hfill
\includegraphics[width=0.45\textwidth]{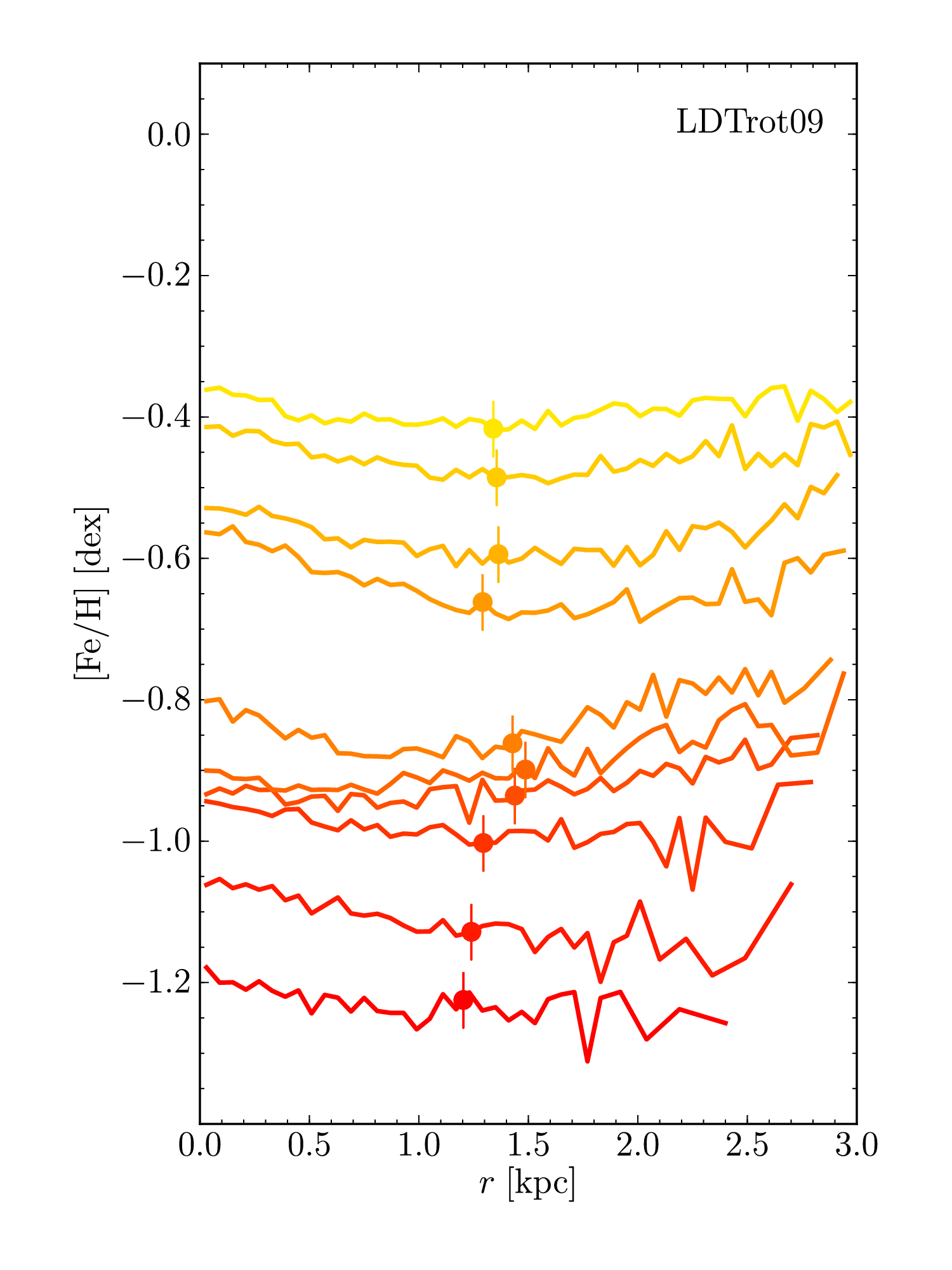}
\hfill

\caption{Mass-weighted metallicity profiles of the LDT models. Plot
configuration as in Fig.~\ref{LDT_metprof_lum}. Here, the metal
contribution of each star particle is weighted by its mass instead of
its luminosity, and therefore the young population does not have the
potential to dominate the measurements. This quantity is not what is
actually observed, but it gives a better view of the true physical
distribution of metals.  \label{LDT_metprof_mass}}

\end{minipage}
\end{figure*}

Figures \ref{LDT_metprof_lum} and \ref{LDT_metprof_mass} show the
evolution of, respectively, the luminosity-weighted and mass-weighted
metallicity profiles in the LDT simulations (Table \ref{table_ldt_sims}).

In the non-rotating cases on the one hand, on the top row of the
figures, negative metallicity gradients can be seen to gradually build
up inside a radius of $\sim 2$ half-light radii during the model's
evolution. This is due to the centrally concentrated, episodic star
formation, which is confined to progressively smaller areas over time
(see Section \ref{section_ldtvshdt} and \citealt{joeri:angmomentum}),
adding to the overall gradient. The more massive model on the top right
plot, however, shows a temporary positive metallicity gradient in its
outskirts between 3 and 4\,Gyr, which is caused by a star formation
episode that starts at larger radii and moves inward (see Fig.
\ref{SFHistogram}). A central negative gradient is quickly restored once
the star formation reaches the center, and remains stable - except for
aging/reddening of the stars which affects the luminosity weighted
values (compare Fig.~\ref{LDT_metprof_lum} to \ref{LDT_metprof_mass}).

The rotating models on the other hand, on the bottom row of the figures,
show metallicity profiles which are flat throughout practically the
whole evolution, out to well past their half-light radius. As shown in
\citet{joeri:angmomentum}, the presence of angular momentum smears out
SF in space and time, leading to a chemically more homogeneous galactic
body.


\subsubsection{Truncated LDT sims}

We truncated the star formation around 8\,Gyr into the simulation. By
then, clear metallicity gradients have built up in all non-rotating
simulations. Furthermore, since the model galaxies have already used up
or dispersed a substantial amount of their gas when forming stars, the
effect of sudden removal of all gas on their structure should be
limited.

\begin{figure*}
\begin{minipage}[t]{2\columnwidth}
\hfill
\includegraphics[width=0.24\textwidth]{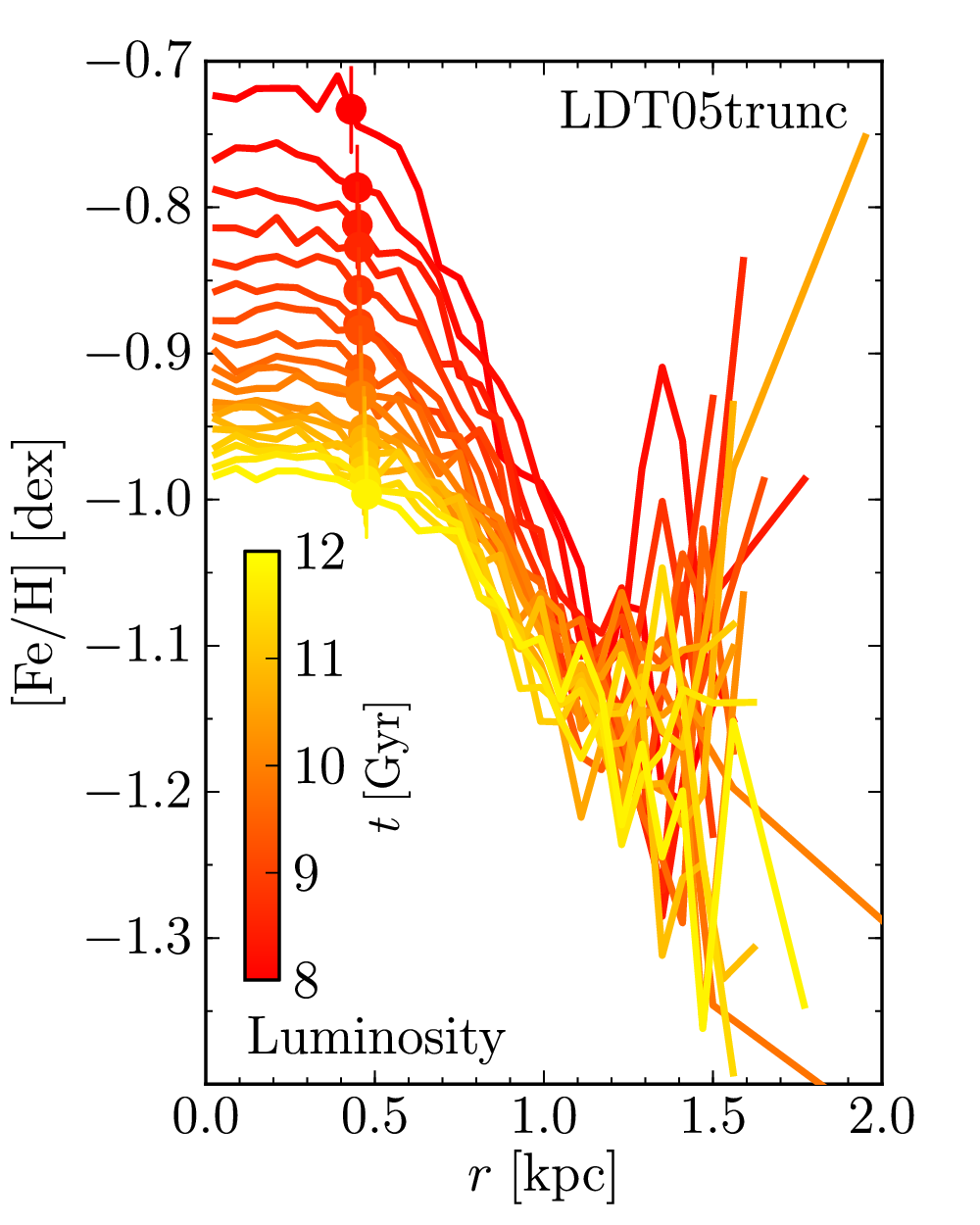}
\hfill
\includegraphics[width=0.24\textwidth]{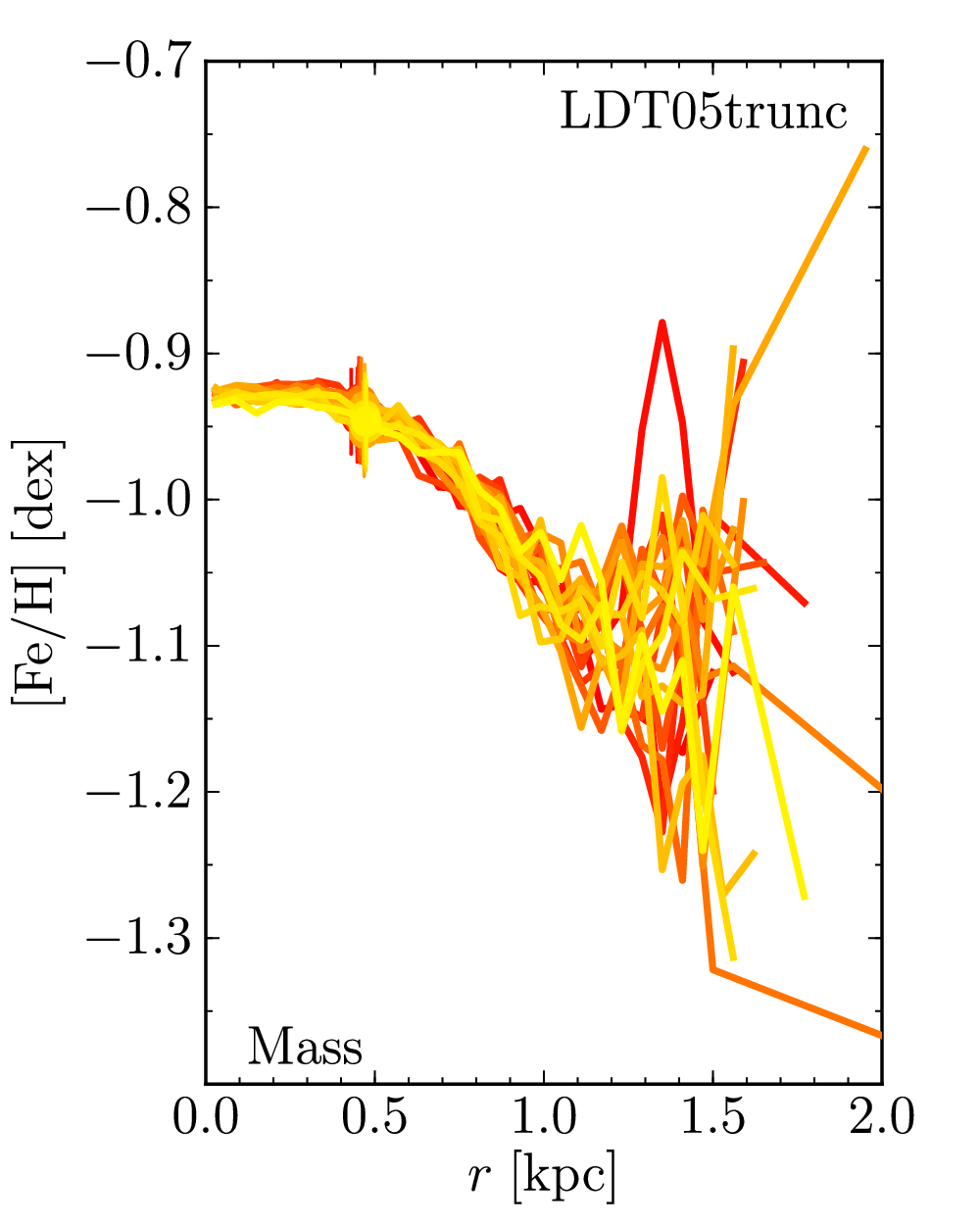}
\hfill
\includegraphics[width=0.24\textwidth]{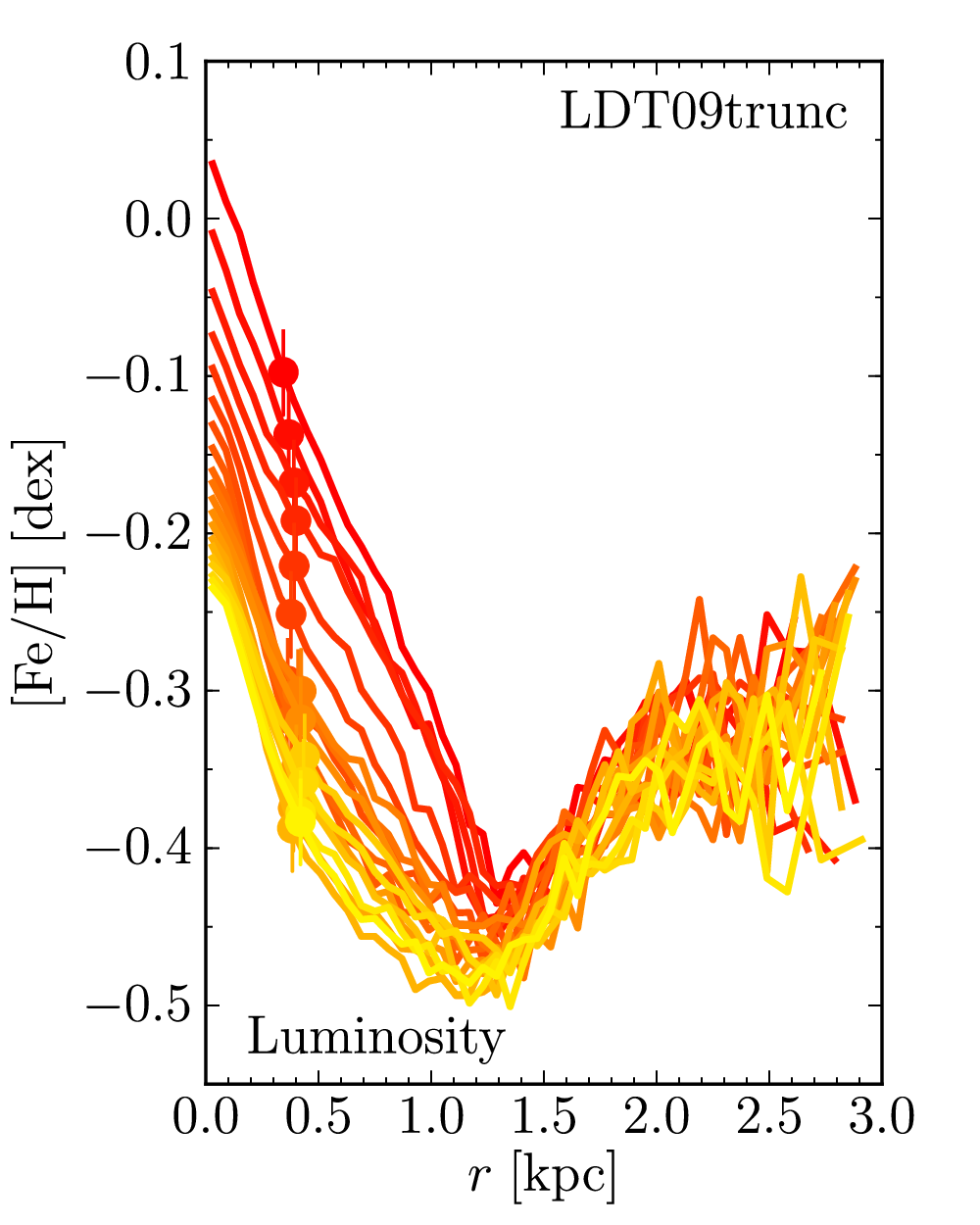}
\hfill
\includegraphics[width=0.24\textwidth]{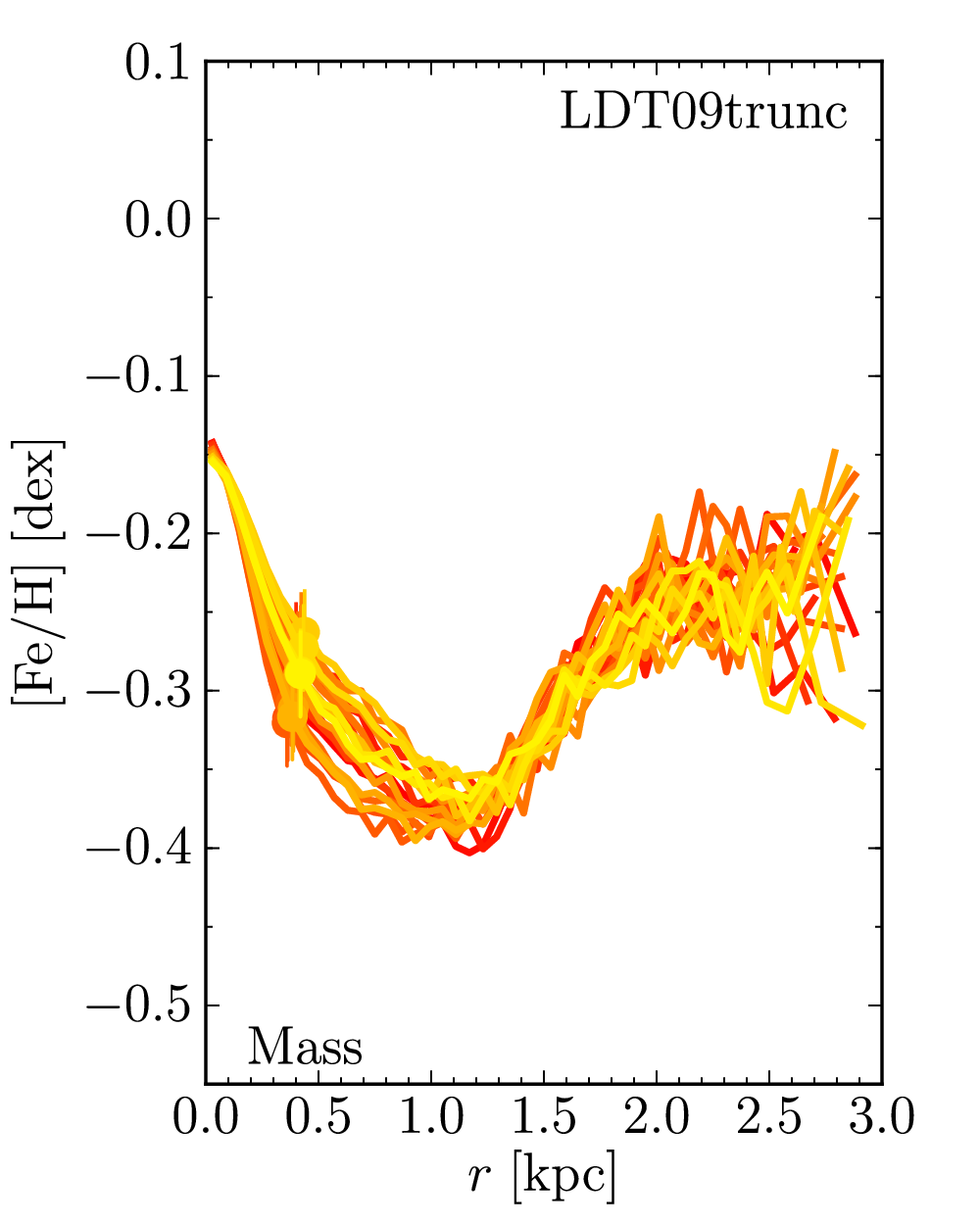}
\hfill

\caption{Metallicity profiles of the truncated LDT sims. The respective
 model and weighting method is indicated on the plot. The time is color
 coded according to the color bar in the leftmost plot.
 \label{LDTtrunc_metprof}}

\end{minipage}
\end{figure*}

Figure~\ref{LDTtrunc_metprof} shows the evolution of the luminosity- and
mass-weighted metallicity profiles in the truncated LDT
simulations. The luminosity-weighted gradients noticeably diminish
over time - or rather, it is mostly the central metallicity that is
dropping - but when looking at the mass-weighted profiles it appears
that there is hardly any evolution in the physical distribution of
metals. The latter can be said to get just slightly shallower over a
time span of 4\,Gyr, and even the gas removal seems to barely have an
effect (the $\mathrm{R_{e}}$ only slightly increases in the first few
timesteps).

\subsection{HDT sims}

\begin{figure*}
\begin{minipage}[t]{2\columnwidth}
\hfill
\includegraphics[width=0.45\textwidth]{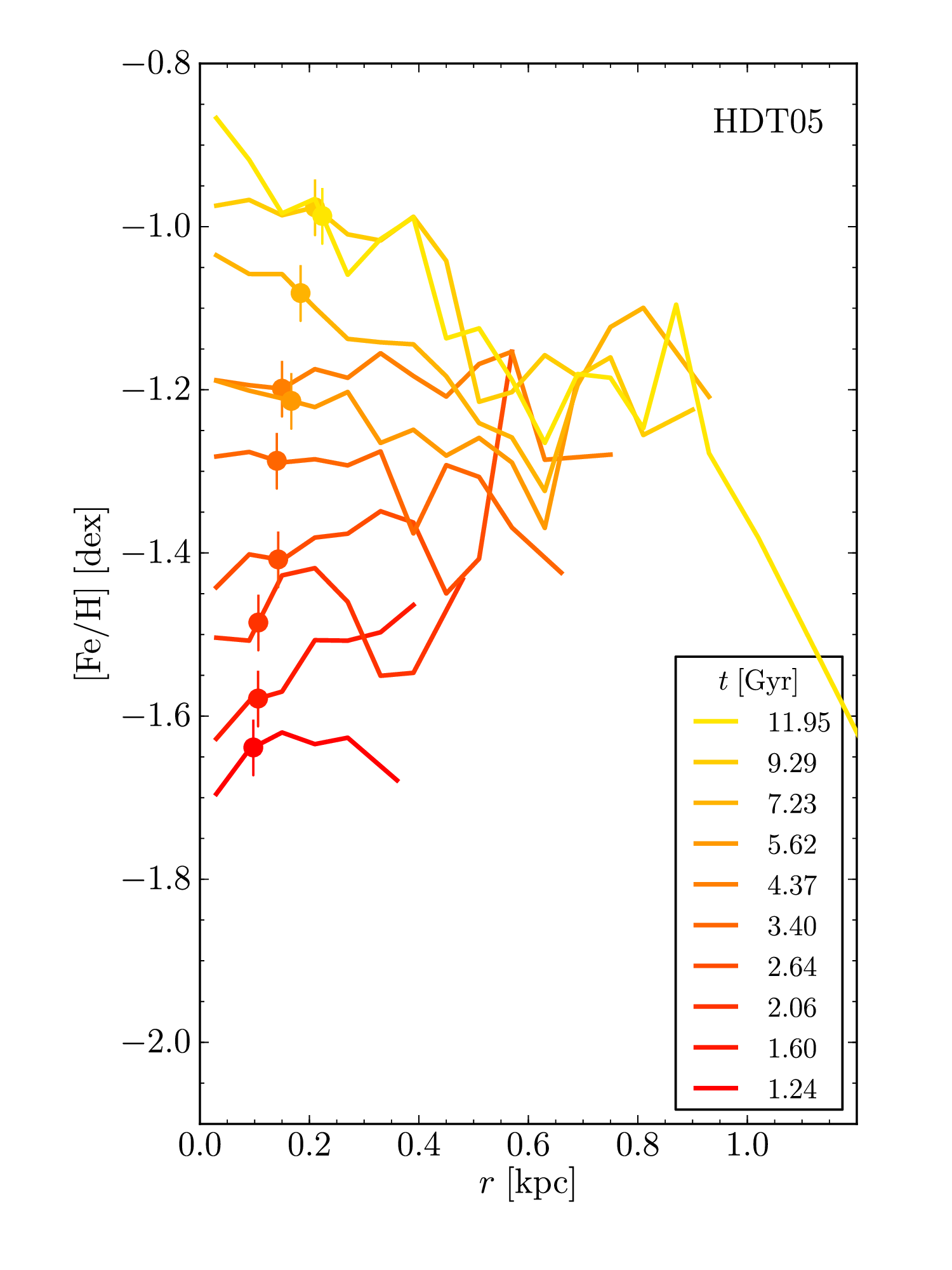}
\hfill
\includegraphics[width=0.45\textwidth]{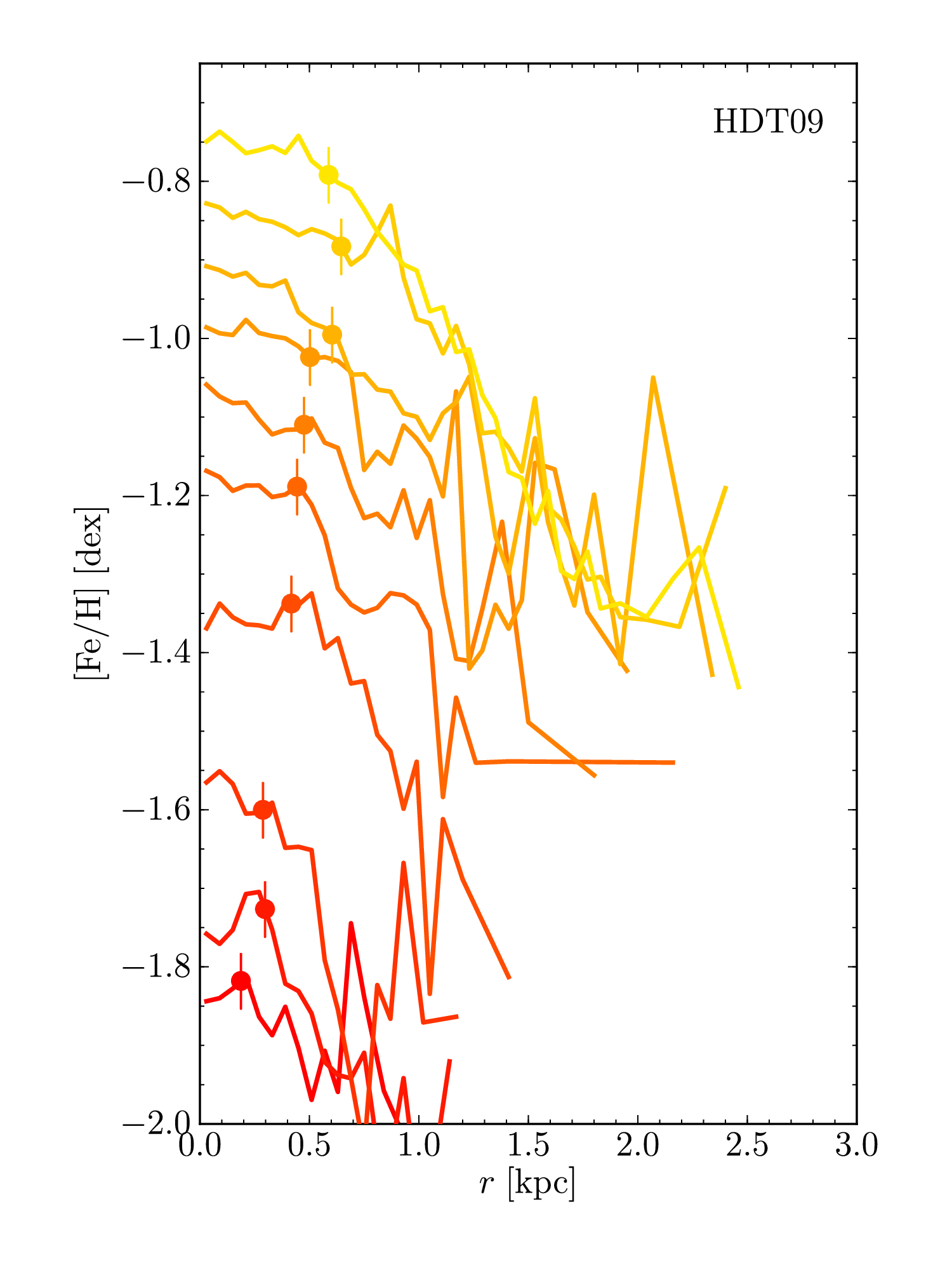}
\hfill
\\
 \hspace*{0mm} 
\hfill
\includegraphics[width=0.45\textwidth]{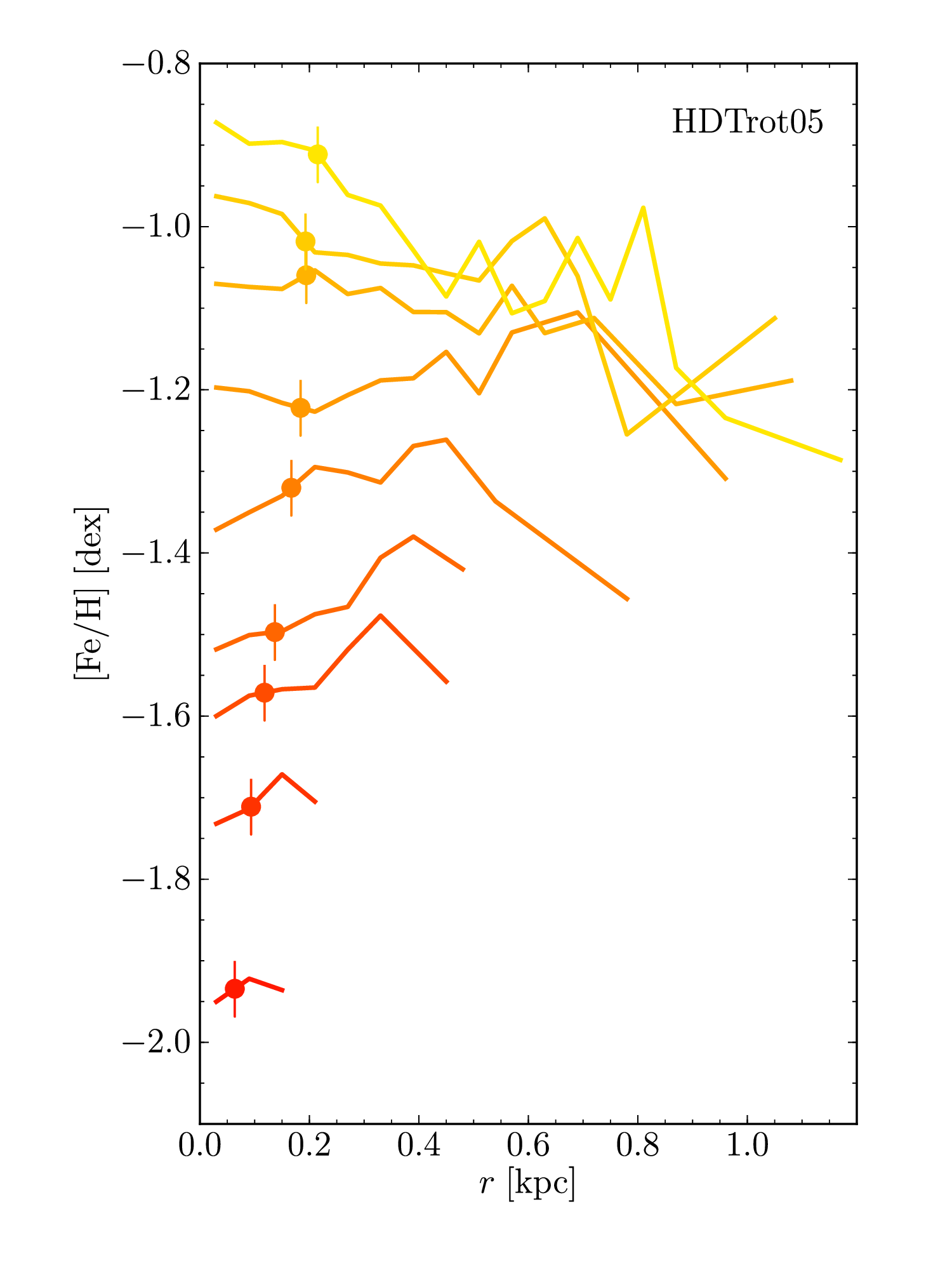}
\hfill
\includegraphics[width=0.45\textwidth]{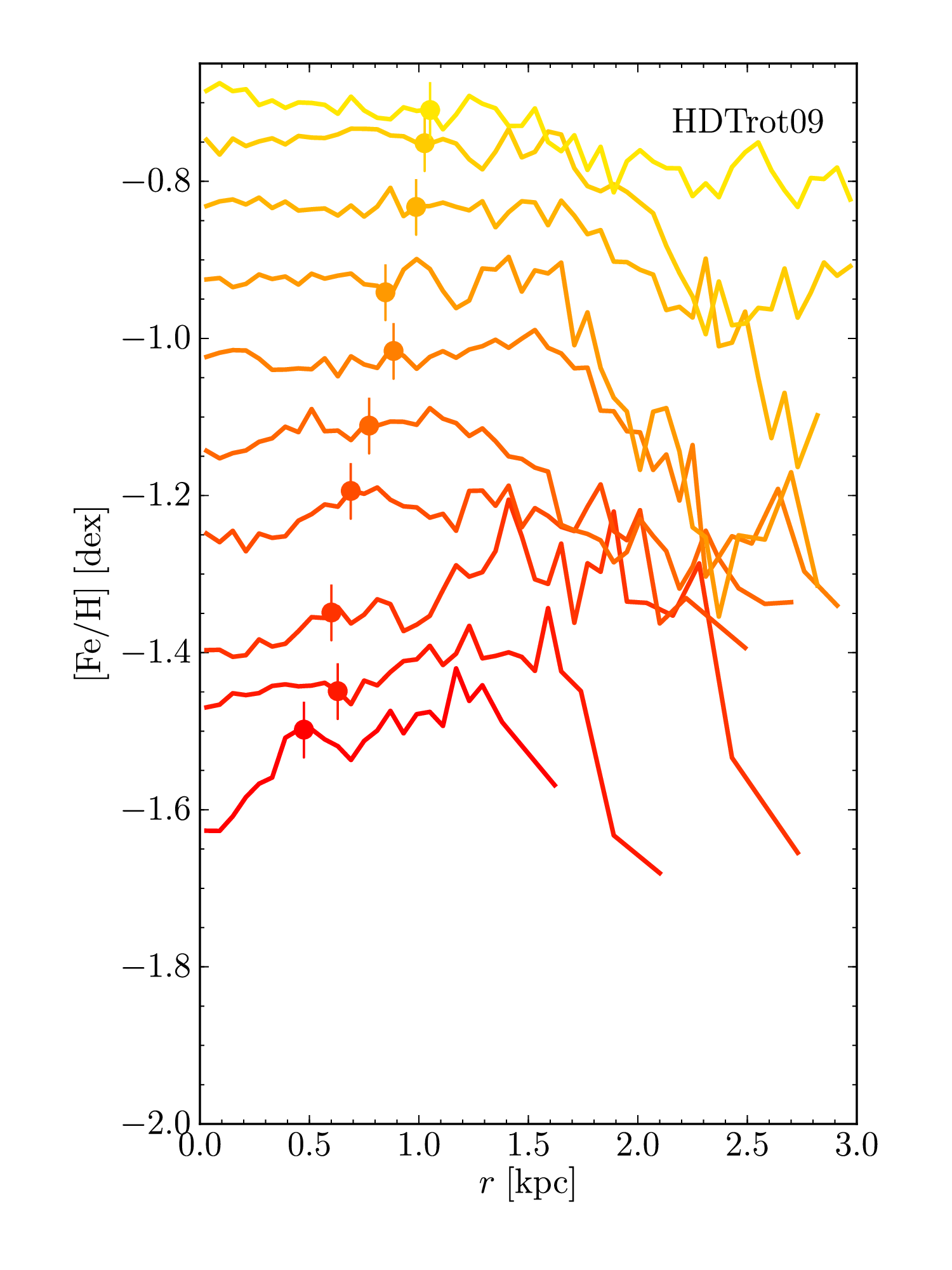}
\hfill 

\caption{Luminosity-weighted (B-band) metallicity profiles of the HDT
models. Plot setup similar to Fig.~\ref{LDT_metprof_lum}, the
simulations can be found in Table \ref{table_hdt_sims}.
\label{HDT_metprof_lum}}

\end{minipage}
\end{figure*}

\begin{figure*}
\begin{minipage}[t]{2\columnwidth}
\hfill
\includegraphics[width=0.45\textwidth]{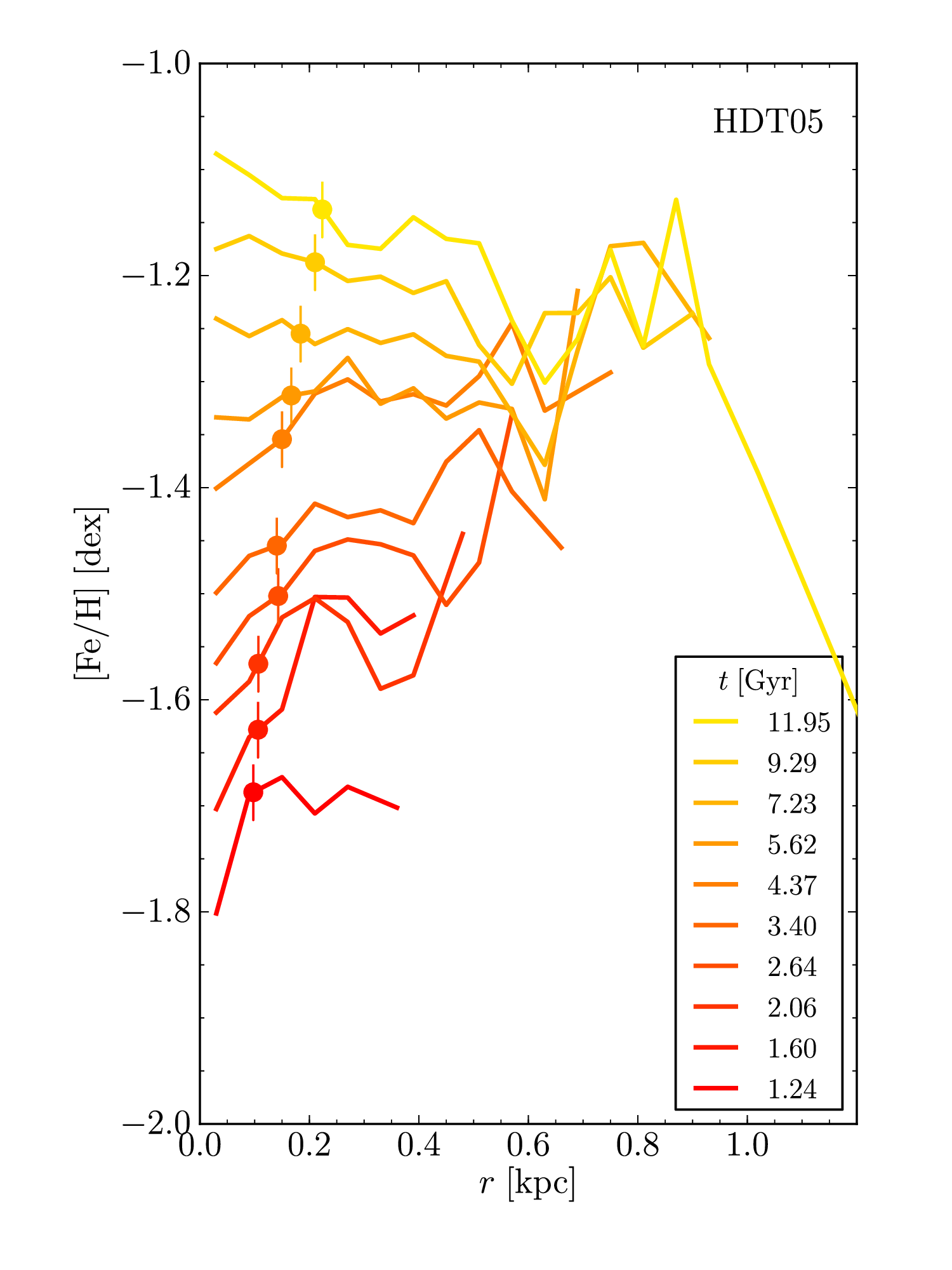}
\hfill
\includegraphics[width=0.45\textwidth]{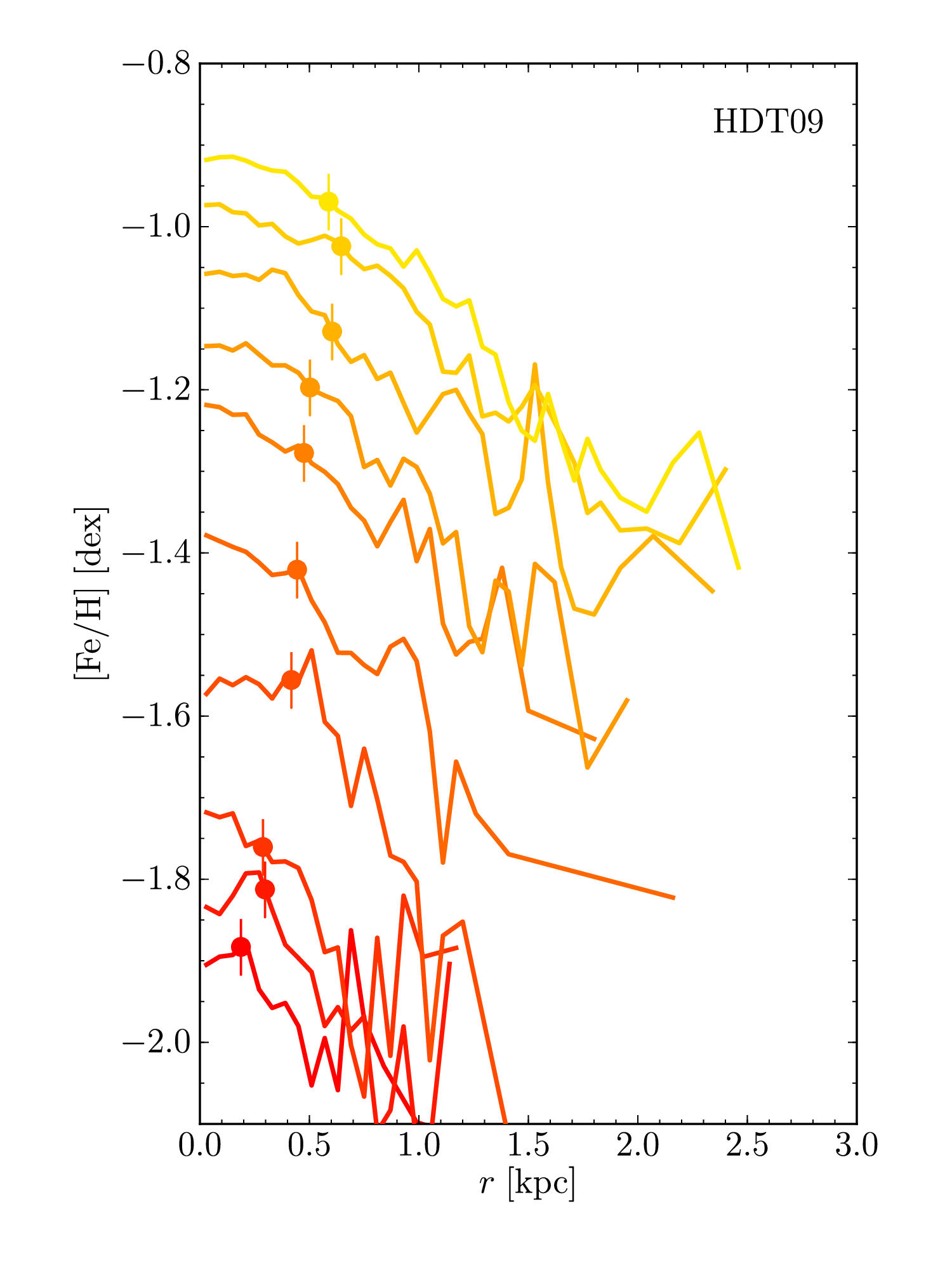}
\hfill
\\
 \hspace*{0mm} 
\hfill
\includegraphics[width=0.45\textwidth]{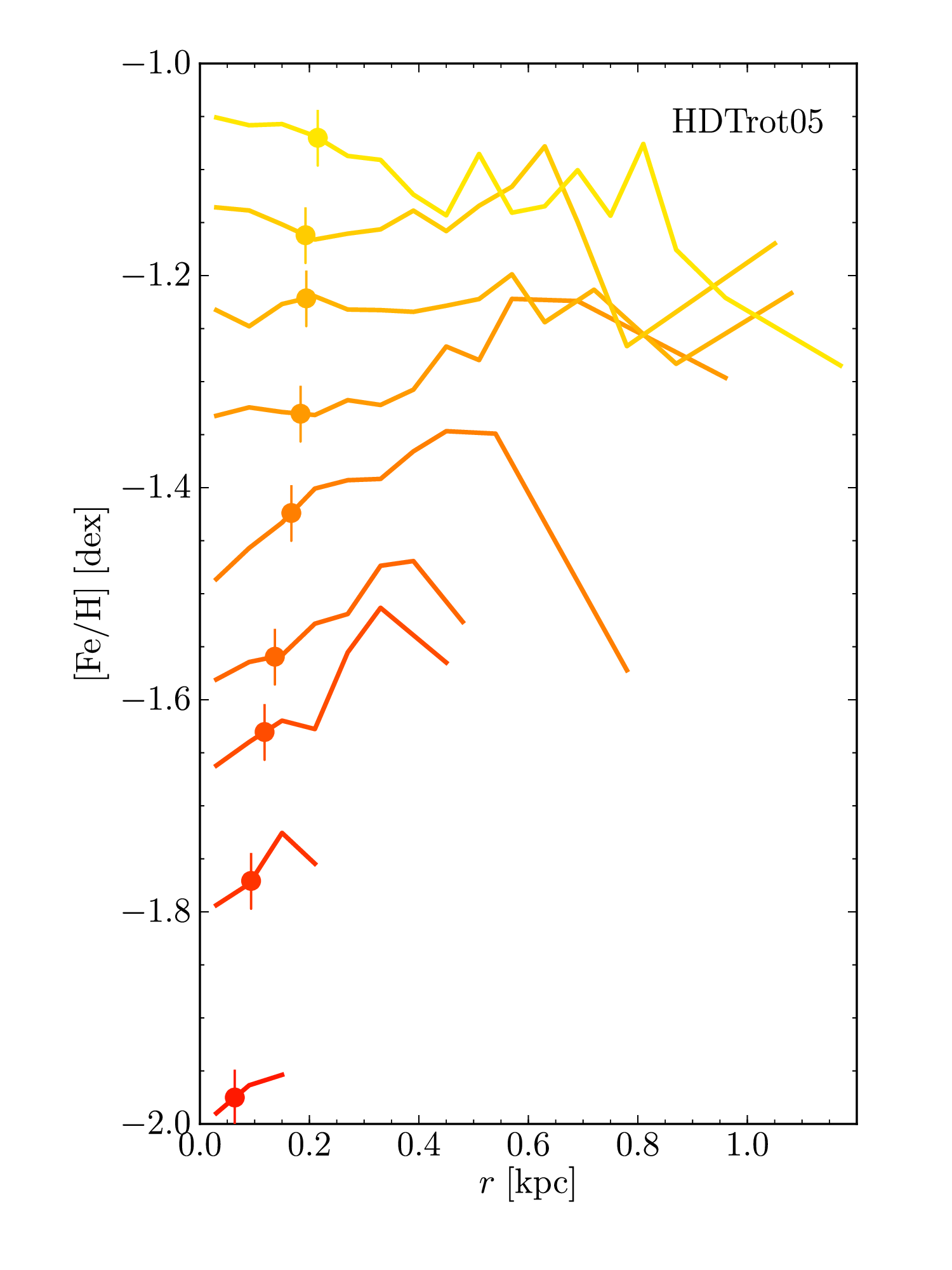}
\hfill
\includegraphics[width=0.45\textwidth]{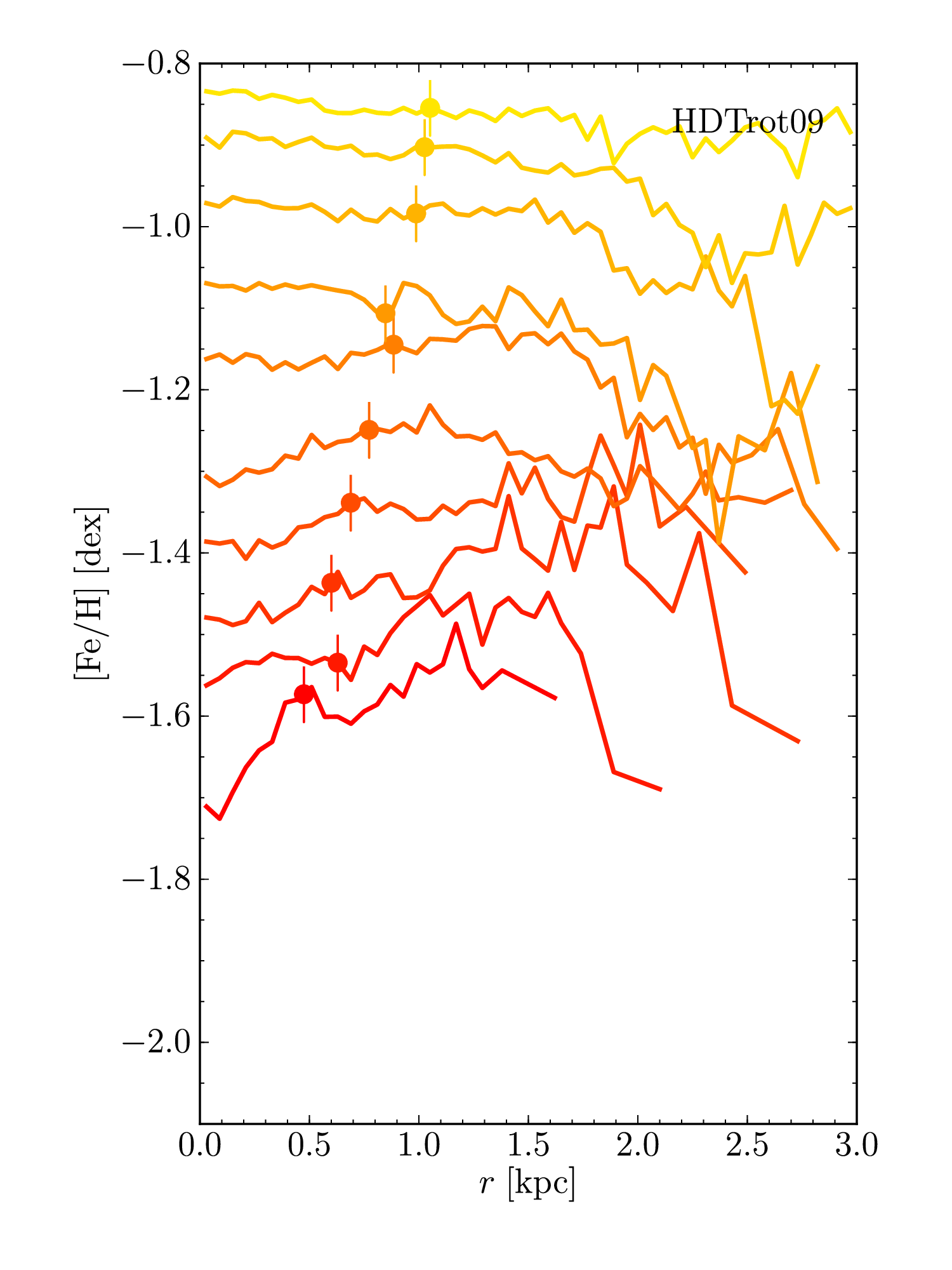}
\hfill

\caption{Mass-weighted metallicity profiles of the HDT models. Plot
 setup similar to Fig.~\ref{LDT_metprof_mass}.
\label{HDT_metprof_mass}}

\end{minipage}
\end{figure*}

Figures \ref{HDT_metprof_lum} and \ref{HDT_metprof_mass} show the
evolution of, respectively, the luminosity-weighted and mass-weighted
metallicity profiles in the HDT simulations (Table \ref{table_hdt_sims}).

Though the SF timescale is now much shorter (Section
\ref{section_ldtvshdt}), there still is a general shrinking of the SF
area over time, and an overall centrally concentrated star formation -
leading to the gradual buildup of negative metallicity gradients in the
non-rotating HDT models.  The high-mass model features a gradient in its
stellar body throughout its entire evolution, while in the low-mass
model it is in place from around 5\,Gyr onwards. The apparent positive
(mass-weighted) gradients during the first few Gyr of the latter, are
partly caused by mass recentering difficulties on an initially low
number of (stochastically generated) stellar particles, and the
stochastic nature of our models in general.

The rotating models show flat(ter) metallicity profiles throughout their
evolution, again explained by the spatially and temporally smeared out
star formation seen in Fig.~\ref{SFHistogram}. For the low-mass model,
however, the flattening is not as strong as in the LDT case, but the
general difference in behaviour with the non-rotating model is still
noticeable. This less pronounced effect is due to the different initial
rotation curves (which go to zero in the center, Section
\ref{section_rotation_curves}, instead of being constant), and the fact
that the HDT scheme produces smaller galaxies (Section
\ref{section_ldtvshdt}), causing the models to receive less initial
angular momentum overall.

\subsubsection{Truncated HDT sims}

\begin{figure*}
\begin{minipage}[t]{2\columnwidth}
\hfill
\includegraphics[width=0.24\textwidth]{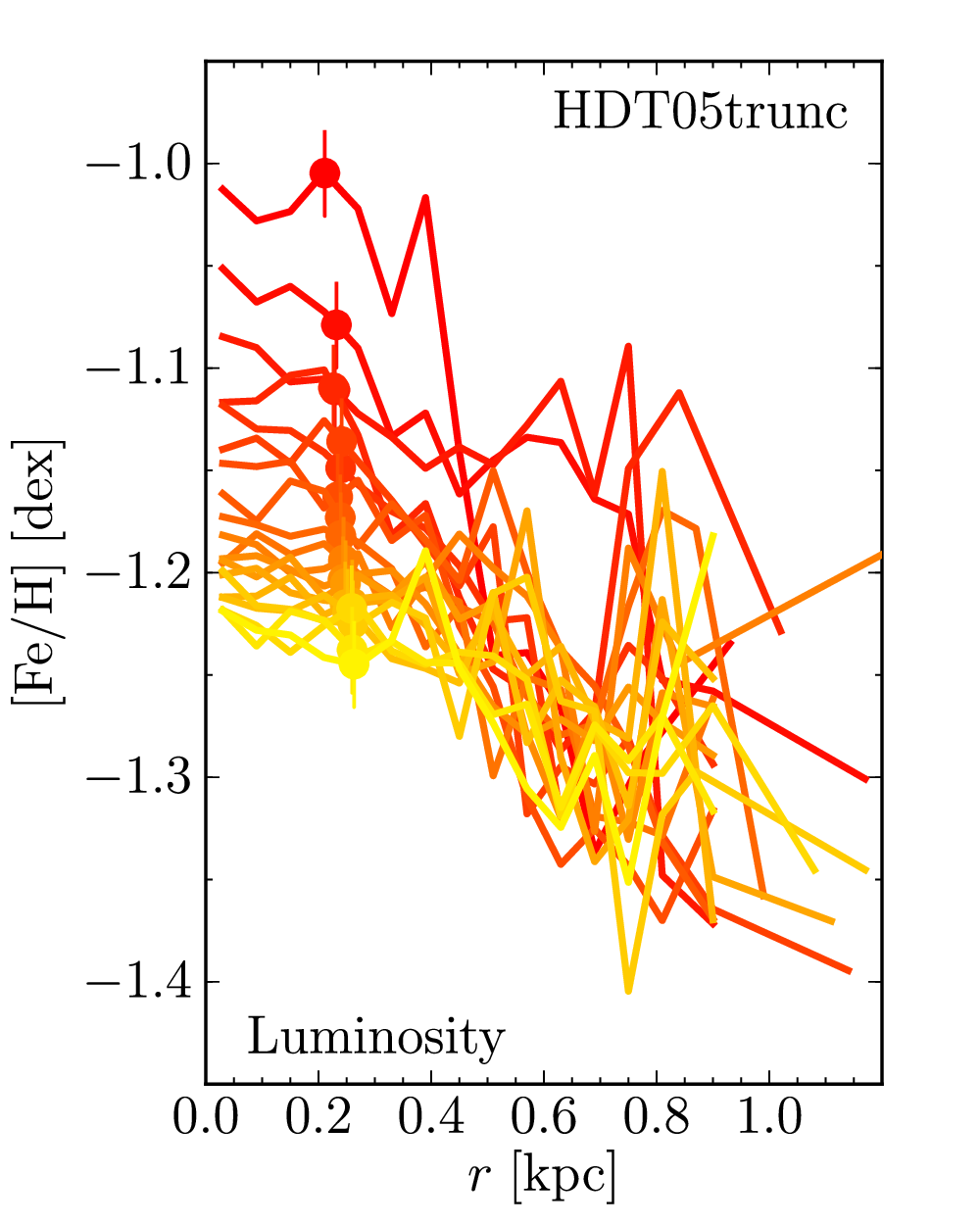}
\hfill
\includegraphics[width=0.24\textwidth]{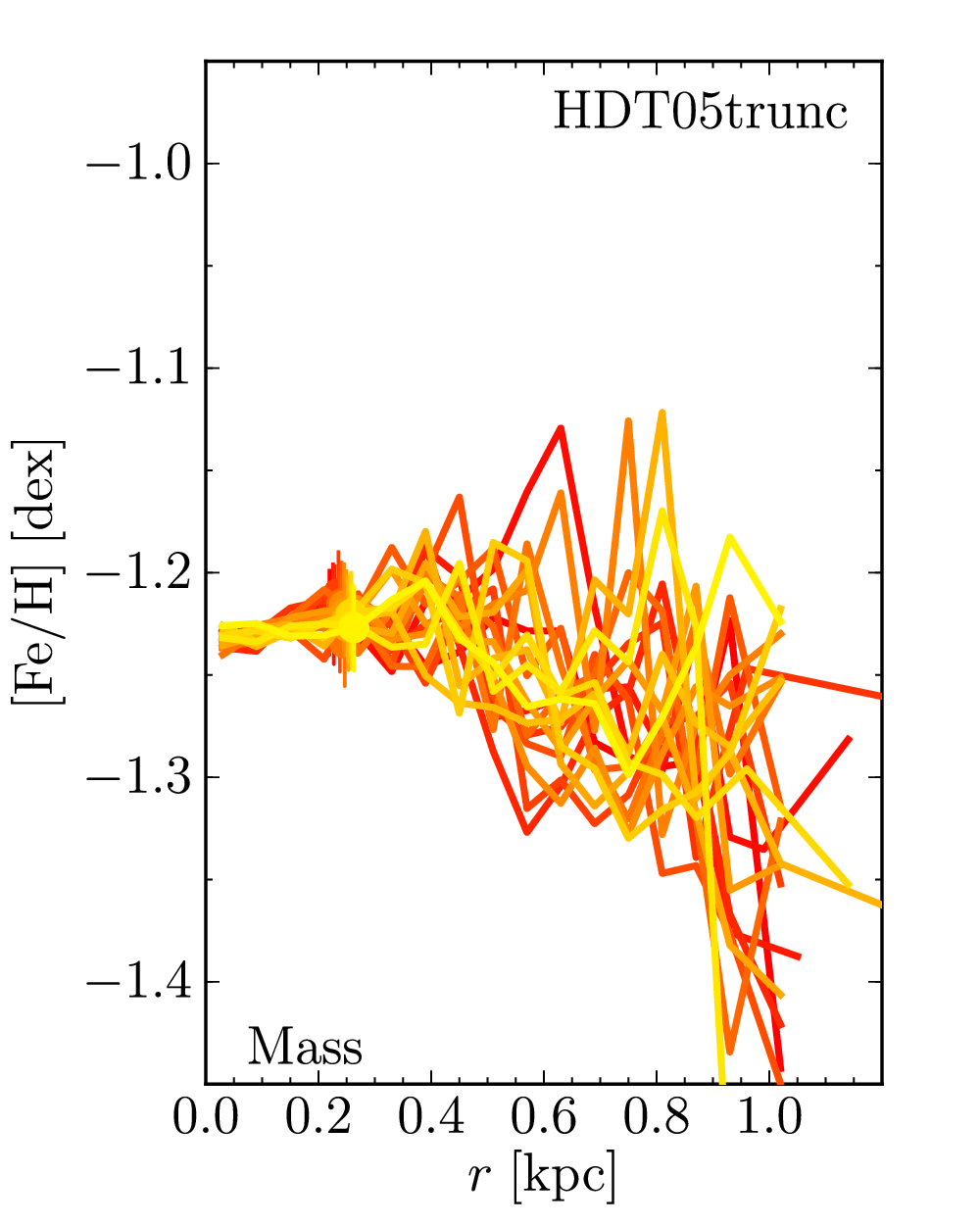}
\hfill
\includegraphics[width=0.24\textwidth]{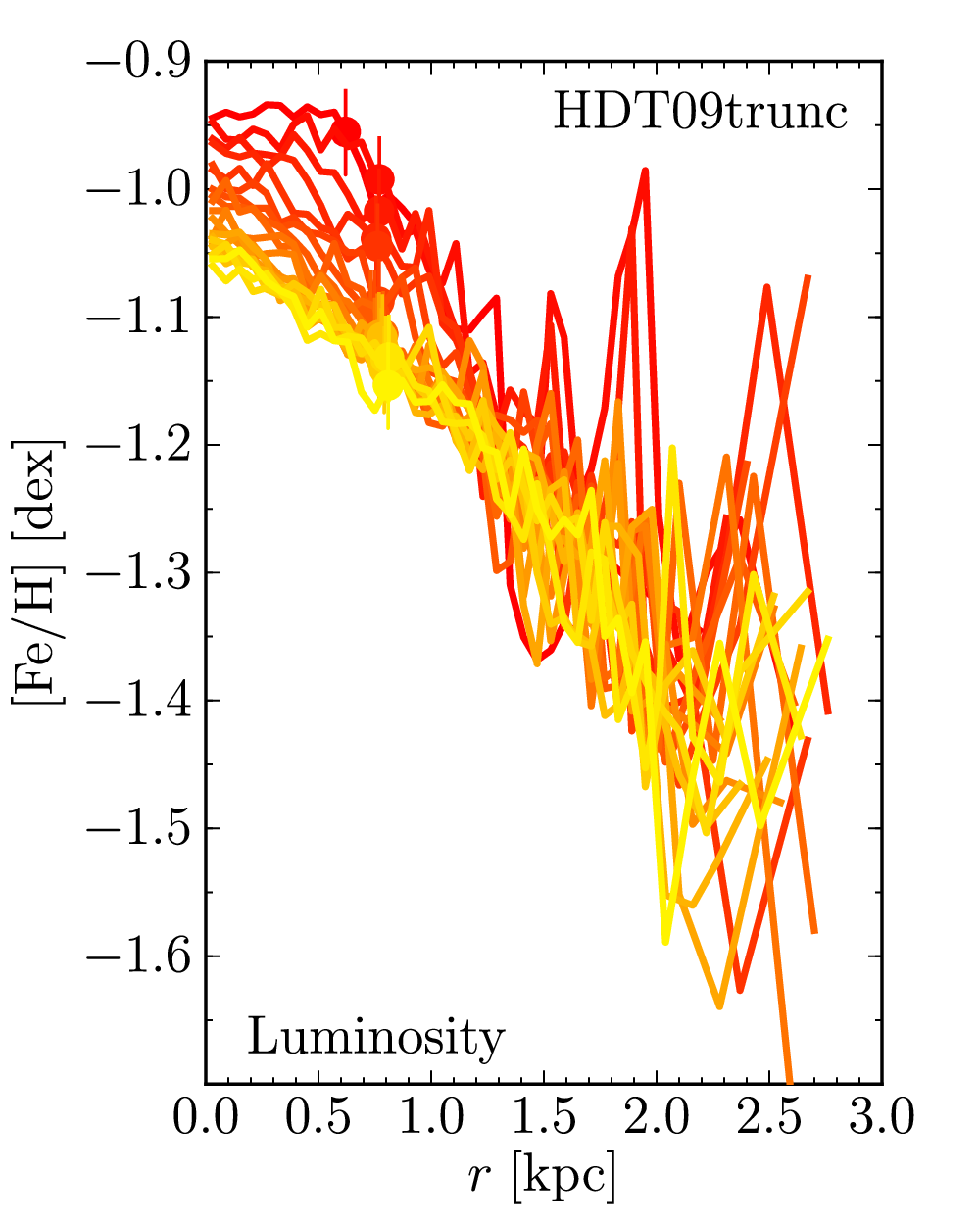}
\hfill
\includegraphics[width=0.24\textwidth]{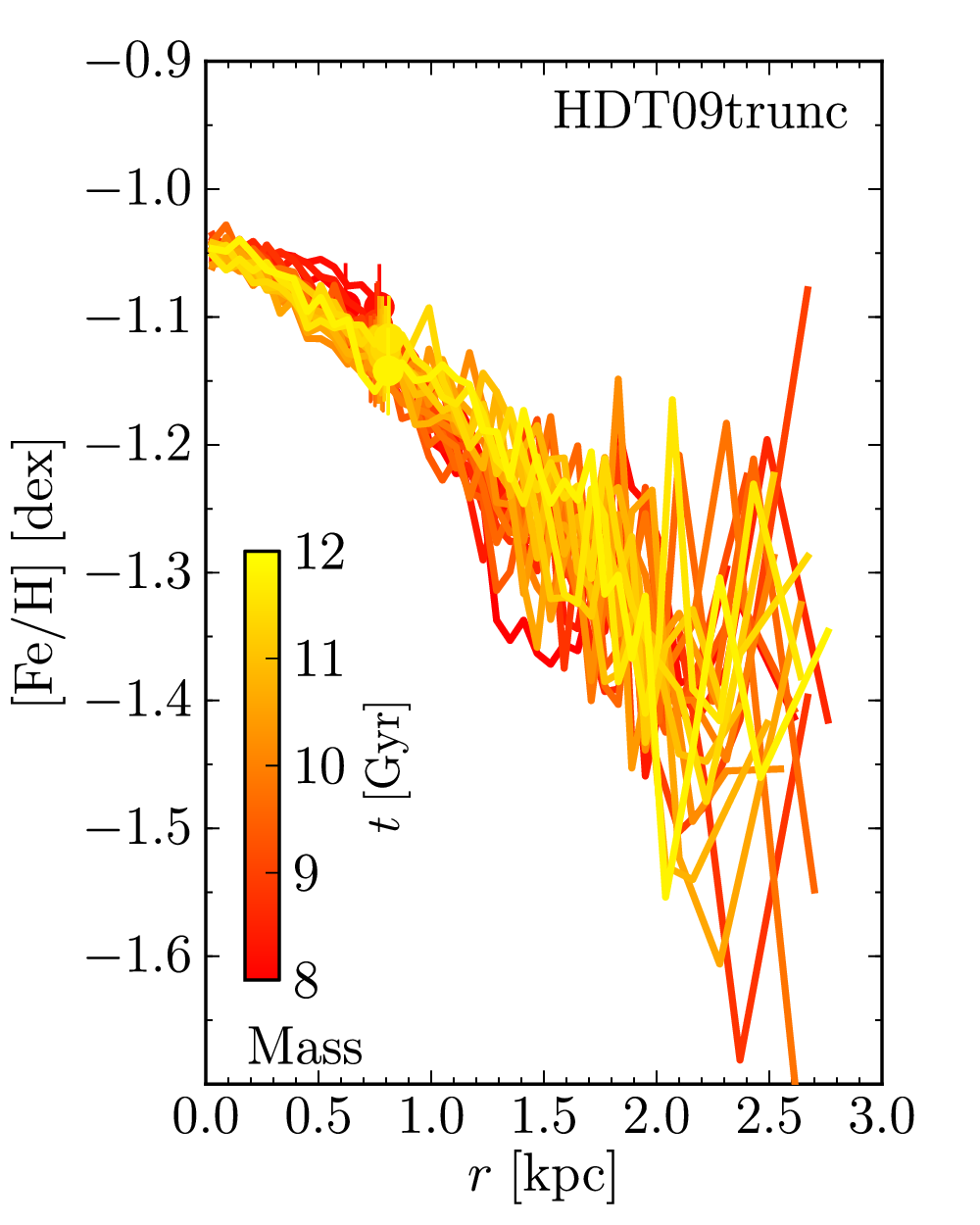}
\hfill

\caption{Metallicity profiles of the truncated HDT sims. Similar to
 Fig.~\ref{LDTtrunc_metprof}. Color bar is in the rightmost plot.
 \label{HDTtrunc_metprof}}

\end{minipage}
\end{figure*}

As in the LDT sims, the truncation time has been chosen to be around 8
Gyr. And similarly, in Fig.~\ref{HDTtrunc_metprof} the
luminosity-weighted profiles can be seen to diminish noticeably (but
survive), while the mass-weighted profiles are much more stable. It
should be noted that the latter vary more significantly than in the LDT
models, meaning the physical distribution of metals changes more, but in
absolute terms this is still very limited. The evolution of the
half-light radius in the first few timesteps also shows that the gas
removal has a bigger effect here, because of the larger amount of gas
present in the central regions (due to higher densities and less gas
used) which is suddenly removed.

The HDT05trunc profiles are particularly noisy, due to the relatively
low number of particles that is actually used for generating the
profile. The stellar mass is very low to begin with (see Table
\ref{table_hdt_sims}, which shows the value at 12\,Gyr), and
additionally, the method used to generate the profiles excludes a large
part of these stellar particles (see beginning of this section).

\subsection{General conclusions on metallicity gradients}

The basic findings on metallicity gradients and the mechanism behind
their evolution from \citet{joeri:angmomentum} hold true in the LDT and
HDT schemes. 

The general conclusions about the evolution of metallicity gradients in
our non-rotating dwarf galaxy models are that
\begin{itemize}
\item metallicity gradients are gradually built up during the evolution
      of the dwarf galaxy, by centrally concentrated star formation
      which additionally gets limited to smaller areas over time,
      progressively adding to the overall metallicity gradient.
\item formed metallicity gradients seem to be robust, and able to
      survive over several Gyr, without significantly changing the
      physical distribution of metals.
\end{itemize}
All this strongly speaks against the possibility of erasing metallicity
gradients in dwarf galaxies without an external disturbance, since even
our most realistic (HDT) models do not show any significant decline in
the gradients.

\subsubsection{Comparison to observed metallicity gradients}

\begin{figure}

\includegraphics[width=0.45\textwidth]{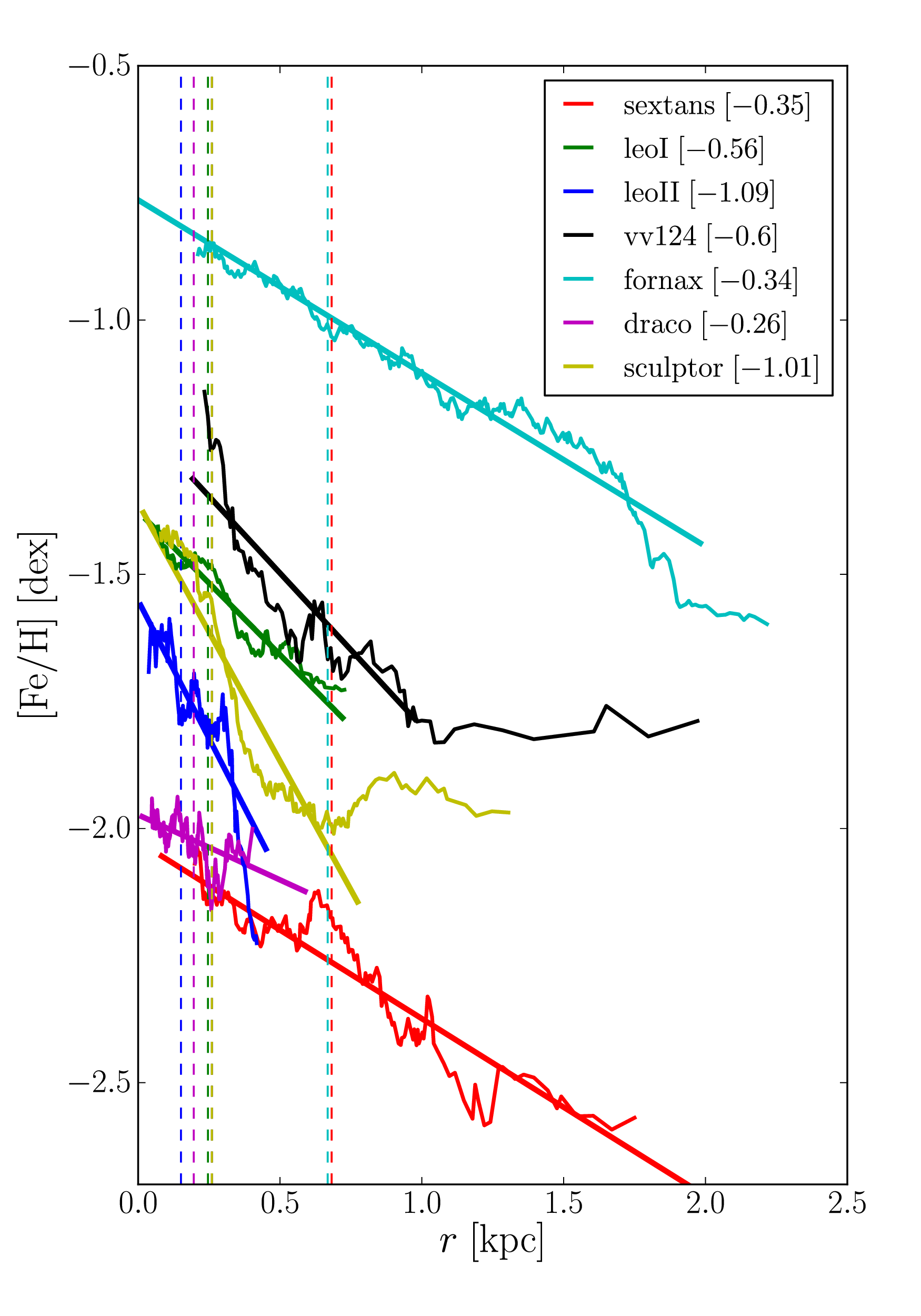}
\caption{Observed radial stellar metallicity profiles of 7 Local Group
dwarf galaxies, in function of radius in kpc. The slopes of the linear
fits to these gradients are indicated in brackets in the legend of the
figure, and are expressed in dex/kpc. Dashed vertical lines indicate the
half-light radii of the galaxies.}

\label{observed_gradients}

\end{figure}

We can compare the stellar metallicity gradients in our model dwarf
galaxies with observed stellar metallicity gradients from dwarf galaxies
in the Local Group, with the aid of Fig.~\ref{observed_gradients} and
Table \ref{obs_sim_comp_table}. We selected 7 dwarf galaxies from the
Local Group, for which the literature provides data on their metallicity
gradients that extend far enough outward (several times their half-light
radius). These galaxies (and corresponding references) are: Sculptor
\citep{tolstoy04}; Fornax \citep{battaglia06}; Sextans
\citep{battaglia11}; LeoI, LeoII, Draco \citep{kirby11}; VV124
\citep{kirby12}. Structural parameters for these object were taken from
the table that \citet{lokas11} compiled from the literature, which were
(for the objects of interest here) mostly obtained from \citet{mateo98}
and \citet{walker10}.

Figure~\ref{observed_gradients} shows the metallicity gradients from the
observed dwarf galaxies, where the radial distance is expressed in,
respectively, kiloparsec and $R_{e}$.  A line is fit to the data points
within $3\,R_{e}$, the slope of which is in brackets in the legend of
the figure and in Table \ref{obs_sim_comp_table}. This collection of
objects displays a wide variety in absolute metallicities, slopes, and
profile shapes, as do the models:
\begin{itemize}
\item Steep, sharply peaked metallicity profiles with a possible
      increase or ``bump'' at larger radii (VV124, Sculptor), which is
      comparable to the metallicity profile of the LDT09 model (Fig.
      \ref{LDT_metprof_lum}, top right, yellow curve). As in the
      simulations, this bump could indicate a significant star formation
      episode in the past that mainly took place in the outer regions of
      the dwarf galaxy, and temporarily enriched these regions more than
      the inner regions. The central metallicity peak is likely
      connected to these dwarf galaxies/models being more centrally
      concentrated.
\item Metallicity profiles that show much less or almost no gradient within their
      $R_{e}$, but get steeper at larger radii (LeoII, Draco), as is the
      case in the LDT05 and HDT09 model (respectively top left of Fig.
      \ref{LDT_metprof_lum} and top right of Fig.
      \ref{HDT_metprof_lum}, yellow curves).
\item Steadily decreasing, almost linear metallicity profiles over their
      entire range (Fornax, Sextans, LeoI), similar to the HDT05 model
      (Fig.~\ref{HDT_metprof_lum}, top left, yellow curve).
\end{itemize}
Table \ref{obs_sim_comp_table} lists the slopes of the metallicity
profiles in both the observations and the simulations, which have all
been calculated in the same way. The simulated dwarf galaxies, which are
set up from only two different mass models, show slopes between -0.3 and
-0.6 dex per kpc (-0.13 and -0.23 dex per $R_{e}$). The observed dwarf
galaxies, that boast a wider range of masses, show slopes between -0.25
and -1.1 dex per kpc (-0.05 and -0.25 dex per $R_{e}$).

Both in shapes and slopes the metallicity profiles of our dwarf galaxy
models fall well within the observed range of metallicity profiles. In
terms of absolute metallicity values the models are located on the high
end of the observed range, which is due to the models generally being
more massive than the observed galaxies. The values for the HDT models
lie around those of Fornax and VV124, while the values for the LDT
models however lie significantly higher, which is in agreement with what
a comparison of the dynamical masses of the observed and simulated
galaxies would imply (as discussed in Section \ref{section_ldtvshdt},
the LDT scheme shifts the galaxy models along the scaling relations
towards higher stellar masses compared to the HDT scheme).

%
%
%

\input{comparison_table}

%% file: comparison_table.tex
\begin{table*}
\begin{minipage}[t]{2\columnwidth}
\caption{Columns: (1) dynamical mass calculated as $167 \beta R_{e}
 \sigma_{c}^{2}$, with $\beta=8$ and $\sigma_{c}$ the central velocity
 dispersion [$10^{6} \mathrm{M_{\odot}}$], (2) V-band magnitude, (3)
 half-light radius [kpc], (4) central velocity dispersion [km/s], (5)
 ratio of V over velocity dispersion, (6) metallicity [dex], (7)(8)
 slope of metallicity profile. Sources: (1) and (6) use data from
 \citet{mateo98}, (2), (3) and (5) use data from \citet{lokas11}, (4)
 uses data from \citet{walker10}. All data for VV124 comes from
 \citet{kirby12} except $M_{V}$ and $V/\sigma$ from \citet{lokas11}.}

 \label{obs_sim_comp_table}

 \begin{center}
 \begin{tabular}{lcccccccc}
  \hline   
  name&mass&$M_{V}$&$R_{e}$&$\sigma_{c}$&$V/\sigma$&[Fe/H]&$\Delta$[Fe/H]&$\Delta$[Fe/H]  \\
  &[$10^{6}M_{\odot}$]&[mag]&[kpc]&[km/s]&&[dex]&[dex/kpc]&[dex/$R_{e}$] \\
  \hline
  Sculptor&15.1&-10.53&0.260&9.2&0.3&-1.8&-1.01&-0.26 \\
  Fornax&98.7&-13.03&0.668&11.7&0.18&-1.3&-0.34&-0.23 \\
  Sextans&38.7&-9.20&0.682&7.9&0.48&-1.7&-0.35&-0.24 \\
  LeoI&25.2&-11.49&0.246&9.2&0.33&-1.5&-0.56&-0.14 \\
  LeoII&9.2&-9.60&0.151&6.6&0.28&-1.9&-1.09&-0.16 \\
  Draco&24.0&-8.74&0.196&9.1&0.21&-2.0&-0.26&-0.05 \\
  VV124&19.5&-12.40&0.260&9.4&0.45&-1.14&-0.60&-0.16 \\
  \hline   
  LDT05&260&-12.51&0.430&21.3&0.18&-0.717&-0.54&-0.23 \\
  LDT09&1090&-15.62&0.391&45.7&0.03&-0.053&-0.39&-0.15 \\

  HDT05&18.2&-10.30&0.224&7.8&0.08&-0.98&-0.57&-0.13 \\
  HDT09&147&-13.34&0.588&13.7&0.57&-0.834&-0.35&-0.20 \\

  LDTrot05&200&-12.7&0.630&15.4&1.45&-0.672&-0.02&-0.01 \\
  LDTrot09&1013&-15.62&1.361&23.6&1.86&-0.281&0.01&0.02 \\

  HDTrot05&16.7&-10.24&0.217&7.6&0.95&-0.922&-0.31& -0.07 \\
  HDTrot09&278&-13.98&1.062&14.0&1.64&-0.736&-0.03&-0.03 \\
  \hline   
\end{tabular}
 \end{center}

\end{minipage}
\end{table*}

%% file: article_stellar_migration.tex
\section{Stellar orbits and kinematics}
\label{section_stellar_migration} 

This section is devoted to a more detailed analysis of the stellar
particles and their orbits and kinematics in our models. We want to have
a measure of how strongly the stellar particles actually move away from
their original orbits, to have an idea what is actually happening with
them throughout the evolution of the models, and see if this could
justify stable stellar metallicity gradients. We do this by looking at
the stellar velocity dispersions of different populations in
Fig.~\ref{dispersion_populations} and at the orbital displacements of
the stellar particles with respect to the mean radius of their orbits at
their time of birth in Fig.~\ref{LDT_stelmig_abs},
\ref{HDT_stelmig_abs}, \ref{LDT_stelmig_diff}, and
\ref{HDT_stelmig_diff}.

As mentioned in Section \ref{section_simulations}, snapshots are made
every $5$ Myr, which is sufficient to resolve each particle's orbit in
detail. The evolution of a stellar particle's radius during the course
of the whole simulation is then re-binned in the time dimension to bins
of 1\,Gyr wide, so that effectively each bin gives the average value of
the stellar particle's radius over several of its orbits. Next, for both
radius and time the averaged ``birth'' values are subtracted, and for
each time bin the average (along with several percentiles) is taken over
all available stellar particles. This gives us a visualization of the
statistical deviation from the original mean orbital radius of the star
particles in our models, in function of time since their birth. In our
simulations we looked at this statistical deviation of both the absolute
difference in mean orbital radius (Fig.~\ref{LDT_stelmig_abs} and
\ref{HDT_stelmig_abs}), and the actual, signed, difference in mean
orbital radius (Fig.~\ref{LDT_stelmig_diff} and
\ref{HDT_stelmig_diff}). The former will give a clearer picture of how
strongly the stellar particles move away from their original mean
orbital radius, the latter will indicate the preferred direction (inward
or outward). For all simulations, only the orbits of a randomly chosen
15 to 25 percent of the stellar particles are extracted from the data
snapshots, to keep the amount of files manageable.  Particles outside of
a radius of $5\times\mathrm{R_{e}}$ are not considered, since they are
in number not important for the metallicity gradients, but could have a
disproportionately large influence on the average values of deviation,
because they can move about easier at the edge of the potential
well. Finally, if a time bin has less than half of the average amount of
orbits of the previous time bins, the time bin is removed (and
subsequently all that come after that as well), to ensure the amount of
stellar particles on which the statistics are done does not become too
low.

\subsection{LDT sims}
\label{section_stelmig_LDT}

\begin{figure*}
\begin{minipage}[t]{2\columnwidth}

\hfill
\includegraphics[width=0.45\textwidth]{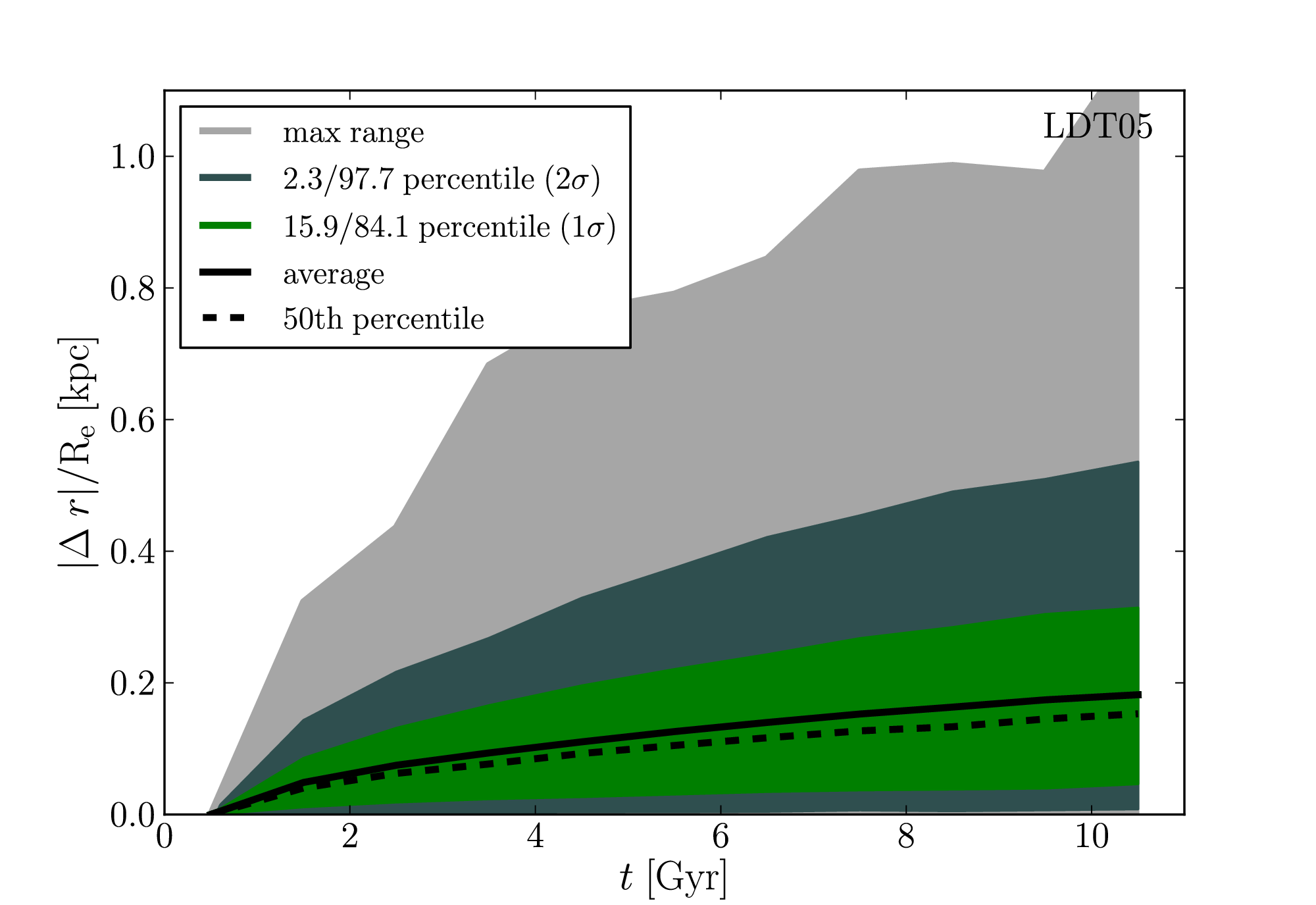}
\hfill
\includegraphics[width=0.45\textwidth]{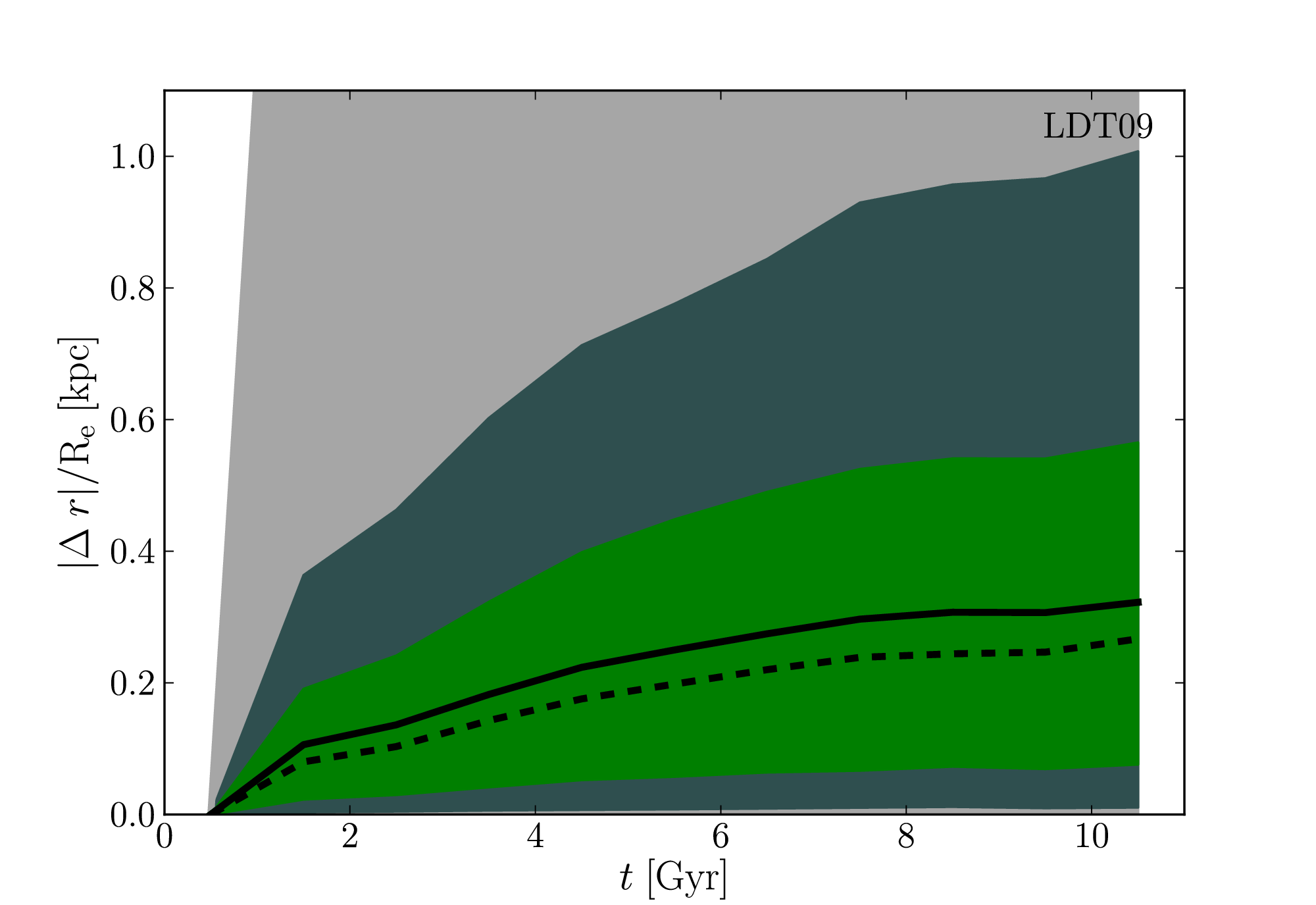}
\hfill
\\
 \hspace*{0mm} 
\hfill
\includegraphics[width=0.45\textwidth]{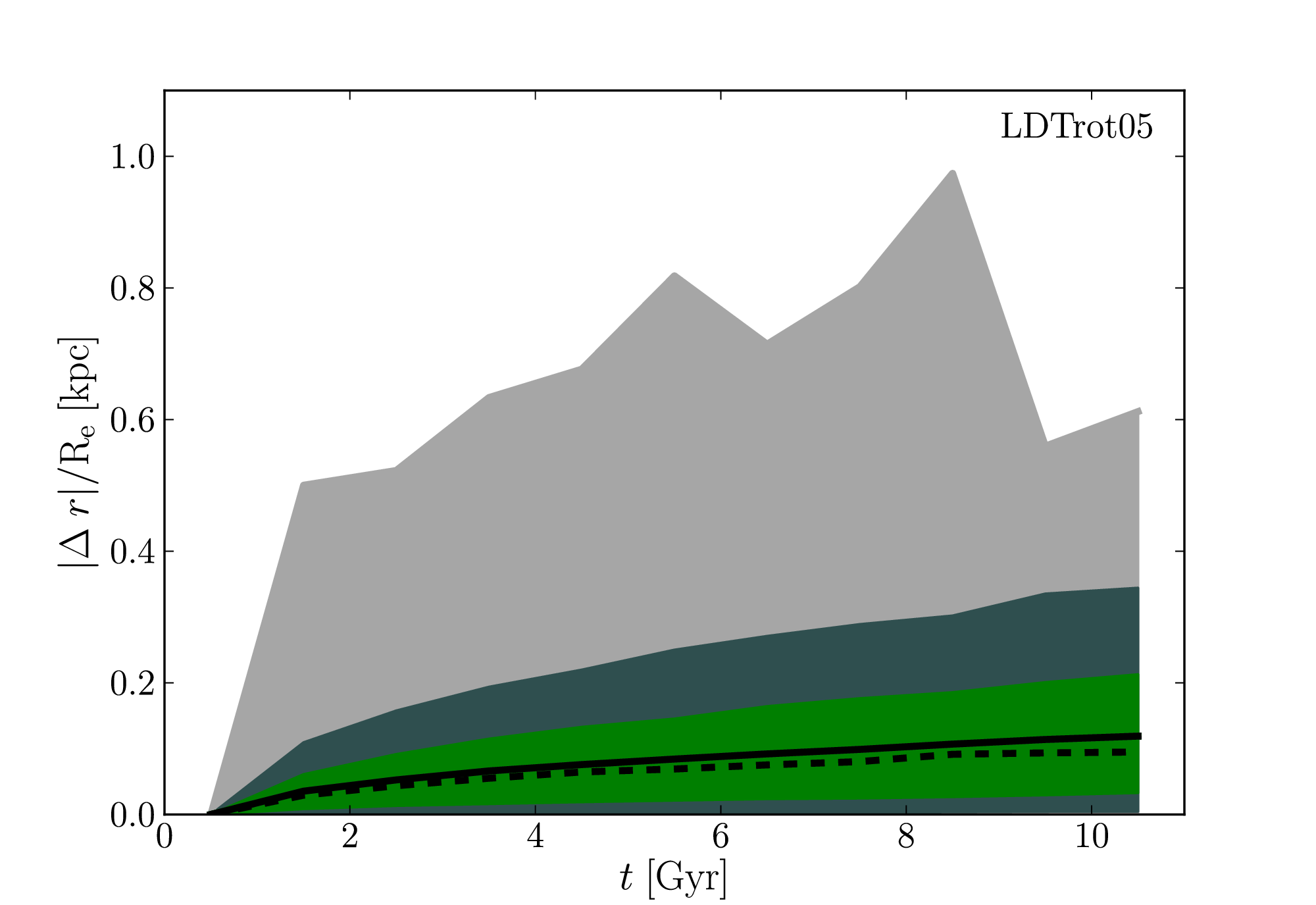}
\hfill
\includegraphics[width=0.45\textwidth]{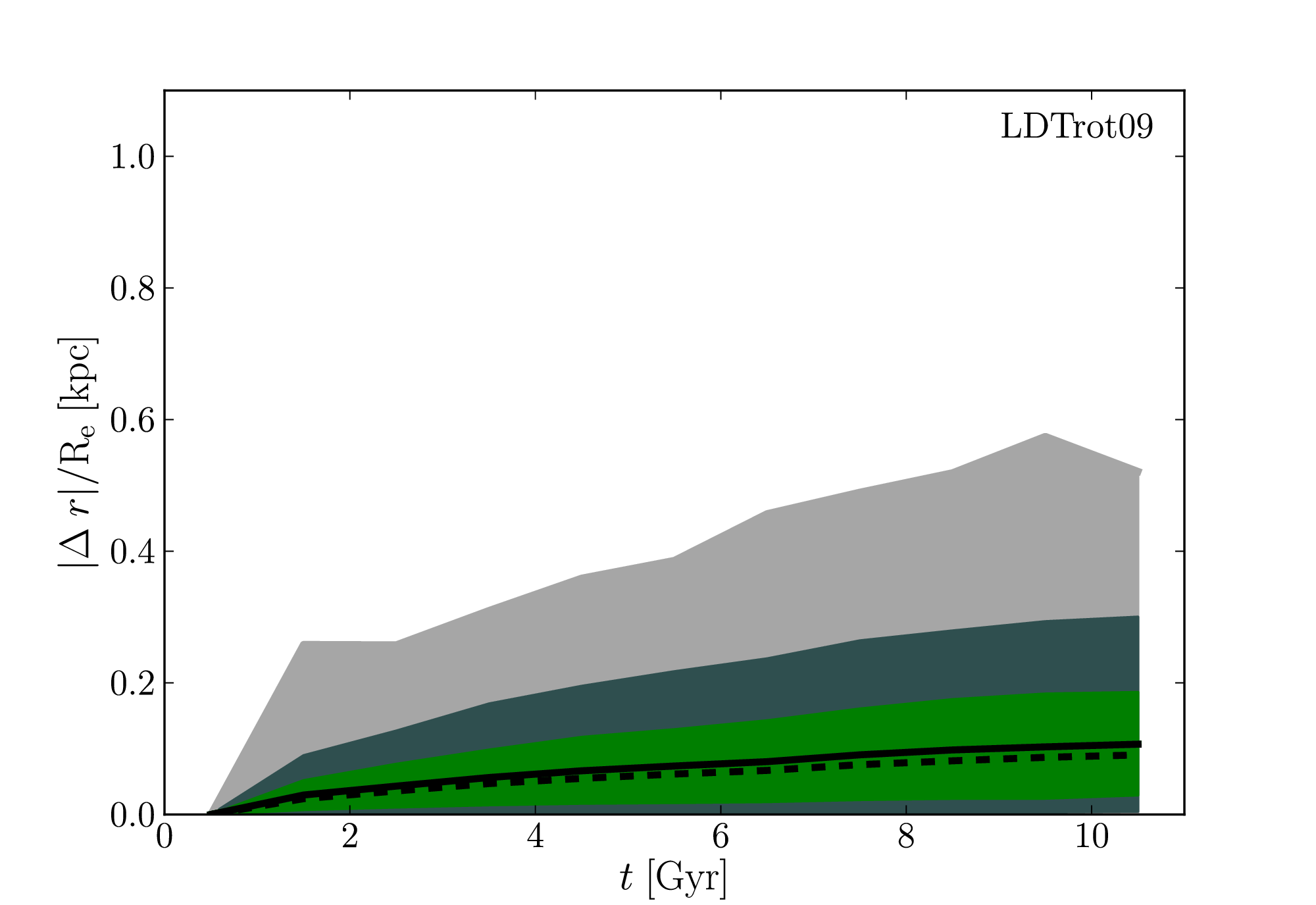}
\hfill

\caption{Radial stellar orbital displacements in the LDT models,
 visualised by the ``absolute'' statistical deviation from the birth
 radius, in function of time since birth (as explained in Section
 \ref{section_stellar_migration}). Radius is expressed in function of
 $\mathrm{R_{e}}$. The grey zone marks the maximum range, while the
 light and darker green respectively represent the 15.9/84.1 and
 2.3/97.7 percentile regions (corresponding to 1$\sigma$ and 2$\sigma$
 if the underlying distribution were Gaussian). The dotted and full line
 show the 50th percentile and the average.  \label{LDT_stelmig_abs}}

\end{minipage}
\end{figure*}

\begin{figure*}
\begin{minipage}[t]{2\columnwidth}
\hfill
\includegraphics[width=0.45\textwidth]{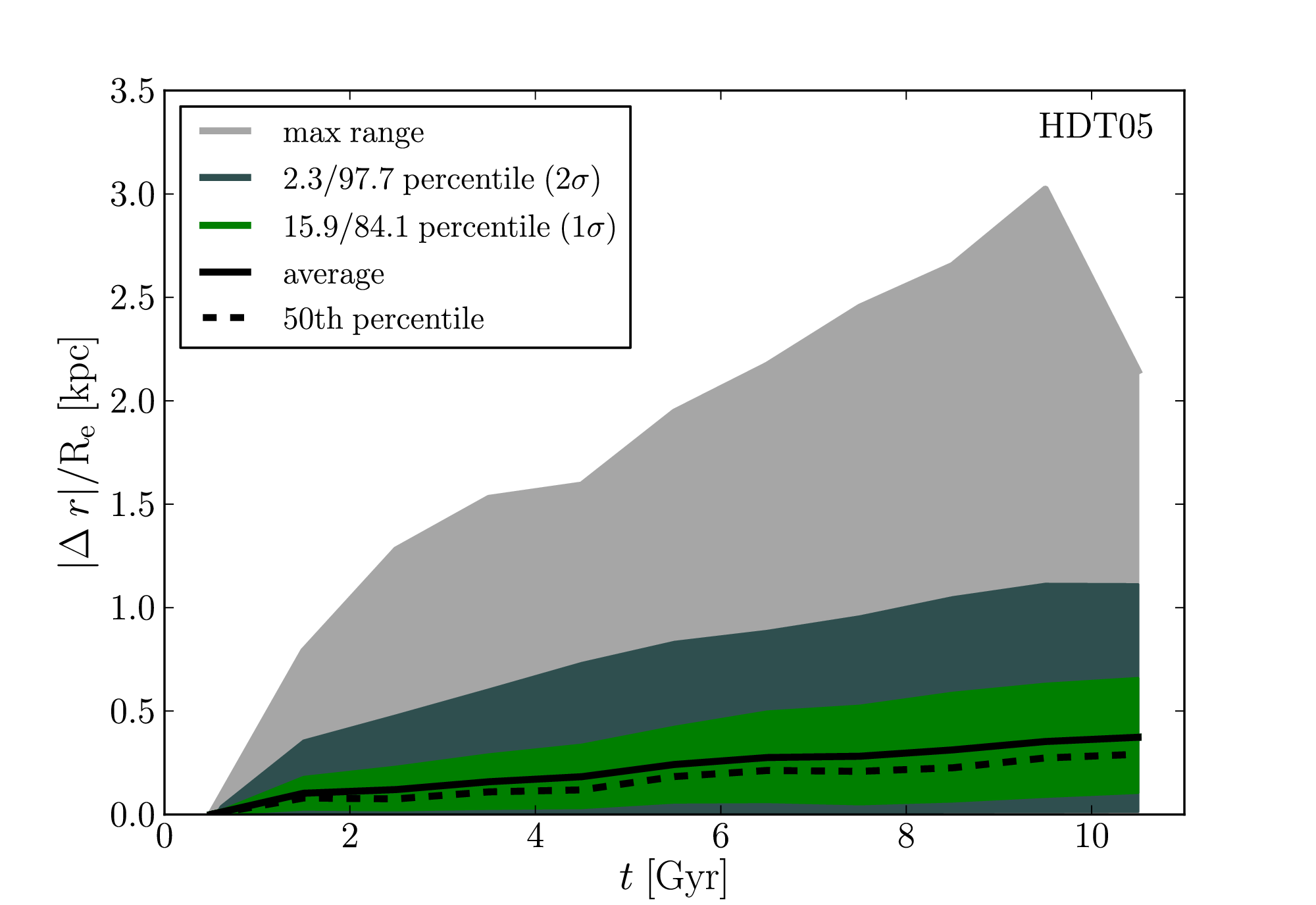}
\hfill
\includegraphics[width=0.45\textwidth]{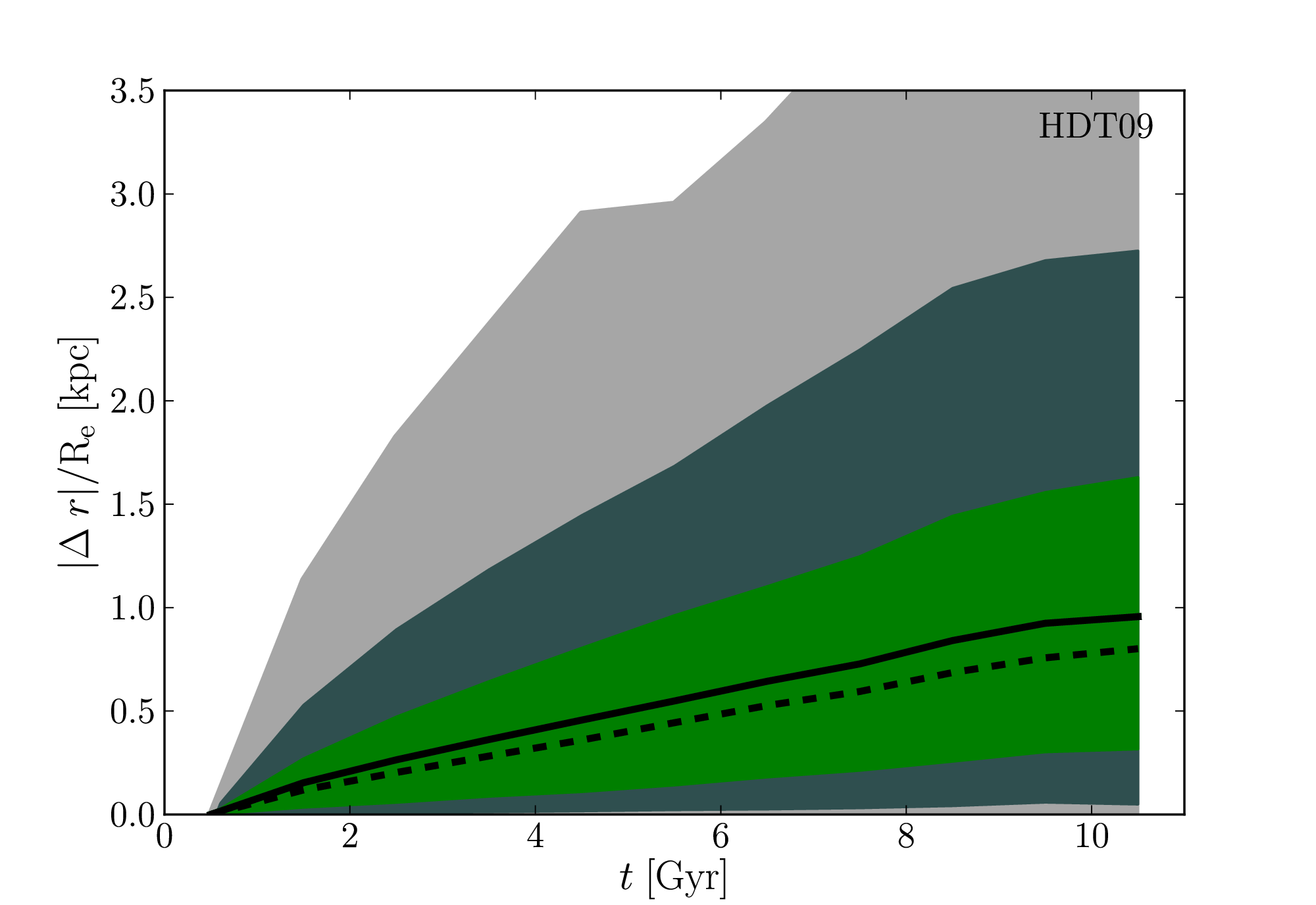}
\hfill
\\
 \hspace*{0mm} 
\hfill
\includegraphics[width=0.45\textwidth]{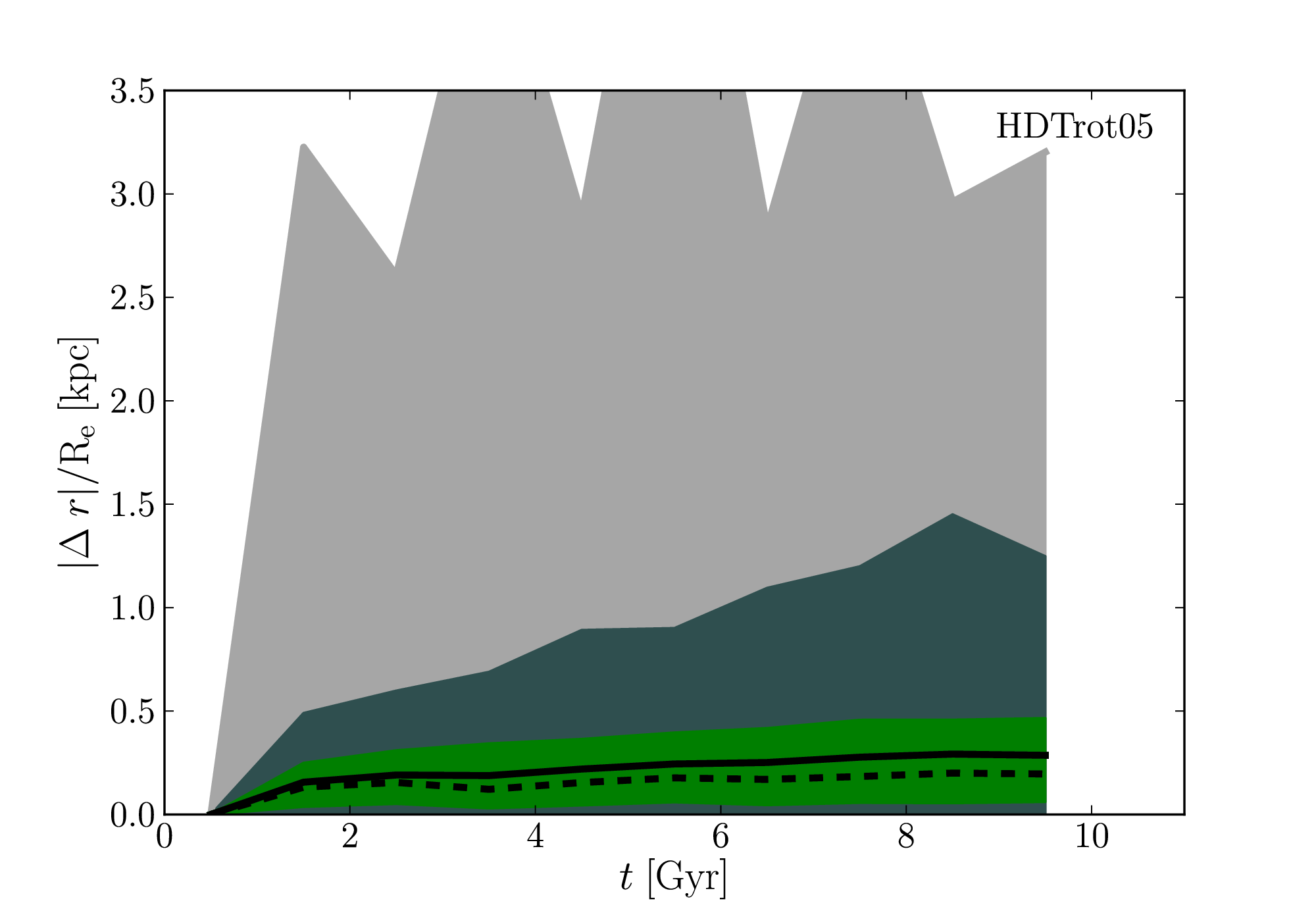}
\hfill
\includegraphics[width=0.45\textwidth]{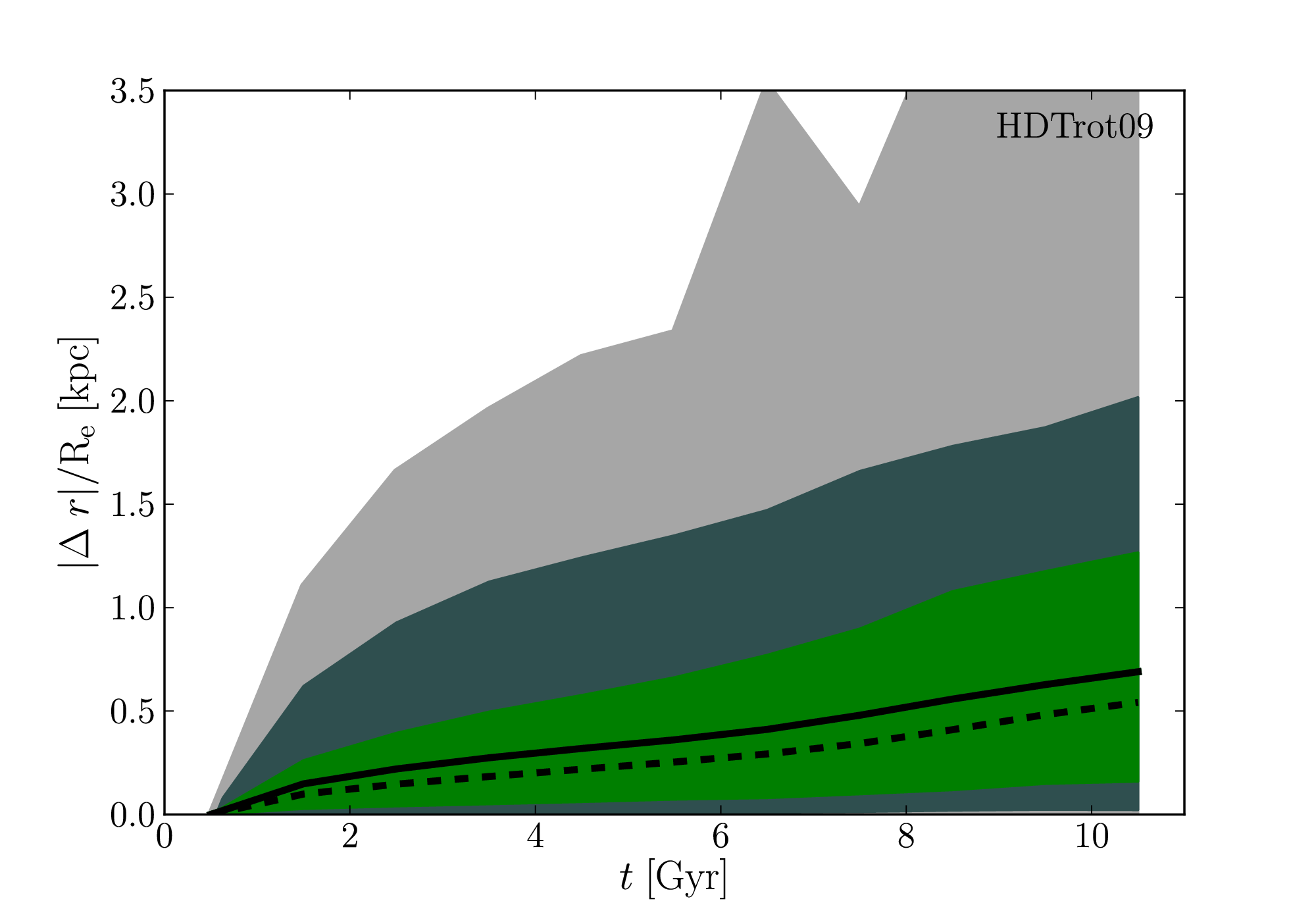}
\hfill

\caption{Radial stellar orbital displacement in the HDT models. Figure
 identical to Fig.~\ref{LDT_stelmig_abs}.  \label{HDT_stelmig_abs}}

\end{minipage}
\end{figure*}

Figure \ref{LDT_stelmig_abs} and \ref{LDT_stelmig_diff} show,
respectively, the ``absolute'' and ``signed'' radial orbital
displacements in our LDT models.

From Fig.~\ref{LDT_stelmig_abs} it is immediately apparent that the
orbital displacement in these models is very limited, to just fractions of
about 0.1 to 0.3 of the $\mathrm{R_{e}}$, on average, over the lifetime
of the simulated dwarf galaxy. Several trends are also noticeable here, :
\begin{itemize}
\item More massive models feature \textit{more} displacement, caused by
      more turbulence and a higher velocity dispersion (properties which
      are set by the mass of the galaxy as the main parameter - see
      Table \ref{obs_sim_comp_table} and \citealt{stinson2007,
      sander:dgmodels, joeri:angmomentum}).
\item Rotating models feature \textit{less} displacement compared to
      non-rotating models with the same mass. The presence of angular
      momentum - and so, ordered motions in the stellar body gaining
      importance over the random motions/turbulence - causes a lower
      velocity dispersion, as can be seen in Table
      \ref{obs_sim_comp_table} and figure 13 of
      \citet{joeri:angmomentum}.
\item The lowering effect of angular momentum gets stronger with
      increasing mass, as evidenced by the fact that the more massive
      LDTrot09 model shows a slightly lower stellar migration than the
      LDTrot05 model. This is because, in higher mass galaxies, ordered
      motions (if they are present) are \textit{inherently} more
      important than the random motions, compared to lower mass
      galaxies. Or in the other direction: the lower the galaxy mass,
      the more dominant random motions/turbulence become over ordered
      motions \citep{kaufmann2007, roy2010, sanchez2010,
      joeri:angmomentum}.
\end{itemize}
In Fig.~\ref{dispersion_populations}, top panel, no clear trends can
be seen in the velocity dispersion over different populations of
stars. The only noticeable property is a possible slight increasing of
the dispersion in all models for the youngest stellar populations, which
can be interpreted as the effect that the (relatively strong) star
formation of the model starts having on its own potential.

\subsection{HDT sims}

Figures \ref{HDT_stelmig_abs} and \ref{HDT_stelmig_diff} show,
respectively, the ``absolute'' and ``signed'' orbital displacements
in the HDT models.

From Fig.~\ref{HDT_stelmig_abs} we can see that in the HDT models the
radial orbital displacement is substantially larger than in the LDT
models - roughly 3 times larger. But still this means only fractions of
the $\mathrm{R_{e}}$, on average, over timespans of several Gyr
(e.g. 0.2 to 0.5~$\mathrm{R_{e}}$ over 5\,Gyr), and in the most massive
model reaching values of the order of the $\mathrm{R_{e}}$ only over the
entire lifespan of the simulated dwarf galaxy. Most of the trends
identified in the LDT models in \ref{section_stelmig_LDT} appear valid
here as well: adding mass increases the radial diffusion, while adding
angular momentum lowers it - although the latter to a slightly lesser
extent than in the LDT models, due to the different rotation curves
(Section \ref{section_rotation_curves}) that deliver less angular
momentum to the central gas. This can be seen in the behaviour of the
velocity dispersion in Table \ref{obs_sim_comp_table} and
Fig.~\ref{dispersion_populations}, bottom frame: there is a clear
difference between the dispersions of models with different mass, but
virtually no difference between models with the same mass but different
angular momenta - contrary to the situation in the top frame for the LDT
models.

\begin{figure*}
\begin{minipage}[t]{2\columnwidth}

\hfill
\includegraphics[width=0.45\textwidth]{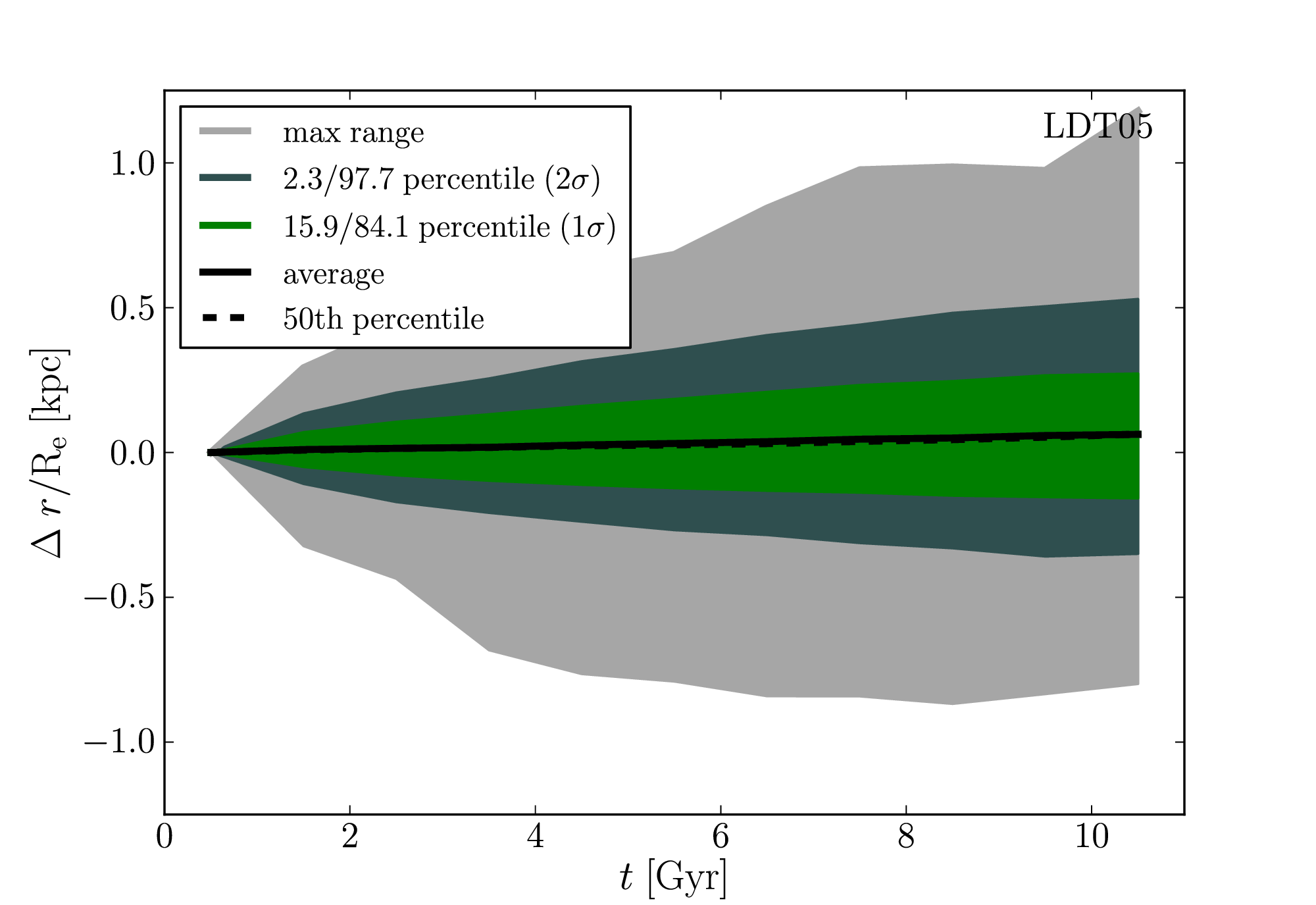}
\hfill
\includegraphics[width=0.45\textwidth]{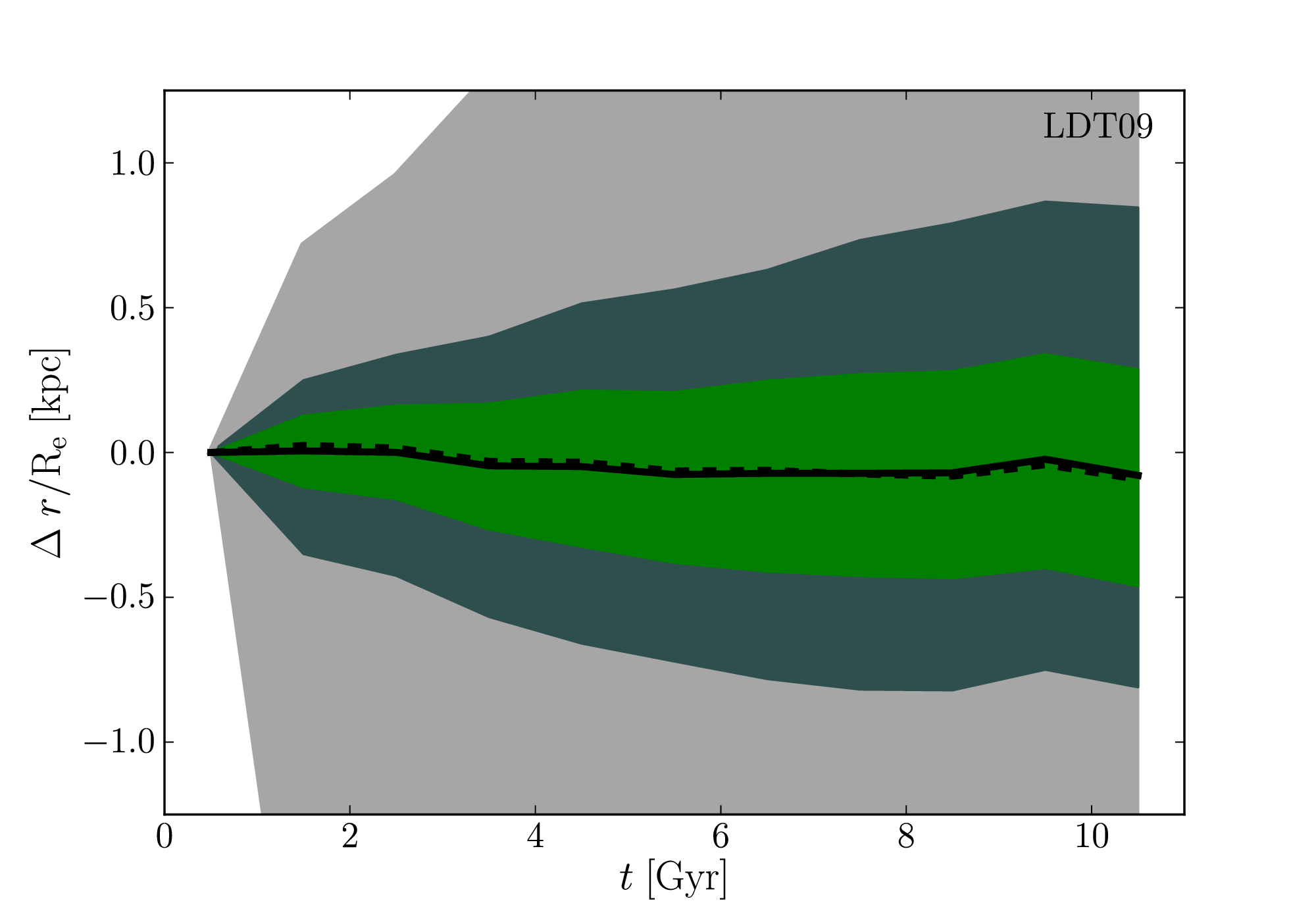}
\hfill
\\
 \hspace*{0mm} 
\hfill
\includegraphics[width=0.45\textwidth]{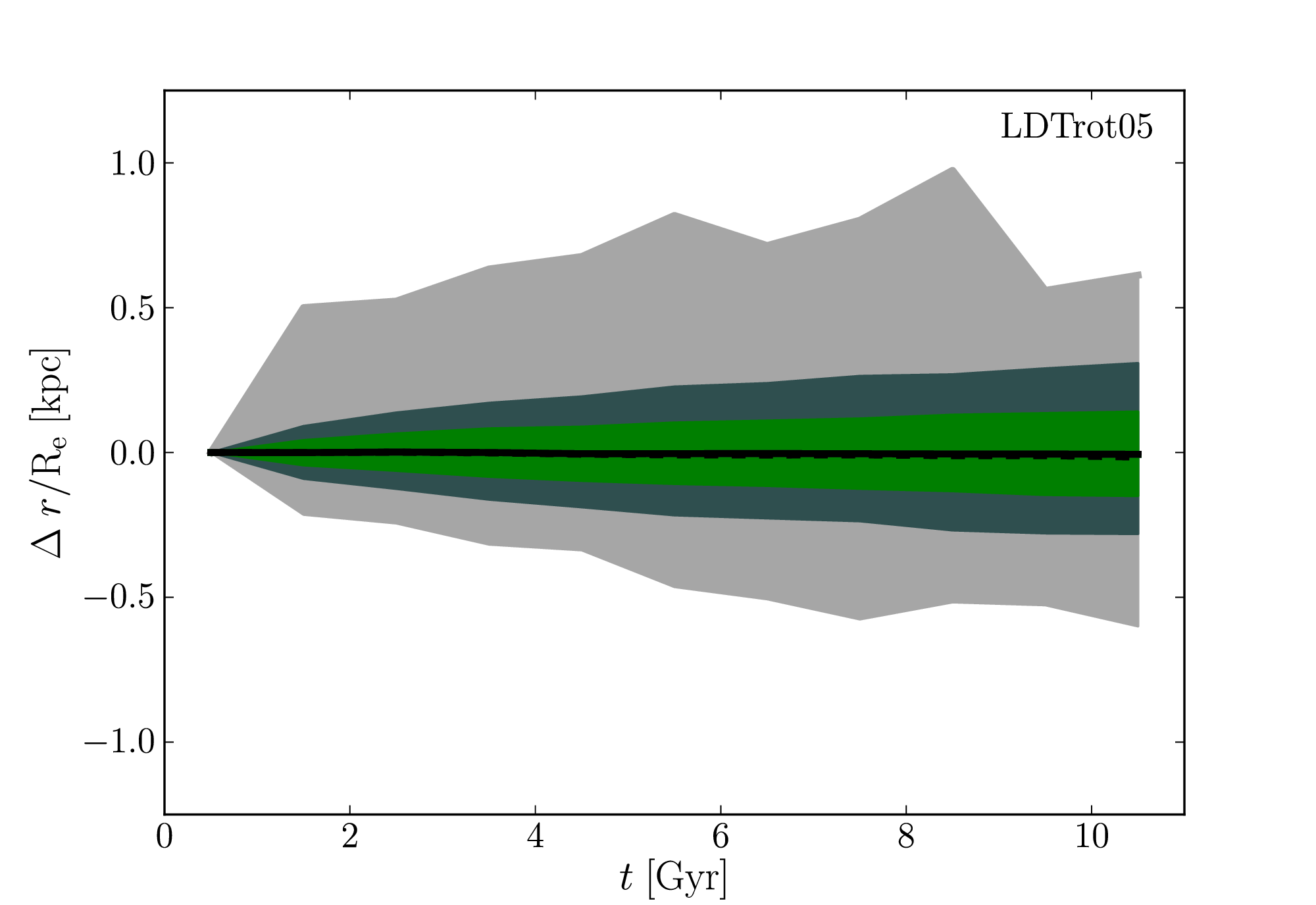}
\hfill
\includegraphics[width=0.45\textwidth]{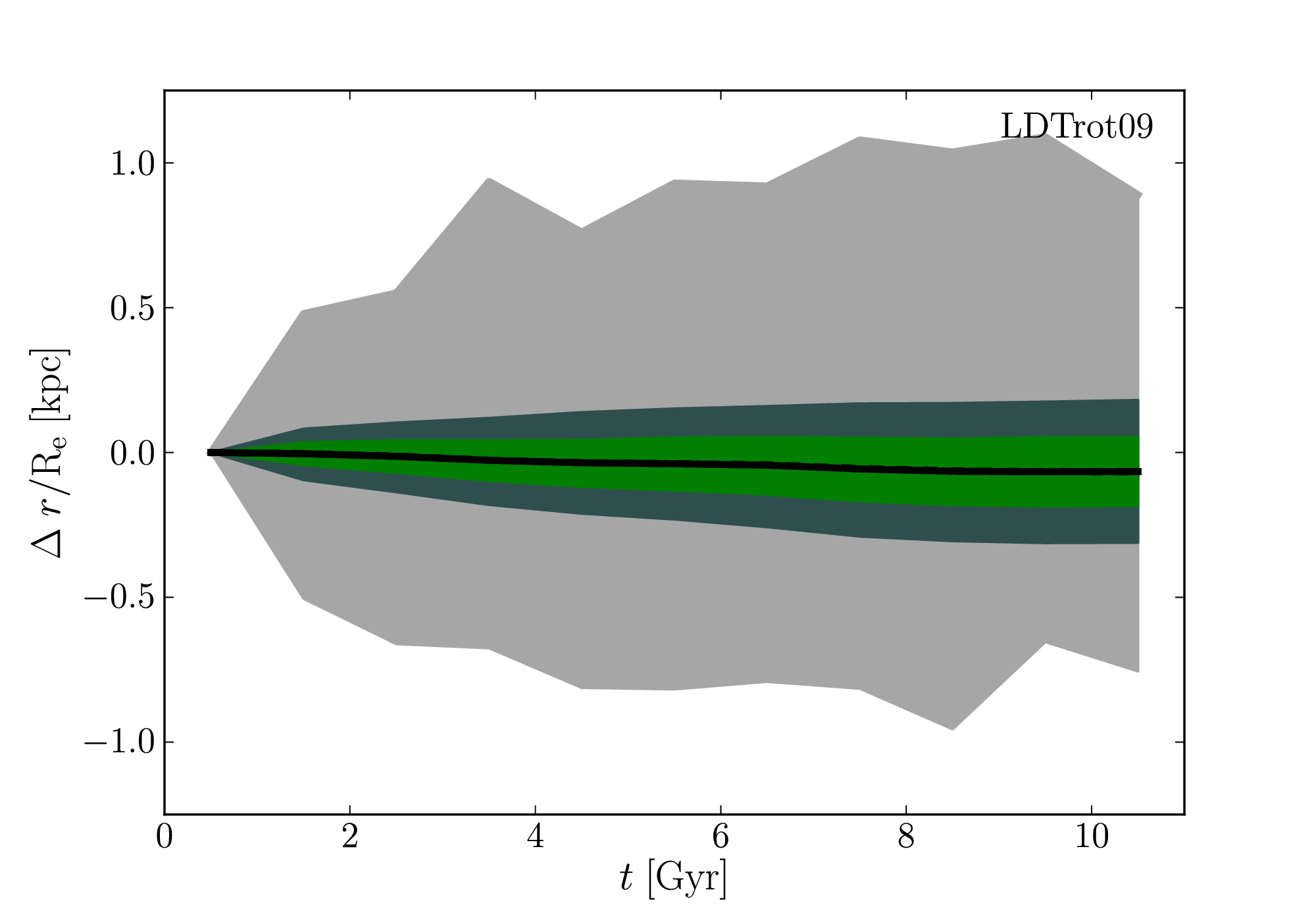}
\hfill

\caption{Radial stellar orbital displacement in the LDT models, here in
 positive or negative distance from the birth radius of the star. Other
 properties of the plots identical to Fig.~\ref{LDT_stelmig_abs}.
 \label{LDT_stelmig_diff}}

\end{minipage}
\end{figure*}

\begin{figure*}
\begin{minipage}[t]{2\columnwidth}
\hfill
\includegraphics[width=0.45\textwidth]{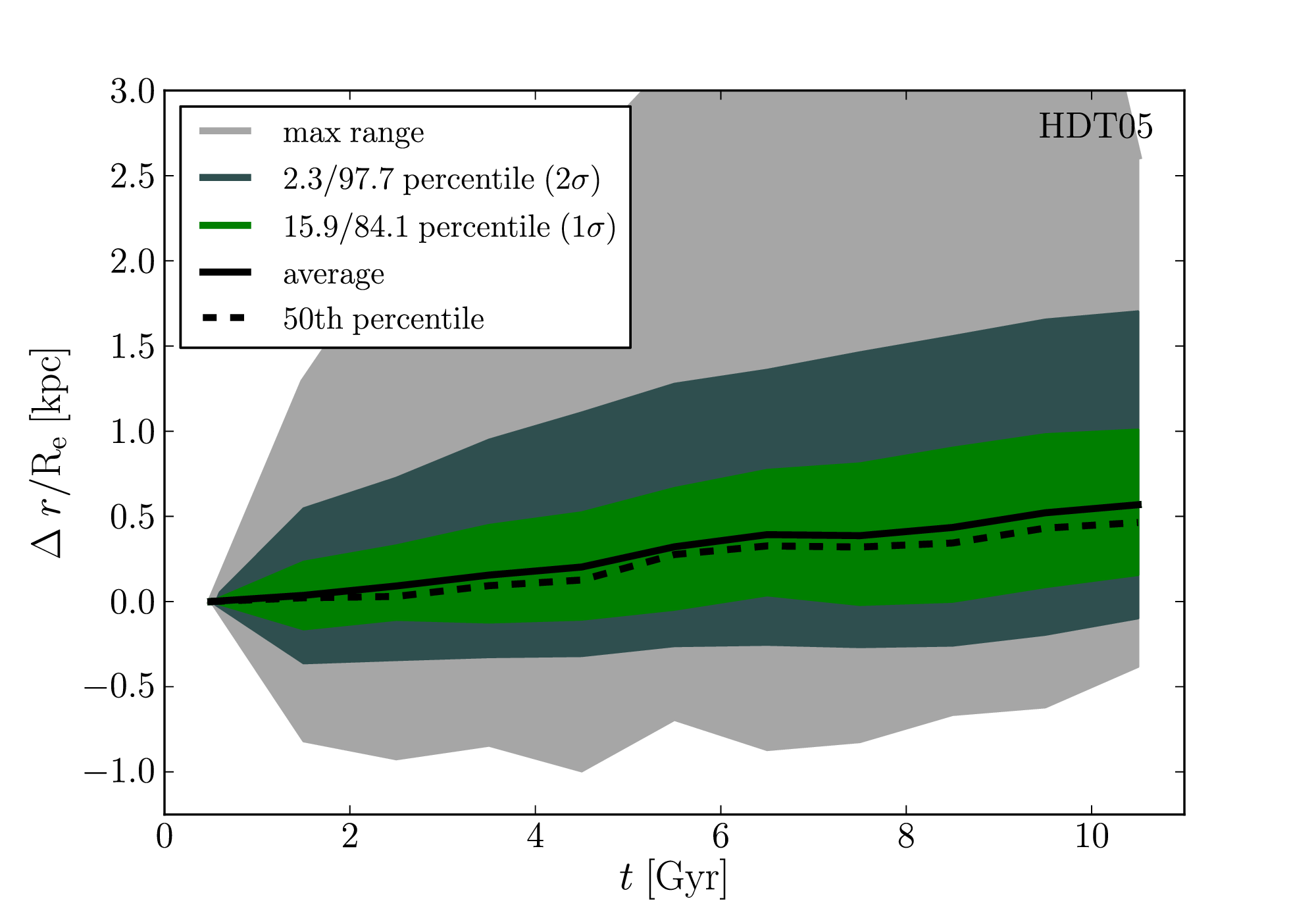}
\hfill
\includegraphics[width=0.45\textwidth]{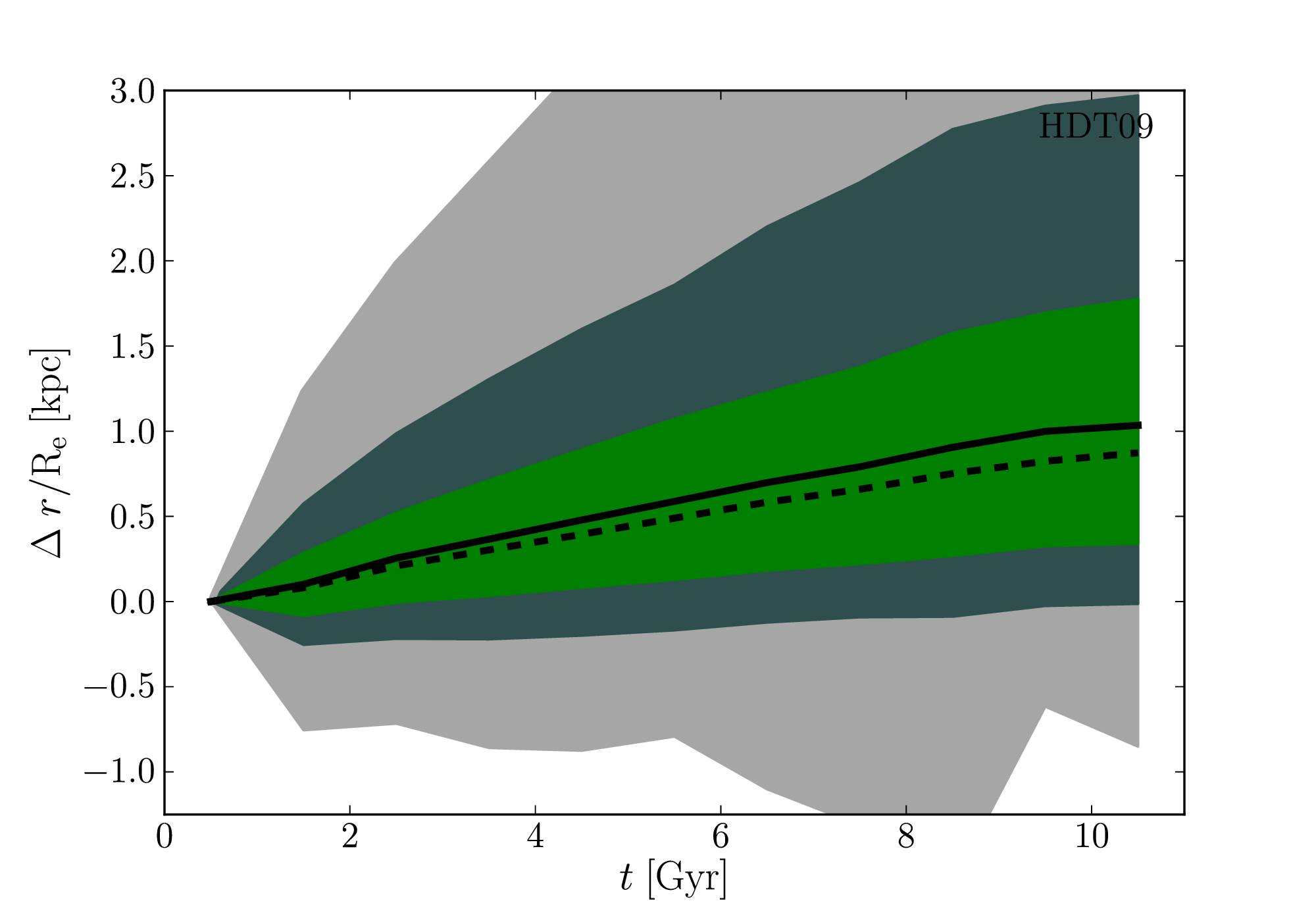}
\hfill
\\
 \hspace*{0mm} 
\hfill
\includegraphics[width=0.45\textwidth]{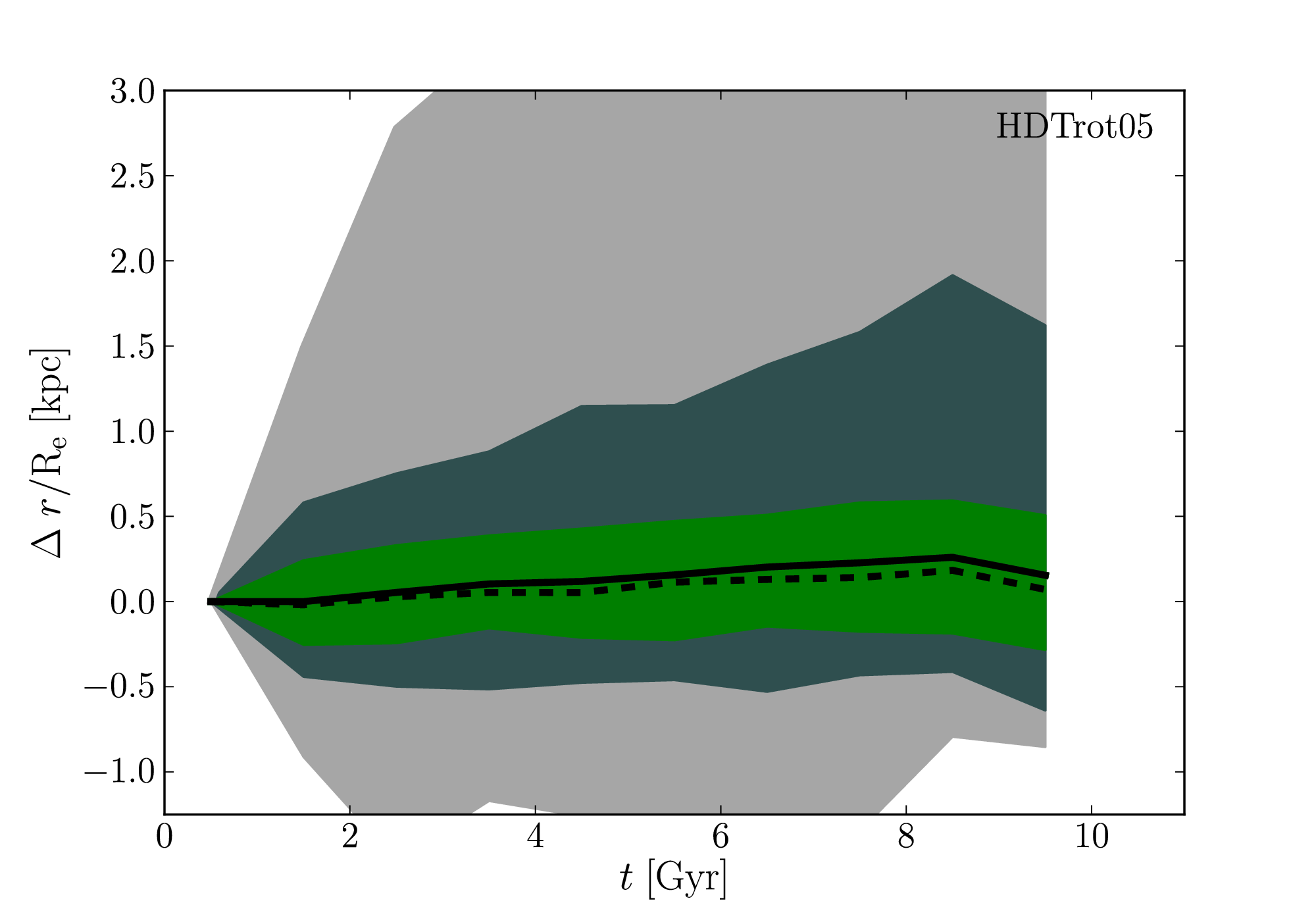}
\hfill
\includegraphics[width=0.45\textwidth]{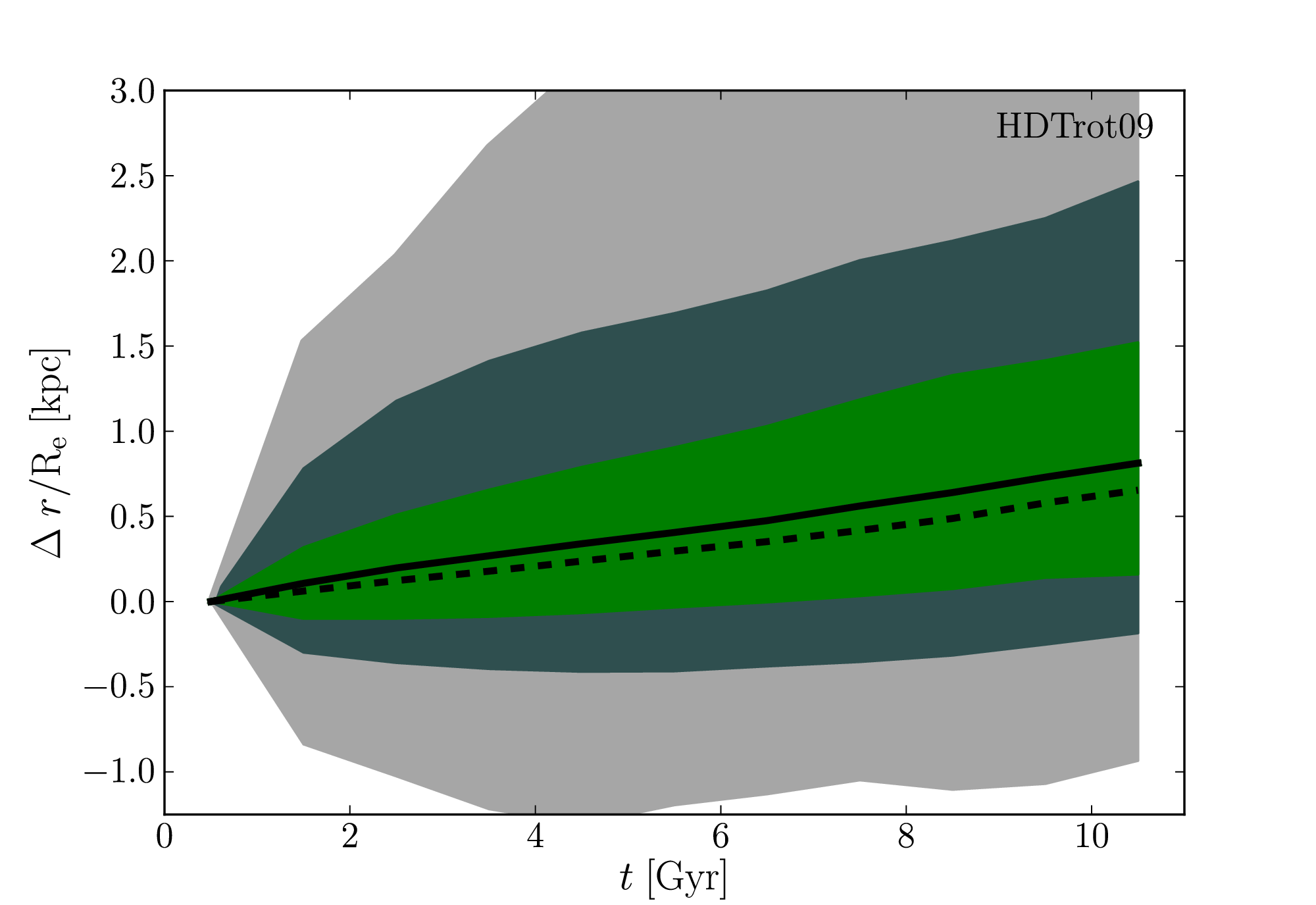}
\hfill

\caption{Radial stellar orbital displacement in the HDT models. Figure identical to
 Fig.~\ref{LDT_stelmig_diff}.
  \label{HDT_stelmig_diff}}

\end{minipage}
\end{figure*}

Figure~\ref{HDT_stelmig_diff} shows that in the gradient-producing
non-rotating HDT models, stellar particles tend to move significantly
outward against the gradient. This is contrary to what is observed in
their LDT counterparts, which show both inward and outward stellar
diffusion (see Fig.~\ref{LDT_stelmig_diff}). There is also an
``equalizing'' effect seen here, where rotation decreases this outward
tendency, though the less massive HDTrot05 model is the only one which
comes anywhere close to showing symmetric (equal inward and outward)
radial orbital displacements.

In Fig.~\ref{dispersion_populations}, bottom panel, there is now a
trend to be seen in the velocity dispersion of the HDT models over
different stellar populations: older populations clearly have larger
velocity dispersions than younger populations.

\subsection{General conclusions on stellar orbits}

The general conclusion here is that radial orbital displacements of
stellar particles are limited in our dwarf galaxy models, being
generally measured in fractions of the half light radius over time spans
of 5 to 10\,Gyr. This qualitative statement is true, independent of the
employed star-formation scheme. This gives us a more solid foundation
for the results concerning the stability of metallicity gradients.

Since this is the case for dwarf galaxy models with either low or high
angular momenta, this strongly indicates that there are no radically
orbit-changing processes at work in dwarf galaxies like there are in
bigger galaxies. In massive spiral galaxies there is the so-called
``radial stellar migration'', which is caused by gravitational
interactions between individual stars and the large-scale spiral
structures rotating in the stellar and gaseous discs that drive stars
away from the corotation radius \citep{sellwood2002, roskar2008,
roskar2012}. This is able to move stars about over large radial
distances while maintaining quasi-circular orbits, and it seems the
evidence, both theoretical and observational, is building up for this
being an effect of broad importance in disc galaxy evolution : it plays
a fundamental role in the forming of thick discs,
\citep{schonrich2009,loebman2011}, the distribution of stellar
populations, \citep{roskar2008}, and has evidently strong implications
for the forming/survival of metallicity gradients
\citep{lepine2003}. None of our models show an appreciable disk or
spiral structures in their stellar body, because it is generally too
thick, dynamically too warm, and influenced by the random motions of the
gas in the low mass regime that we are investigating here
\citep{kaufmann2007, roy2010, sanchez2010}. Only the rotating dwarfs
feature a somewhat flattened stellar body and fast-transient spiral
structures in their gas distribution, but these structures are produced
by outwardly expanding supernova-blown bubbles being deformed by shear,
which are too short-lived and fade away too quickly to have any impact
on the stellar dynamics of the models. What - if anything - is happening
with the orbits in these dwarf galaxy models is most likely linked to
the much more gentle effect of ``dynamical heating'', that changes
stellar orbits in a more gradual manner. It was already suggested by
\citet{spitzer53} that, for instance, massive gas clouds could have a
noticeable influence on the random velocities of stellar orbits.

\begin{figure}

\includegraphics[width=0.45\textwidth]{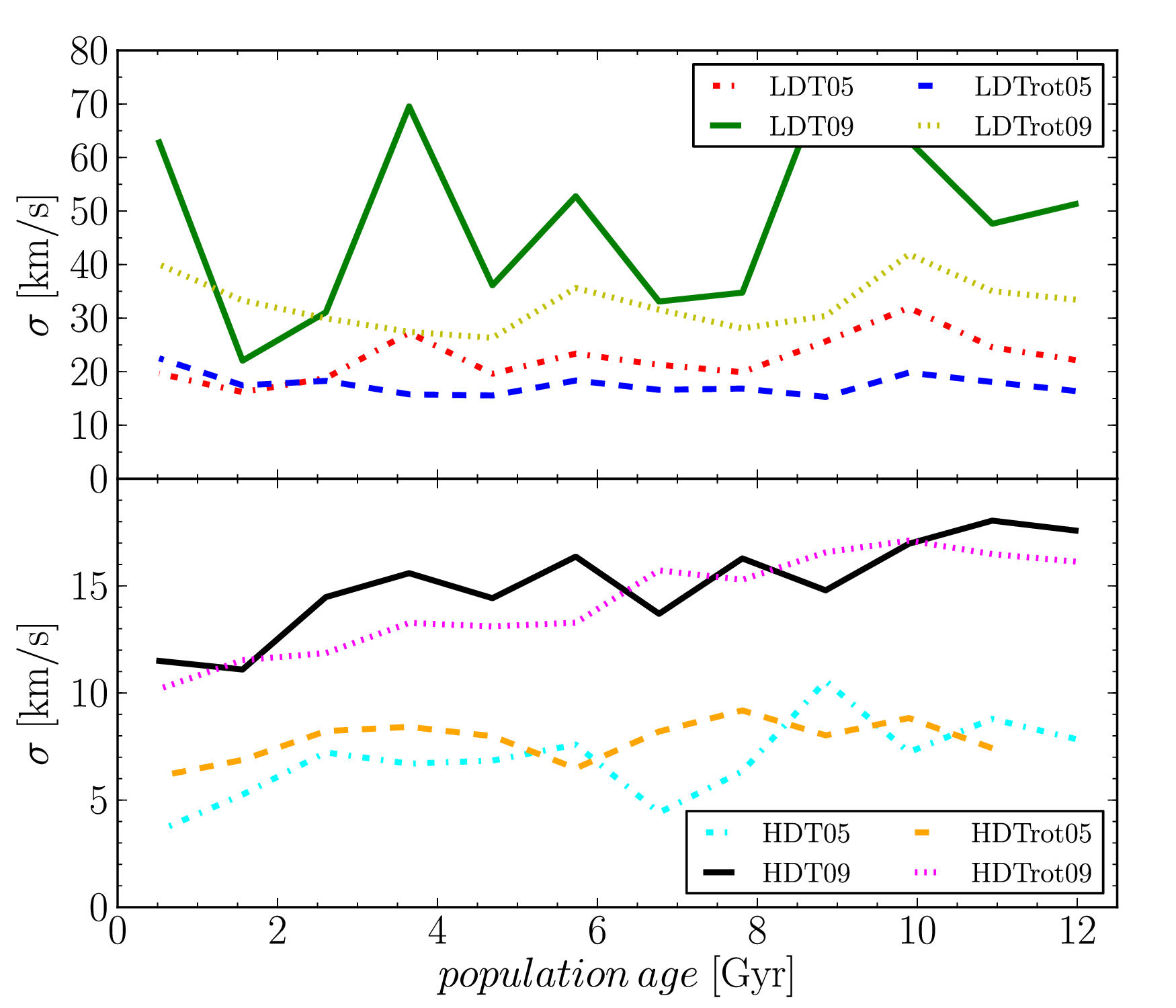}
\caption{Velocity dispersions of stellar populations of different ages
in the dwarf galaxy models, the upper and lower panel respectively show
the LDT models and the HDT models. For all models the last snapshots
were taken, their stellar populations divided into 12 equal age bins,
and for each age bin the velocity dispersion was calculated in a box
with width $R_{e}$ around the center of the galaxy. }

\label{dispersion_populations}

\end{figure}

Quantitatively, the star formation criteria do play a role, however. The
HDT models show noticeably more stellar diffusion than the LDT models,
who barely undergo any orbital displacement at all over their entire
simulation time.  This increased diffusion is also more strongly aimed
outward - agains the gradient - though it is still limited in absolute
terms. Two likely causes are the more turbulent character of the gas in
general (which is adopted by the stellar body), and the increased
scattering of stellar particles off dense gas clumps. Both effects are
expected to become more important when the density threshold for star
formation increases, and will therefore increase the strength of
dynamical heating of the stellar body in the HDT models. This is all
visualized in Fig.~\ref{dispersion_populations}, where the velocity
disperions of different stellar populations in the last snapshot of each
simulation are shown. Gradual dynamical heating of the stellar body
would be expected to leave a footprint here, by causing the older
populations (that have been dynamically heated longer) to have larger
velocity dispersions than the younger populations. The LDT models,
having extremely small orbital displacements, do not show any trend like
this in the upper panel of Fig.~\ref{dispersion_populations}. Besides
the scatter on the plots, which is probably due to the model's star
formation peaks that influence its own potential, the velocity
dispersion is roughly similar for populations of all ages.  The HDT
models, on the other hand, all clearly show this dynamical heating
footprint on their velocity dispersions in the lower panel of
Fig.~\ref{dispersion_populations}.  


%% file: article_conclusion.tex
\section{Conclusion}
\label{section_conclusion}

In this paper we investigated how, in simulated dwarf galaxies in
isolation, metallicity gradients are formed, how they evolve and how,
once formed, they can be maintained. Furthermore, we adressed the
importance of dynamical orbit-changing processes in the dwarf galaxy
regime, and their potential in erasing or weakening existing population
gradients. We hereby also investigated the role of the density threshold
for star formation.

Firstly, in Section \ref{section_metallicity_profiles}, we found that
metallicity gradients are gradually built up during the evolution of
non-rotating dwarf galaxy models by ever more centrally concentrated
star formation adding to the overall gradient. On themselves, the formed
gradients easily survive and their strength hardly declines over several
Gyr, indicating that only external disturbances would be able to
significantly weaken or erase population gradients in dwarf
galaxies. The metallicity gradients produced by our dwarf galaxy
models are found to agree well with observed metallicity gradients of
dwarf galaxies in the Local Group, both in shapes and slopes.

Secondly, from Section \ref{section_stellar_migration} we conclude
that the orbital displacements of the stars are quite limited in our
models, of the order of only fractions of the half light radius over
time-spans of 5 to 10\,Gyr in both our rotating and non-rotating
models. This is contrary to what is found in massive disc galaxies,
where scattering of stars off the corotation resonance of large-scale
spiral structure can cause significant radial migration. The absence of
long-lived, major spiral structures in the simulated dwarf galaxies
leaves only turbulent gas motions and scattering off dense gas clouds as
scattering agents of stars, leading to an only mild dynamical heating of
the stellar body that allows for the long-term survival of population
gradients.

Finally, increasing the density threshold for star formation from $ 0.1$
to $ 100~ \mathrm{amu/cm^{-3}} $, which - together with increased
feedback efficiency and novel cooling curves below $10^{4}$\,K -
represents a much more realistic description of star forming regions,
has profound influences on the mode of star formation in our models. It
produces high density, cold, star forming clumps, shorter star formation
timescales, and lower stellar masses. On the matter of population
gradient evolution and orbital displacements it also has a clear
influence, producing stronger dynamical heating of the stellar particles
which is seen through larger orbital displacements and clear trends in
the velocity dispersion over different stellar populations. In absolute
terms, however, the effect of this dynamical heating remains very
limited.